\documentclass{article}
\usepackage{citesort} 

\usepackage{times}
\usepackage[latin1]{inputenc}
\usepackage{amsmath}
\usepackage{amssymb}
\usepackage{latexsym}
\usepackage{theorem}
\usepackage{url}

\makeatletter
\newcommand{\pname}[1]{\item[{#1}.]\def\@currentlabel{#1}}
\makeatother

\newcommand{\ab}{\allowbreak}

\newcommand{\node}{\eta}

\newcommand{\fbegin}[1]{#1^{\mathrm{m-event}}}
\newcommand{\cutlast}[1]{{#1}\lceil}
\newcommand{\onep}{p_{\mathrm{c}}}
\newcommand{\oneq}{q_{\mathrm{c}}}
\newcommand{\indp}[2]{\mathrm{makejk}({#1},{#2})}

\newcommand{\sand}{\mathop\bigwedge}
\newcommand{\sor}{\mathop\bigvee}
\newcommand{\inj}{\mathrm{inj}}
\newcommand{\injopt}{[\mathrm{inj}]}

\newcommand{\host}{\kw{host}}
\newcommand{\getkey}{\kw{getkey}}
\newcommand{\nmrsign}{\kw{nmrsign}}
\newcommand{\nmrtrue}{\kw{true}}
\newcommand{\nmrchecksign}{\kw{nmrchecksign}}
\newcommand{\mac}{\kw{mac}}
\newcommand{\cto}{\kw{ct}}
\newcommand{\pred}{\mathit{pred}}
\newcommand{\patts}{\kw{patterns}}
\newcommand{\subg}{\kw{sub}}
\newcommand{\taggen}{\kw{tagGen}}

\newcommand{\Bp}{{\cal R}_{\mathrm{ProtAdv}}}
\newcommand{\Bpr}{{\cal R}_{\mathrm{Protocol}}}
\newcommand{\Bpconstr}{{\cal R}_{\mathrm{Constr}}}
\newcommand{\Bpdestr}{{\cal R}_{\mathrm{Destr}}}

\newcommand{\Bzero}{{\cal R}_0}
\newcommand{\params}[1]{\kw{Params}_{#1}}
\newcommand{\tset}{{\cal M}}
\newcommand{\idc}{\mathrm{Id}_0}
\newcommand{\ct}[1]{\kw{ct}_{#1}}
\newcommand{\size}{\kw{size}}
\newcommand{\condense}{\kw{condense}}

\newcommand{\Id}{\mathrm{Id}}

\newcommand{\trace}{{\cal T}}

\newtheorem{lemma}{Lemma}
\newtheorem{theorem}{Theorem}
\newtheorem{corollary}{Corollary}
\newtheorem{proposition}{Proposition}
\theorembodyfont{\rmfamily}
\newtheorem{definition}{Definition}
\newtheorem{example}{Example}
\newtheorem{Remark}{Remark}
\newtheorem{proof}{Proof}
\newtheorem{proofsk}{Proof sketch}

\def\und{\leavevmode{\kern0.03em\vbox{\hrule width0.47em}\kern0.03em}}

\makeatletter
\def\judg{\@ifnextchar [{\@judgAB}{\@judgB}}
\def\@judgAB[#1]#2{#1\vdash#2\@isC}
\def\@judgB#1{\emset\vdash#1\@isC}
\def\@isC{\@ifnextchar [{\@yesC}{}}
\def\@yesC[#1]{\undefinedcase #1}
\makeatother

\newcommand{\emset}{\emptyset}
\newcommand{\ty}{\mathbin{:}}

\newenvironment{restate}[1]%
  {\begin{trivlist}\item[]{\normalsize\bf Proof of #1}\hspace*{4mm}\it}%
  {\end{trivlist}}

\newcommand{\proofcomplete}{\hspace*{\fill}$\Box$}

\newenvironment{defn}{\begin{tabbing}
  \hspace{1.5em} \= \hspace{.27\linewidth - 1.5em} \= \hspace{1.5em} \= \kill
  }{
  \end{tabbing}}

\newcommand{\entry}[2]{\>$#1$\>\>#2}
\newcommand{\clause}[2]{$#1$\>\>#2}
\newcommand{\categ}[2]{\clause{#1::=}{#2}}

\newcommand{\kw}[1]{\mathit{#1}}

\newcommand{\fn}{\kw{fn}}               % free names
\newcommand{\fv}{\kw{fv}}               % free variables
\newcommand{\dom}{\kw{dom}}

\newcommand{\IF}{\kw{if}\ }
\newcommand{\THEN}{\ \kw{then}\ }
\newcommand{\ELSE}{\ \kw{else}\ }
\newcommand{\LET}{\kw{let}\ }
\newcommand{\IN}{\ \kw{in}\ }

\newcommand{\Res}[1]{(\nu#1)}

\newcommand{\parpop}{\mid}
\newcommand{\Repl}[1]{\mathord{! #1}}
\newcommand{\aguard}[2]{\IF{#1} = {#2}\THEN}
\newcommand{\guard}[4]{\aguard{#1}{#2}{#3}\ELSE{#4}}
\newcommand{\cinput}[2]{{#1}({#2})}
\newcommand{\coutput}[2]{\overline{#1}\langle{#2}\rangle}
\newcommand{\interv}[2]{\{ {#1}, \ldots, {#2}\}}
\newcommand{\aletfun}[2]{\LET{#1}={#2}\IN}
\newcommand{\letfun}[4]{\aletfun{#1}{#2}{#3}\ELSE{#4}}
\newcommand{\aletdef}[2]{\LET{#1}={#2}\IN}
\newcommand{\letdef}[3]{\aletdef{#1}{#2}{#3}}
\newcommand{\asevent}{\mathtt{event}}
\newcommand{\pasevent}{\mathrm{event}}

\newcommand{\skA}{\mathit{sk}_A}
\newcommand{\pkA}{\mathit{pk}_A}
\newcommand{\skB}{\mathit{sk}_B}
\newcommand{\pkB}{\mathit{pk}_B}

\newcommand{\sAa}{\mathit{sAa}}
\newcommand{\sAb}{\mathit{sAb}}
\newcommand{\sBa}{\mathit{sBa}}
\newcommand{\sBb}{\mathit{sBb}}

\newcommand{\session}{\mathrm{session}}

\newcommand{\last}{\mathrm{last}}

\newcommand{\rw}{\kw{Init}}

\newcommand{\ntuple}[1]{{#1}\kw{tuple}}
\newcommand{\nth}[2]{{#1}\kw{th}_{#2}}

\newcommand{\sencrypt}{\kw{sencrypt}}
\newcommand{\sdecrypt}{\kw{sdecrypt}}
\newcommand{\hash}{h}

\newcommand{\sencryptp}{\kw{sencrypt}_{\kw{p}}}
\newcommand{\sdecryptp}{\kw{sdecrypt}_{\kw{p}}}

\newcommand{\pk}{\kw{pk}}
\newcommand{\pencrypt}{\kw{pencrypt}_{\kw{p}}}
\newcommand{\pdecrypt}{\kw{pdecrypt}_{\kw{p}}}

\newcommand{\sign}{\kw{sign}}
\newcommand{\checksign}{\kw{checksignature}}
\newcommand{\getmess}{\kw{getmessage}}

\newcommand{\defg}{\mathrm{def}(g)}

\newcommand{\gdh}{\mathit{b}}

\newcommand{\rename}{\theta}

\newcommand{\act}{\alpha}
\newcommand{\acti}{F}  

\newcommand{\pasbegin}{\mathrm{m\text{-}event}}
\newcommand{\pasend}{\mathrm{event}}
\newcommand{\attacker}{\mathrm{attacker}}
\newcommand{\mess}{\mathrm{message}}
\newcommand{\comp}{\mathrm{comp}}

\newcommand{\rewrite}{\Rightarrow}
\newcommand{\lp}{[\![}
\newcommand{\rp}{]\!]}
\newcommand{\emptyseq}{\emptyset}
\newcommand{\der}{{\cal F}_{P'_0, \rw}}
\newcommand{\comment}[1]{\tag{\textrm{#1}}}

\newcommand{\equal}{\mathit{equal}}
\newcommand{\envsigmarho}{\sigma\rho}
\newcommand{\labelset}{\mathit{Label}}
\newcommand{\clabel}{\mathit{SId}}
\newcommand{\inject}{\phi^{\mathrm{i}}}

\newcommand{\saturate}{\mathsf{saturate}}
\newcommand{\sel}{\mathsf{sel}}
\newcommand{\impl}{\sqsupseteq}
\newcommand{\deriv}{\mathsf{deriv}}

\newcommand{\reach}{\mathsf{derivable}}
\newcommand{\solve}[1]{\mathsf{solve}_{#1}}
\newcommand{\solvep}[1]{\mathsf{solve}'_{#1}}
\newcommand{\decomp}{\kw{decomp}}
\newcommand{\decomphyp}{\kw{decomphyp}}
\newcommand{\elimdup}{\kw{elimdup}}
\newcommand{\elimtaut}{\kw{elimtaut}}
\newcommand{\elimattx}{\kw{elimattx}}
\newcommand{\elimnot}{\kw{elimnot}}
\newcommand{\fnot}{{\cal F}_{\mathrm{not}}}
\newcommand{\elimredundanthyp}{\kw{elimredundanthyp}}
\newcommand{\simplify}{\kw{simplify}}
\newcommand{\elim}{\kw{elim}}

\newcommand{\satstset}{{\cal R}_0}
\newcommand{\sattmp}{{\cal R}}
\newcommand{\satend}{{\cal R}_1}
\newcommand{\derivstset}{{\cal R}_1}
\newcommand{\derivtmp}{{\cal R}}
\newcommand{\derivtmpp}{{\cal R}'}
\newcommand{\beginset}{{\cal F}_{\mathrm{me}}}
\newcommand{\eventset}{{\cal E}}
\newcommand{\rset}[1]{{\cal R}_{#1}}
\newcommand{\rsetp}[1]{{\cal R}'_{#1}}

\newcommand{\ResInstr}[2]{\Res{#1\ty #2}}
\newcommand{\ReplInstr}[2]{\Repl{^{#1} {#2}}}

\newcommand{\instr}[1]{\mathrm{instr}({#1})}
\newcommand{\instradv}[1]{\mathrm{instrAdv}({#1})}
\newcommand{\instrinj}[1]{\mathrm{instr}'({#1})}
\newcommand{\rlbl}{\ell}
\newcommand{\advnfs}{b_0}

\newcommand{\rlabel}{\mathit{label}}

\newcommand{\delete}{\mathrm{unInstr}}

\newcommand{\env}{E}
\newcommand{\pset}{{\cal P}}
\newcommand{\cfalse}{\mathrm{false}}

\newcommand{\step}{\tau}

\title{Automatic Verification of Correspondences\\
for Security Protocols\thanks{This paper is an updated and extended version of~\cite{Blanchet2001} and~\cite{Blanchet2002}.}
}

\author{Bruno Blanchet\\
CNRS, {\'E}cole Normale Sup{\'e}rieure, INRIA\thanks{This research has been done within the INRIA ABSTRACTION project-team (common with the CNRS and the {\'E}NS).}\\
{\tt Bruno.Blanchet\string@ens.fr}
}

\begin{document}

\maketitle
\sloppy

\begin{abstract}
We present a new technique for verifying correspondences in security
protocols. 
In particular, correspondences can be used to formalize authentication.
Our technique is fully automatic, it can handle an
unbounded number of sessions of the protocol, and it is efficient in
practice. 
It significantly extends a previous technique for the verification
of secrecy. 
The protocol is represented in an extension of the pi calculus with
fairly arbitrary cryptographic primitives. This protocol
representation includes the specification of the correspondence to be
verified, but no other annotation. This representation is then
translated into an abstract representation by Horn clauses, which is
used to prove the desired correspondence.  Our technique has
been proved correct and implemented. We have tested it on various
protocols from the literature. The experimental results show that
these protocols can be verified by our technique in less than 1~s.
\end{abstract}

\section{Introduction}

The verification of security protocols has already been the
subject of numerous research works. It is particularly
important since the design of protocols is error-prone, and
errors cannot be detected by testing, since they appear only in the
presence of a malicious adversary.  An important trend in this area aims
to verify protocols in the so-called Dolev-Yao model~\cite{Dolev83}, 
with an unbounded number of sessions, while
relying as little as possible on human intervention. While
protocol insecurity is NP-complete for a bounded number of sessions~\cite{Rusinowitch03}, it is undecidable for an unbounded number of sessions~\cite{Durgin04}. Hence, automatic verification for an unbounded number of sessions cannot be achieved for all protocols. It is typically 
achieved using language-based techniques such as typing or abstract
interpretation, which can handle infinite-state systems thanks to safe
approximations. These techniques are not complete (a correct protocol
can fail to typecheck, or false attacks can be found by abstract
interpretation tools), but they are sound (when they do not find
attacks, the protocol is guaranteed to satisfy the considered
property). This is important for the certification of protocols.

Our goal in this paper is to extend previous work in this line of
research by providing a fully automatic technique for verifying
correspondences in security protocols, without bounding the number
of sessions of the protocol. Correspondences are properties
of the form: if the protocol executes some event, then it must have
executed some other events before\footnote{In the CSP terminology, our events correspond to CSP signal events.}. 
We consider a rich language of correspondences, in which the events
that must have been executed can be described by a logical formula
containing conjunctions and disjunctions. Furthermore, we consider
both non-injective correspondences (if the protocol executes some
event, then it must have executed some other events at least once) and
injective correspondences (if the protocol executes some event $n$
times, then it must have executed some other events at least $n$
times).
Correspondences, initially named correspondence assertions~\cite{Woo93}, 
and the similar notion
of agreement~\cite{Lowe97} were first introduced to model
authentication. Intuitively, a protocol authenticates 
$A$ to $B$ if, when $B$ thinks he talks to $A$, then he actually talks
to $A$. 
When $B$ thinks he has run the protocol with $A$, he
executes an event $e(A,B)$. When $A$ thinks she runs the
protocol with $B$, she executes another event
$e'(A,B)$. Authentication is satisfied when, if $B$ executes his
event $e(A,B)$, then $A$ has executed her event $e'(A,B)$. 
Several variants along this scheme appear in the
literature and, as we show below, our
technique can handle most of them. 
Our correspondences can also encode secrecy, as follows. A protocol
preserves the secrecy of some value $M$ when the adversary cannot
obtain $M$. We associate an ``event'' $\attacker(M)$ to the fact that
the adversary obtains $M$, and represent the secrecy of $M$ as
``$\attacker(M)$ cannot be executed'', that is, ``if $\attacker(M)$
has been executed, then false.''
More complex properties can also be specified by our correspondences,
for example that all messages of the protocol have been sent
in order; this feature was used in~\cite{Abadi07}.

Our technique is based on a
substantial extension of a previous verification technique for
secrecy~\cite{Abadi04c,Blanchet2001,Weidenbach99}.
More precisely, the protocol is represented in the process calculus
introduced in~\cite{Abadi04c}, which is an extension of the pi
calculus with fairly arbitrary cryptographic primitives. This process
calculus is extended with events, used in the statement of
correspondences. These events are the only required annotation of the
protocol; no annotation is needed to help the tool proving
correspondences.
The protocol is then automatically translated into a set of Horn
clauses. This translation requires significant extensions with respect
to the translation for secrecy given in~\cite{Abadi04c}, and can be
seen as an implementation of a type system, as
in~\cite{Abadi04c}. Some of these extensions improve the precision of
the analysis, in particular to avoid merging different nonces.
Other extensions define the translation of events. 
Finally, this set of Horn clauses is passed to a resolution-based 
solver,
similar to that of~\cite{Blanchet2001,Blanchet04e,Weidenbach99}. Some minor
extensions of this solver are required to prove correspondences. This
solver does not always terminate, but we show in
Section~\ref{sect:taggedterm} that it terminates for a large class of
well-designed protocols, named \emph{tagged protocols}. Our
experiments also demonstrate that, in practice, it terminates on many 
examples of protocols.

The main advantages of our method can be summarized as follows. It
is fully automatic; the user only has to code the protocol and the
correspondences to prove. It puts no bounds on the number of
sessions of the protocol or the size of terms that the adversary can
manipulate. It can handle fairly general cryptographic primitives,
including shared-key encryption, public-key encryption, signatures,
one-way hash functions, and Diffie-Hellman key agreements.  It relies
on a precise semantic foundation. One limitation of the technique is
that, in rare cases, the solving algorithm does not terminate.  The
technique is also not complete: the translation into Horn clauses
introduces an abstraction, which forgets the number of repetitions
of each action~\cite{Blanchet05b}. This abstraction is key to the
treatment of an unbounded number of sessions. 
Due to this abstraction, the tool provides sufficient conditions for
proving correspondences, but can fail on correct protocols.
Basically, it fails to prove protocols that 
first need to keep some value secret and later reveal it (see 
Section~\ref{sec:protclauses}).
In practice, the tool is still very precise and, in our 
experiments, it always succeeded in proving protocols that
were correct.

Our technique is implemented in the protocol verifier ProVerif,
available at \url{http://www.proverif.ens.fr/}.

\paragraph{Comparison with Other Papers on ProVerif}
As mentioned above, this paper extends previous work on the
verification of secrecy~\cite{Abadi04c} in order to prove
correspondences. Secrecy (defined as the impossibility for the adversary to 
compute the secret) and correspondences are trace properties.
Other papers deal with the proof of certain
classes of observational equivalences, \emph{i.e.},
that the adversary cannot distinguish certain processes:
\cite{Blanchet04,Blanchet04c} deal with the proof of strong secrecy, \emph{i.e.},
that the adversary cannot see when the value of a secret changes;
\cite{Blanchet07b} deals with the proof of equivalences between processes
that differ only by the terms that they contain. Moreover, \cite{Blanchet07b}
also explains how to handle cryptographic primitives defined by equational 
theories (instead of rewrite rules) and how to deal with guessing attacks
against weak secrets.

As shown in~\cite{Blanchet04e}, the resolution algorithm terminates
for tagged protocols. The present paper extends this result in 
Section~\ref{sect:taggedterm}, by providing a characterization of
tagged protocols at the level of processes instead of at the level
of Horn clauses.

ProVerif can also reconstruct an attack using a derivation from the
Horn clauses, when the proof of a secrecy property
fails~\cite{Allamigeon05}. Although the present paper does not detail
this point, this work has also been extended to the reconstruction of
attacks against non-injective correspondences.

Finally, \cite{Abadi04f}, \cite{Abadi07}, and~\cite{Blanchet08b} present 
three case studies done at least partly using ProVerif:
\cite{Abadi04f} studies a certified email protocol, \cite{Abadi07}
studies the Just Fast Keying protocol, and~\cite{Blanchet08b}
studies the Plutus secure file system. These case studies rely partly
on the results presented in this paper.

\paragraph{Related Work} 
We mainly focus on the works that automatically verify correspondences
and authentication for security protocols, without bounding the
number of sessions.

The NRL protocol analyzer~\cite{Escobar06,Meadows96b}, based on
narrowing in rewriting systems, can verify correspondences defined in
a rich language of logical formulae~\cite{Syverson96}. It
is sound and complete, but does not always terminate. Our Horn clause
representation is more abstract than the representation of NRL, which
should enable us to terminate more often and be more efficient, while
remaining precise enough to prove most desired properties.

Gordon and Jeffrey designed a system named Cryptic for verifying authentication
by typing in security protocols~\cite{Gordon2002b,Gordon2003,Gordon2004}.
They handle shared-key and public-key cryptography. Our system allows
more general cryptographic primitives (including hash functions and
Diffie-Hellman key agreements).
Moreover, in our system, no annotation is needed, whereas, in Cryptic,
explicit type casts and checks have to be manually added. However,
Cryptic has the advantage that type checking always terminates,
whereas, in some rare cases, our analyzer does not.

Bugliesi et al.~\cite{Bugliesi06} define another type system
for proving authentication in security protocols. The
main advantage of their system is that it is compositional:
it allows one to prove independently the correctness of the code
of each role of the protocol. However, the form of messages
is restricted to certain tagged terms. This approach is compared
with Cryptic in~\cite{Bugliesi05}.

Backes et al.~\cite{Backes07} prove secrecy and authentication for
security protocols, using an abstract-interpretation-based analysis.
This analysis builds a causal graph, which captures the causality
among program events; the security properties are proved by traversing
this graph.  This analysis can handle an unbounded number of sessions
of the protocol; it always terminates, at the cost of additional
abstractions, which may cause false attacks. It handles shared-key and
public-key cryptography, but not Diffie-Hellman key agreements. It
assumes that the messages are typed, so that names can be
distinguished from other terms.

Bodei et al.~\cite{Bodei05} show message authentication via a control
flow analysis on a process calculus named Lysa. Like~\cite{Backes07},
they handle shared-key and public-key cryptography, and their analysis
always terminates, at the cost of additional abstractions. The notion
of authentication they prove is different from ours: they show message
authentication rather than entity authentication.

Debbabi et al.~\cite{Debbabi97} also verify authentication thanks to a
representation of protocols by inference rules, very similar to our
Horn clauses. However, they verify a weaker notion of
authentication (corresponding to aliveness: if $B$ terminates the
protocol, then $A$ must have been alive at some point before),
and handle only shared-key encryption.

A few other methods require little human effort, while supporting an
unbounded number of runs: the verifier 
of~\cite{Heather05}, based on rank functions, can prove the
correctness of or find attacks against protocols with atomic symmetric
or asymmetric keys. Theorem proving~\cite{Paulson98} often requires
manual intervention of the user. An exception to this
is~\cite{Cortier2001}, but it deals only with secrecy. The theorem
prover TAPS~\cite{Cohen2003} often succeeds without or with little
human intervention.

Model checking~\cite{Lowe96,Mitchell97} in general implies a
limit on the number of sessions of the protocol. This problem has been
tackled by~\cite{Broadfoot2000,Broadfoot2004,Roscoe99}.  They recycle
nonces, to use only a finite number of them in an infinite number of
runs. The technique was first used for sequential runs, then
generalized to parallel runs in~\cite{Broadfoot2004}, but with the
additional restriction that the agents must be ``factorisable''.
(Basically, a single run of the agent has to be split into several
runs such that each run contains only one fresh value.) 

Strand spaces~\cite{Thayer99} are a formalism for 
reasoning about security protocols. They have been used for elegant
manual proofs of authentication~\cite{Guttman02}. The automatic tool
Athena~\cite{Song01} combines model checking and theorem proving, and
uses strand spaces to reduce the state space.
Scyther~\cite{Cremers06} uses an extension of Athena's method
with trace patterns to analyze simultaneously a group of traces.
These tools still sometimes
limit the number of sessions to guarantee termination.

Amadio and Prasad~\cite{Amadio1999b} note that authentication
can be translated into secrecy, by using a judge process. The
translation is limited in that only one message can be registered by
the judge, so the verified authentication property is not exactly the
same as ours.

\paragraph{Outline} 
Section~\ref{sec:calculus} introduces our process calculus. 
Section~\ref{sec:secrecyauth} defines
the correspondences that we verify, including
secrecy and various notions of authentication. 
Section~\ref{sec:veriffirst} outlines the main ideas behind
our technique for verifying correspondences. 
Section~\ref{sec:Horn} explains the construction of Horn
clauses and shows its correctness, Section~\ref{sec:solv} describes our 
solving algorithm and shows its correctness, and
Section~\ref{sec:verif} applies these results to the proof of correspondences. 
Section~\ref{sect:termination} discusses the termination of our algorithm:
it shows termination for tagged protocols and how to obtain termination
more often in the general case.
Section~\ref{sect:extensions} presents some extensions to our framework.
Section~\ref{sec:results} gives our experimental results on
a selection of security protocols of the literature, and 
Section~\ref{sec:concl} concludes. 
The proofs of our results
are grouped in the appendices.

\section{The Process Calculus}\label{sec:calculus}

In this section, we present the process calculus that we use to represent
security protocols: we give its syntax, semantics, and illustrate
it on an example protocol. 

\subsection{Syntax and Informal Semantics}

\begin{figure}[t]
\begin{center}
\begin{minipage}{15cm}
\begin{defn}
\categ{M, N}{terms}\\ 
\entry{x, y, z}{variable}\\ 
\entry{a, b, c, k}{name}\\ 
\entry{f(M_1, \ldots, M_n)}{constructor application}
\end{defn}
\begin{defn}
\categ{P, Q}{processes}\\
\entry{\coutput{M}{N}.P}{output}\\
\entry{\cinput{M}{x}.P}{input}\\
\entry{0}{nil}\\
\entry{P\parpop Q}{parallel composition}\\
\entry{\Repl{P}}{replication}\\
\entry{\Res{a}P}{restriction}\\
\entry{\letfun{x}{g(M_1, \ldots, M_n)}{P}{Q}}{destructor application}\\
\entry{\guard{M}{N}{P}{Q}}{conditional}\\
\entry{\asevent(M).P}{event}
\end{defn}
\end{minipage}
\end{center}
\caption{Syntax of the process calculus}\label{fig:syntax}
\end{figure}

Figure~\ref{fig:syntax} gives the syntax of terms (data) and processes
(programs) of our calculus.  The identifiers $a$, $b$, $c$, $k$,
and similar ones range over names, and $x$, $y$, and $z$ range over
variables.  The syntax also assumes a set of symbols for constructors
and destructors; we often use $f$ for a constructor and $g$ for a
destructor.

Constructors are used to build terms. Therefore, the terms are
variables, names, and constructor applications of the form $f(M_1,
\ldots, M_n)$; the terms are untyped.  
On the other hand, destructors do not appear in terms,
but only manipulate terms in processes. They are partial functions on
terms that processes can apply.  The process $\letfun{x}{g(M_1,
\ldots, M_n)}{P}{Q}$ tries to evaluate $g(M_1, \ldots, M_n)$; if this
succeeds, then $x$ is bound to the result and $P$ is executed, else
$Q$ is executed. More precisely, the semantics of a destructor $g$ of
arity $n$ is given by a set $\defg$ of rewrite rules of the form
$g(M_1, \ldots, M_n) \rightarrow M$ where $M_1, \ldots, M_n, M$ are
terms without names, and the variables of $M$ also occur in 
$M_1, \ldots, M_n$. We extend these rules by $g(M'_1, \ldots,
M'_n) \rightarrow M'$ if and only if there exist a substitution
$\sigma$ and a rewrite rule $g(M_1, \ldots, M_n) \rightarrow M$ in
$\defg$ such that $M'_i = \sigma M_i$ for all $i \in \interv{1}{n}$,
and $M' = \sigma M$. We assume that the set $\defg$ is
finite. (It usually contains one or two rules in 
examples.) We define destructors by rewrite rules
instead of the equalities used in~\cite{Abadi04c}. This definition allows
destructors to yield several different results
non-deterministically. (Non-deterministic rewrite rules are used in our 
modeling of Diffie-Hellman key agreements; see Section~\ref{sec:DiffieHellman}).
\begin{figure}[t]
{\bf Tuples:}\\
Constructor: tuple $\ntuple{n}(x_1, \ldots, x_n)$\\
Destructors: projections $\nth{i}{n}(\ntuple{n}(x_1, \ldots, x_n)) \rightarrow x_i$\\
{\bf Shared-key encryption:}\\
Constructor: encryption of $x$ under the key $y$, $\sencrypt(x, y)$\\
Destructor: decryption $\sdecrypt(\sencrypt(x, y), y) \rightarrow x$\\
{\bf Probabilistic shared-key encryption:}\\
Constructor: encryption of $x$ under the key $y$ with random coins $r$, $\sencryptp(x, y, r)$\\
Destructor: decryption $\sdecryptp(\sencryptp(x, y, r), y) \rightarrow x$\\
{\bf Probabilistic public-key encryption:}\\
Constructors: encryption of $x$ under the key $y$ with random coins $r$, $\pencrypt(x, y, r)$\\
%NOTE cut ``public''
\phantom{Constructors: }public key generation from a secret key $y$, $\pk(y)$\\
Destructor: decryption $\pdecrypt(\pencrypt(x, \pk(y), r) , y) \rightarrow x$\\
{\bf Signatures:}\\
Constructors: signature of $x$ with the secret key $y$, $\sign(x,y)$\\
\phantom{Constructors: }public key generation from a secret key $y$, $\pk(y)$\\
Destructors: signature verification $\checksign(\sign(x, y), \pk(y)) \rightarrow x$\\
\phantom{Destructors: }message without signature $\getmess(\sign(x, y)) \rightarrow x$\\
{\bf Non-message-revealing signatures:}\\
Constructors: signature of $x$ with the secret key $y$, $\nmrsign(x,y)$\\
\phantom{Constructors: }public key generation from a secret key $y$, $\pk(y)$\\
\phantom{Constructors: }constant $\nmrtrue$\\
Destructor: verification $\nmrchecksign(\nmrsign(x, y), \pk(y), x) \rightarrow \nmrtrue$\\
{\bf One-way hash functions:}\\
Constructor: hash function $\hash(x)$\\
{\bf Table of host names and keys}\\
Constructor: host name from key $\host(x)$\\
Private destructor: key from host name $\getkey(\host(x)) \rightarrow x$

\caption{Constructors and destructors}\label{fig:cryptoop}
\end{figure}
Using constructors and destructors, we can represent data structures
and cryptographic operations as summarized in
Figure~\ref{fig:cryptoop}. (We present only probabilistic public-key
encryption because, in the computational model, a secure public-key
encryption algorithm must be probabilistic. We have chosen to present
deterministic signatures; we could easily model probabilistic
signatures by adding a third argument $r$ containing the random coins,
as for encryption.  The coins should be chosen using a restriction
$\Res{a}$ which creates a fresh name $a$, representing a fresh random
number.)

Constructors and destructors can be public or private. The public
ones can be used by the adversary, which is
the case when not stated otherwise. 
The private ones can be used only by
honest participants. They are useful in practice to model tables of
keys stored in a server, for instance.  A public
constructor $\host$ computes a host name from a long-term secret key,
and a private destructor $\getkey$ returns the key from the host name,
and simulates a lookup in a table of pairs (host name, key). Using a
public constructor $\host$ allows the adversary to create and register
any number of host names and keys. However, since $\getkey$ is
private, the adversary cannot compute a key from the host name, which
would break all protocols: host names are public while keys of honest
participants are secret.

The process calculus provides additional instructions for executing
events, which will be used for specifying correspondences. The process
$\asevent(M).P$ executes the event $\asevent(M)$, then executes
$P$. 

The other constructs in the syntax of Figure~\ref{fig:syntax} are
standard; most of them come from the pi calculus. The input process
$\cinput{M}{x}.P$ inputs a message on channel $M$, and executes $P$
with $x$ bound to the input message. The output process
$\coutput{M}{N}.P$ outputs the message $N$ on the channel $M$ and then
executes $P$. We allow communication on channels that can be arbitrary 
terms. (We could adapt our work to the case in which channels are 
only names.)  Our
calculus is monadic (in that the messages are terms rather than tuples
of terms), but a polyadic calculus can be simulated since tuples are
terms. It is also synchronous (in that a process $P$ is executed after
the output of a message).  The nil process $0$ does nothing. The
process $P \parpop Q$ is the parallel composition of $P$ and $Q$. The
replication $\Repl{P}$ represents an unbounded number of copies of $P$
in parallel. The restriction $\Res{a} P$ creates a new name $a$ and
then executes $P$. The conditional $\guard{M}{N}{P}{Q}$ executes $P$
if $M$ and $N$ reduce to the same term at runtime; otherwise, it
executes~$Q$. We define $\letdef{x}{M}{P}$ as syntactic sugar for
$P\{M/x\}$. As usual, we may omit an $\kw{else}$ clause when it
consists of~$0$.

The name $a$ is bound in the process $\Res{a}P$. The variable $x$ is bound in $P$
in the processes $\cinput{M}{x}.P$ and $\letfun{x}{g(M_1, \ldots,
M_n)}{P}{Q}$.  We write $\fn(P)$ and $\fv(P)$
for the sets of names and variables free in $P$, respectively.  A
process is closed if it has no free variables; it may have free names.
We identify processes up to renaming of bound names and variables. We
write $\{M_1/x_1, \ldots, M_n/x_n\}$ for the substitution that
replaces $x_1$, \ldots, $x_n$ with $M_1$, \ldots, $M_n$, respectively.

\subsection{Operational Semantics}\label{subsec:opsem}

\begin{figure}[t]
\begin{align}
&\env,\pset \cup \{ \,0\, \} \rightarrow \env, \pset\tag{Red Nil}\\
&\env,\pset \cup \{ \,\Repl{P}\, \} \rightarrow \env, \pset \cup \{ \,P, \Repl{P}\, \}\tag{Red Repl}\\
&\env,\pset \cup \{ \,P \parpop Q\, \} \rightarrow \env, \pset \cup \{ \,P, Q\, \}\tag{Red Par}\\
&\env,\pset \cup \{ \,\Res{a}P\, \} \rightarrow \env \cup \{ a'\}, \pset \cup \{ \,P\{ a'/a \}\,\}\tag{Red Res}\\
&\qquad\text{where $a' \notin \env$.}\notag\\
&\env,\pset \cup \{ \,\coutput{N}{M}.Q, \cinput{N}{x}.P\, \}\rightarrow \env, \pset \cup \{ \,Q, P\{ M/x \}\, \}\tag{Red I/O}\\
&\env,\pset \cup \{ \,\letfun{x}{g(M_1, \ldots, M_n)}{P}{Q}\,\}\rightarrow \env, \pset \cup \{ \, P\{ M' / x\}\, \}\notag\\
&\qquad \text{if $g(M_1, \ldots, M_n) \rightarrow M'$}\tag{Red Destr 1}\\
&\env,\pset \cup \{ \,\letfun{x}{g(M_1, \ldots, M_n)}{P}{Q}\,\}\rightarrow \env, \pset \cup \{ \, Q\, \}\tag{Red Destr 2}\\
&\qquad \text{if there exists no $M'$ such that $g(M_1, \ldots, M_n) \rightarrow M'$}\notag\\
&\env,\pset \cup \{\,\guard{M}{M}{P}{Q}\, \}\rightarrow \env, \pset \cup \{ \,P\,\}\tag{Red Cond 1}\\
&\env,\pset \cup \{\,\guard{M}{N}{P}{Q}\, \}\rightarrow \env, \pset \cup \{ \,Q\,\}\tag{Red Cond 2}\\
&\qquad \text{if $M\neq N$}\notag\\
&\env,\pset \cup \{\,\asevent(M).P\, \}\rightarrow \env, \pset \cup \{ \, P\,\}\tag{Red Event}
\end{align}
\caption{Operational semantics}\label{fig:reduction}
\end{figure}

A semantic configuration is a pair $\env, \pset$ where the environment
$\env$ is a finite set of names and $\pset$ is a finite multiset of
closed processes. The environment $\env$ must contain at least all
free names of processes in $\pset$. The configuration $\{ a_1, \ab \ldots,\ab
a_n \}, \ab \{ P_1, \ab \ldots, \ab P_n \}$ corresponds intuitively to the process
$\Res{a_1}\ldots\Res{a_n}(P_1 \parpop \ldots \parpop P_n)$. The
semantics of the calculus is defined by a reduction relation
$\rightarrow$ on semantic configurations, shown in
Figure~\ref{fig:reduction}. The rule (Red Res) is the only
one that uses renaming. This is important so that the parameters of
events are not renamed after the execution of the event, to
be able to compare them with the parameters of events
executed later.  This semantics is superficially different from those
of~\cite{Abadi04c,Blanchet2002}, which were defined using a structural
congruence relation and a reduction relation on
processes. 
The new semantics (in particular the renaming point mentioned above)
provides simplifications in the definitions of correspondences
(Definitions~\ref{def:exectrace}, \ref{def:corresp}, \ref{def:execevent}, 
\ref{def:injcorresp}, and~\ref{def:nestedquery}) and in the proofs that 
correspondences hold.

\subsection{Example}\label{sec:example}

As a running example, we consider a simplified version of the
Needham-Schroeder public-key protocol~\cite{Needham78}, with the correction by
Lowe~\cite{Lowe96}, in which host names are replaced by public keys,
which makes interaction with a server useless. (The version tested in
the benchmarks is the full version. Obviously, our tool can verify
much more complex protocols; we use this simple example for
illustrative purposes.) The protocol contains the following messages:
\begin{center}
\begin{tabular}{l l l}
Message 1. &$A \rightarrow B :$&$\{ a,\pkA\}_{\pkB}$\\
Message 2. &$B \rightarrow A :$&$\{ a, b, \pkB\}_{\pkA}$\\
Message 3. &$A \rightarrow B :$&$\{ b \}_{\pkB}$\\
\end{tabular}
\end{center}
$A$ first sends to $B$ a nonce (fresh name) $a$ encrypted under the
public key of $B$. $B$ decrypts this message using his secret key
$\skB$ and replies with the nonce $a$, a fresh nonce he chooses $b$,
and its own public key $\pkB$, all encrypted under $\pkA$.  When $A$
receives this message, she decrypts it. When $A$ sees the nonce $a$, she
is convinced that $B$ answered since only $B$ can decrypt the first
message and obtain $a$. Then $A$ replies with the nonce $b$ encrypted
under $\pkB$.  $B$ decrypts this message. When $B$ sees the nonce $b$,
he is convinced that $A$ replied, since only $A$ could decrypt the
second message and obtain $b$.
The presence of $\pkA$ in the first message and $\pkB$ in the second message 
makes explicit that these messages are for sessions between $A$ and $B$,
and so avoids man-in-the-middle attacks, such as the well-known
attack found by Lowe~\cite{Lowe96}.
This protocol can be represented in our calculus by the process $P$,
explained below:
{\allowdisplaybreaks\begin{align*}
&P_A(\skA, \pkA, \pkB) = \Repl{} \cinput{c}{x\_\pkB}.\Res{a}
\asevent(e_1(\pkA, x\_\pkB, a)).\\*
&\qquad \Res{r_1} \coutput{c}{\pencrypt((a, \pkA), x\_\pkB, r_1)}.\\*
&\qquad \cinput{c}{m}.
\aletfun{(=a, x\_b, =x\_\pkB)}{\pdecrypt(m, \skA)}\\*
&\qquad \asevent(e_3(\pkA, x\_\pkB, a, x\_b)).
 \Res{r_3} \coutput{c}{\pencrypt(x\_b, x\_\pkB, r_3)}\\*
&\qquad \aguard{x\_\pkB}{\pkB}\\*
&\qquad \asevent(e_A(\pkA, x\_\pkB, a, x\_b)).
\coutput{c}{\sencrypt(\sAa, a)}.
\coutput{c}{\sencrypt(\sAb, x\_b)}\\
&P_B(\skB, \pkB, \pkA) = \Repl{} \cinput{c}{m'}.
\aletfun{(x\_a, x\_\pkA)}{\pdecrypt(m', \skB)}
\Res{b}\\*
&\qquad\asevent(e_2(x\_\pkA, \pkB, x\_a, b)).\Res{r_2}
\coutput{c}{\pencrypt((x\_a, b, \pkB), x\_\pkA, r_2)}.\\*
&\qquad \cinput{c}{m''}.
\aletfun{(=b)}{\pdecrypt(m'', \skB)}\\*
&\qquad \aguard{x\_\pkA}{\pkA}\\*
&\qquad \asevent(e_B(x\_\pkA, \pkB, x\_a, b)).
\coutput{c}{\sencrypt(\sBa, x\_a)}.
\coutput{c}{\sencrypt(\sBb, b)}\\
&P = \Res{\skA}\Res{\skB}\aletdef{\pkA}{\pk(\skA)}\aletdef{\pkB}{\pk(\skB)}\\*
&\qquad\coutput{c}{\pkA}\coutput{c}{\pkB}.(P_A(\skA, \pkA, \pkB) \parpop P_B(\skB, \pkB, \pkA))
\end{align*}}%
The channel $c$ is public: the adversary can send and listen on it.
We use a single public channel and not two or more channels because
the adversary could take a message from one channel and relay it on
another channel, thus removing any difference between the channels.
The process $P$ begins with the creation of the secret and public keys
of $A$ and $B$. The public keys are output on channel $c$ to model
that the adversary has them in its initial knowledge.  Then the
protocol itself starts: $P_A$ represents $A$, $P_B$ represents $B$.
Both principals can run an unbounded number of sessions, so $P_A$ and
$P_B$ start with replications.

We consider that $A$ and $B$ are both willing to talk to any
principal. So, to determine to whom $A$ will talk, we consider that
$A$ first inputs a message containing the public key $x\_\pkB$ of its
interlocutor. (This interlocutor is therefore chosen by the adversary.)
Then $A$ starts a protocol run by choosing a nonce $a$, and executing
the event $e_1(\pkA, x\_\pkB, a)$. Intuitively, this event records that
$A$ sent Message~1 of the protocol, for a run with the participant of
public key $x\_\pkB$, using the nonce $a$.  
Event $e_1$ is placed before the actual output of Message~1; this 
is necessary for the desired correspondences to hold:
if event $e_1$ followed the output of Message~1, 
one would not be able to prove that event $e_1$ must have been 
executed, even though Message~1 must have been sent, because Message~1 could
be sent without executing event $e_1$. The situation is similar
for events $e_2$ and $e_3$ below.
Then $A$ sends the first message
of the protocol $\pencrypt((a, \pkA), x\_\pkB, r_1)$, where $r_1$ 
are fresh coins, used to model that public-key encryption is 
probabilistic. $A$ waits for the
second message and decrypts it using her secret key $\skA$.  If
decryption succeeds, $A$ checks that the message has the right form
using the pattern-matching construct 
$\aletfun{(=a, x_b, =x\_\pkB)}{\pdecrypt(m, \skA)}\ldots$
This construct is syntactic sugar for
$\aletfun{y}{\pdecrypt(m,\skA)} \aletfun{x_1}{\nth{1}{3}(y)}
\aletfun{x_b}{\nth{2}{3}(y)} \aletfun{x_3}{\nth{3}{3}(y)}
\aguard{x_1}{a} \aguard{x_3}{x\_\pkB}\ldots$
Then $A$ executes the event $e_3(\pkA, x\_\pkB, a, x\_b)$, to record
that she has received Message~2 and sent Message~3 of the protocol, 
in a session with the 
participant of public key $x\_\pkB$, and nonces $a$ and $x\_b$. 
Finally, she sends the last message of the protocol
$\pencrypt(x\_b, x\_\pkB, r_3)$. 
After sending this message, $A$ executes some actions needed only for
specifying properties of the protocol.  When $x\_\pkB = \pkB$, that
is, when the session is between $A$ and $B$, $A$ executes the event
$e_A(\pkA, x\_\pkB, a, x\_b)$, to record that $A$ ended a session of
the protocol, with the participant of public key $x\_\pkB$ and nonces
$a$ and $x\_b$. $A$ also outputs the secret name $\sAa$ encrypted
under the nonce $a$ and the secret name $\sAb$ encrypted under the nonce
$x\_b$.  These outputs are helpful in order to formalize the secrecy of
the nonces.  Our tool can prove the secrecy of free names, but not the
secrecy of bound names (such as $a$) or of variables (such as $x\_b$).
In order to overcome this limitation, we publish the encryption of a
free name $\sAa$ under $a$; then $\sAa$ is secret if and only if the
nonce $a$ chosen by $A$ is secret. Similarly, $\sAb$ is secret if and
only if the nonce $x\_b$ received by $A$ is secret.

The process $P_B$ proceeds similarly: it executes the protocol, with
the additional event $e_2(x\_\pkA, \ab \pkB, \ab x\_a, \ab b)$ to record that
Message~1 has been received and Message~2 has been sent by $B$, 
in a session with the participant
of public key $x\_\pkA$ and nonces $x\_a$ and $b$. After finishing
the protocol itself, when $x\_\pkA = \pkA$, that is, when the session
is between $A$ and $B$, $P_B$ executes the event $e_B(x\_\pkA, \pkB, 
x\_a, b)$, to record that $B$ finished the protocol, and 
outputs $\sBa$ encrypted under $x\_a$ and $\sBb$ encrypted under $b$,
to model the secrecy of $x\_a$ and $b$ respectively.

The events will be used in order to formalize authentication.
For example, we formalize that, if $A$ ends a session of the protocol,
then $B$ has started a session of the protocol with the same nonces by
requiring that, if $e_A(x_1, x_2, x_3,x_4)$ has been executed, then
$e_2(x_1, x_2, x_3, x_4)$ has been executed.\footnote{For this purpose, 
the event $e_A$ must not be
executed when $A$ thinks she talks to the adversary. Indeed, in this
case, it is correct that no event has been executed by the
interlocutor of $A$, since the adversary never executes events.}

\vspace*{-1mm}%
\section{Definition of Correspondences}\label{sec:secrecyauth}

In this section, we formally define the correspondences that we verify.
We prove correspondences of the form ``if an event $e$ has been 
executed, then events $e_{11}$, \ldots, $e_{1l_1}$ have been executed, 
or \ldots, or $e_{m1}$, \ldots, $e_{ml_m}$ have been executed''. 
These events may include arguments, which allows one to relate the
values of variables at the various events.
Furthermore, we can replace the event $e$ with
the fact that the adversary knows some term (which allows us to prove
secrecy properties), or that a certain message has been sent on 
a certain channel.
We can prove that each execution of $e$ corresponds to a distinct
execution of some events $e_{jk}$ (injective correspondences, defined
in Section~\ref{sec:definjcorresp}),
and we can prove that the events $e_{jk}$ have been executed in
a certain order (general correspondences, defined in 
Section~\ref{sec:defgencorresp}).

We assume that the protocol is executed in the presence of an
adversary that can listen to all messages, compute, and send all
messages it has, following the so-called Dolev-Yao model~\cite{Dolev83}. 
Thus, an adversary can be
represented by any process that has a set of public names $\rw$ in its
initial knowledge and that does not contain events. (Although the initial
knowledge of the adversary contains only names in $\rw$, one can give any terms
to the adversary by sending them on a channel in $\rw$.)

\begin{definition}
Let $\rw$ be a finite set of names. The closed process $Q$ is an
\emph{$\rw$-adversary} if and only if $\fn(Q) \subseteq \rw$ and $Q$ does not
contain events.
\end{definition}

\subsection{Non-injective Correspondences}

Next, we define when a trace satisfies an atom $\act$, generated by the following grammar:
\begin{defn}
\categ{\act}{atom}\\
\entry{\attacker(M)}{attacker knowledge}\\
\entry{\mess(M,M')}{message on a channel}\\
\entry{\pasevent(M)}{event}
\end{defn}
Intuitively, a trace satisfies $\attacker(M)$ when the attacker has
$M$, or equivalently, when $M$ has been sent on a public channel in
$\rw$. It satisfies $\mess(M,M')$ when the message $M'$ has been sent on
channel $M$.  Finally, it satisfies $\pasevent(M)$ when the event
$\asevent(M)$ has been executed.

\begin{definition}\label{def:exectrace}
We say that a trace $\trace = \env_0, \pset_0 \rightarrow^* \env', \pset'$
satisfies $\attacker(M)$ if and only if $\trace$ contains a
reduction $\env,\pset \cup \{ \,\coutput{c}{M}.Q, \cinput{c}{x}.P\,
\}\rightarrow \env, \pset \cup \{ \,Q, P\{ M/x \}\, \}$ for some
$\env$, $\pset$, $x$, $P$, $Q$, and $c \in \rw$.

We say that a trace $\trace = \env_0, \pset_0 \rightarrow^* \env', \pset'$
satisfies $\mess(M, M')$ if and only if $\trace$ contains a reduction
$\env,\pset \cup \{ \,\coutput{M}{M'}.Q, \cinput{M}{x}.P\,
\}\rightarrow \env, \pset \cup \{ \,Q, P\{ M'/x \}\, \}$ for some
$\env$, $\pset$, $x$, $P$, $Q$.

We say that a trace $\trace = \env_0, \pset_0 \rightarrow^* \env', \pset'$
satisfies $\pasevent(M)$ if and only if $\trace$ contains a
reduction $\env,\pset \cup \{\,\asevent(M).P\, \}\rightarrow \env,
\pset \cup \{ \, P\,\}$ for some $\env$, $\pset$, $P$.
\end{definition}

The correspondence $\act \Rightarrow \sor_{j=1}^{m} \left(\act_j \rightsquigarrow 
\sand_{k = 1}^{l_j} \pasevent(M_{jk})\right)$, formally defined below, means intuitively that, if an
instance of $\act$ is satisfied, then for some $j \in \{ 1,
\ldots, m \}$, the considered instance of $\act$ is an instance of $\act_j$
and a corresponding instance of the each of the events 
$\asevent(M_{j1})$, \ldots,
$\asevent(M_{jl_j})$ has been executed.\footnote{The implementation
in ProVerif uses a slightly different notation: $\act_j$ is omitted,
but additionnally equality tests are allowed on the right-hand side of
$\rightsquigarrow$, so that one can check that $\act$ is actually an
instance of $\act_j$.}

\begin{definition}\label{def:corresp}
The closed process $P_0$ \emph{satisfies the correspondence} 
\[\act \Rightarrow \sor_{j=1}^{m} \left(\act_j \rightsquigarrow 
\sand_{k = 1}^{l_j} \pasevent(M_{jk})\right)\]
against $\rw$-adversaries if and only if, for any
$\rw$-adversary $Q$, for any $\env_0$ containing $\fn(P_0) \cup \rw
\cup \fn(\act) \cup \bigcup_j \fn(\act_j) \cup \bigcup_{j,k} \fn(M_{jk})$, for any substitution 
$\sigma$, for any trace $\trace = \env_0, \{ P_0, Q \}
\rightarrow^* \env', \pset'$, if $\trace$ satisfies $\sigma \act$, 
then there exist $\sigma'$ and $j \in \{ 1,
\ldots, m \}$ such that $\sigma' \act_j = \sigma \act$ and,
for all $k \in \{ 1, \ldots, l_j \}$,
$\trace$ satisfies $\pasevent(\sigma' M_{jk})$ as well.
\end{definition}

This definition is very general; we detail some interesting particular 
cases below.
When $m=0$, the disjunction $\sor_{j=1}^{m} \ldots$
is denoted by $\cfalse$. When $\act = \act_j$ for all $j$, 
we abbreviate the correspondence
by
$\act \rightsquigarrow \sor_{j=1}^{m}
\sand_{k = 1}^{l_j} \pasevent(M_{jk})$.
This correspondence means that, if an instance of $\act$ is satisfied, then
for some $j \leq m$, a corresponding instance of $\asevent(M_{j1})$,
\ldots, $\asevent(M_{jl_j})$ has been executed. The variables in $\act$
are universally quantified (because, in Definition~\ref{def:corresp},
$\sigma$ is universally quantified). The variables in $M_{jk}$ that
do not occur in $\act$ are existentially quantified (because
$\sigma'$ is existentially quantified).

\begin{example}\label{exa:q1}
In the process of Section~\ref{sec:example}, the correspondence 
$\pasevent(e_B(x_1, \ab x_2, \ab x_3, \ab x_4)) \rightsquigarrow
\pasevent(e_1(x_1, \ab x_2, \ab x_3)) \wedge 
\pasevent(e_2(x_1, \ab x_2, \ab x_3, \ab x_4)) \wedge 
\pasevent(e_3(x_1, \ab x_2, \ab x_3, \ab x_4))$ means that, if the event 
$e_B(x_1, \ab x_2, \ab x_3, \ab x_4)$ has been executed, then the events
$e_1(x_1, \ab x_2, \ab x_3)$, $e_2(x_1, \ab x_2, \ab x_3, \ab x_4)$, and 
$e_3(x_1, \ab x_2, \ab x_3, \ab x_4)$
have been executed, with the same value of the arguments $x_1, \ab x_2, 
\ab x_3, \ab x_4$.

% Event identifiers

\newcommand{\kwp}[1]{\mathit{#1}}

\newcommand{\TTPsend}{\kwp{TTP\_send}}
\newcommand{\Shas}{\kwp{S\_has}}
\newcommand{\Rreceived}{\kwp{R\_received}}

% Functions 

\newcommand{\kwf}[1]{\mathit{#1}}
\newcommand{\fMessage}{\kwf{msg}}

\newcommand{\fE}{\sencrypt}
\newcommand{\fS}{\sign}

% Constants

\newcommand{\kwc}[1]{\mathit{#1}}

\newcommand{\cAuth}{\kwc{Auth}}
\newcommand{\cNoAuth}{\kwc{NoAuth}}

\newcommand{\cTTPSigKey}{\kwc{sk}_{\kwc{TTP}}}

The correspondence
\[\begin{split}
&\pasevent(\Rreceived(\fMessage(x,z)))
\Rightarrow \\
&\quad\! (\pasevent(\Rreceived(\fMessage(x,(z', \cAuth)))) 
\rightsquigarrow\\
&\quad\! \phantom{(} \pasevent(\Shas(k, \fMessage(x,(z', \cAuth)))) \wedge\\
&\quad\! \phantom{(} \pasevent(\TTPsend(\fS(
(\fE(\fMessage(x,(z', \cAuth)), k), x), \cTTPSigKey))))\\
&\!\vee (\pasevent(\Rreceived(\fMessage(x,(z', \cNoAuth)))) 
\rightsquigarrow\\
&\quad\! \phantom{(} \pasevent(\Shas(k, \fMessage(x,(z', \cNoAuth)))) \wedge\\
&\quad\! \phantom{(} \pasevent(\TTPsend(\fS(\fE(\fMessage(x,(z', \cNoAuth)),k),\cTTPSigKey))))
\end{split}\]
means that, if the event $\Rreceived(\fMessage(x,z))$ has been executed, 
then two cases can happen: either $z = (z', \cAuth)$ or $z = (z', \cNoAuth)$
for some $z'$.
In both cases, the events 
$\TTPsend(\mathit{certificate})$
and $\Shas(k, \fMessage(x,z))$ 
have been executed for some $k$, but with a different value of $\mathit{certificate}$:
$\mathit{certificate} = \fS((\mathit{S2TTP}, \ab x), \ab \cTTPSigKey)$ when $z =
(z',\cAuth)$, and $\mathit{certificate} =
\fS(\mathit{S2TTP}, \ab \cTTPSigKey)$
when $z = (z', \cNoAuth)$, with $\mathit{S2TTP} = \fE(\fMessage(x, \ab  \ab z), \ab k)$.
A similar correspondence was used in our study of a certified email
protocol, in collaboration with Mart{\'\i}n 
Abadi~\cite[Section~5, Proposition~4]{Abadi04f}. 
We refer to that paper for additional details.
\end{example}

The following definitions are particular cases of 
Definition~\ref{def:corresp}.

\begin{definition}\label{def:secr}
The closed process $P$ \emph{preserves the secrecy of all instances
of} $M$ from $\rw$ if and only if it satisfies the correspondence
$\attacker(M) \rightsquigarrow \cfalse$ against 
$\rw$-adversaries.
\end{definition}
When $M$ is a free name, this definition is equivalent to that 
of~\cite{Abadi04c}.

\begin{example}\label{exa:q2}
The process $P$ of Section~\ref{sec:example} preserves the secrecy of $\sAa$ when 
the correspondence $\attacker(\sAa) \rightsquigarrow \cfalse$
is satisfied. In this case, intuitively, $P$ preserves the secrecy of the 
nonce $a$ that $A$ chooses. The situation is similar for $\sAb$, $\sBa$,
and $\sBb$.
\end{example}

\begin{definition}\label{def:noninjag}
\emph{Non-injective agreement} is a correspondence of the form
$\pasevent(e(x_1, \ab \ldots, \ab x_n)) \rightsquigarrow 
\pasevent(e'(x_1, \ab \ldots, \ab x_n))$.
\end{definition}
Intuitively, the correspondence $\pasevent(e(x_1, \ldots, x_n))
\rightsquigarrow \pasevent(e'(x_1, \ldots, x_n))$ means that, if an
event $e(M_1, \ldots, M_n)$ is executed, then
the event $e'(M_1, \ldots, M_n)$ has also been executed.
This definition can be used to represent Lowe's notion of
non-injective agreement~\cite{Lowe97}.

\begin{example}\label{exa:q3}
In the example of Section~\ref{sec:example}, the
correspondence $\pasevent(e_A(x_1, \ab x_2, \ab x_3, \ab x_4)) \rightsquigarrow
\pasevent(e_2(x_1, \ab x_2, \ab x_3, \ab x_4))$ means that, if $A$ executes an
event $e_A(x_1, \ab x_2, \ab x_3, \ab x_4)$, then $B$ has executed
the event $e_2(x_1, \ab x_2, \ab x_3, \ab x_4)$. So, if $A$ terminates
the protocol thinking she talks to $B$, then $B$ is actually
involved in the protocol. Moreover, the agreement on the parameter of
the events, $\pkA = x\_\pkA$, $x\_\pkB = \pkB$, $a = x\_a$, and $x\_b
= b$ implies that $B$ actually thinks he talks to $A$, and that $A$
and $B$ agree on the values of the nonces.

The correspondence $\pasevent(e_B(x_1, \ab x_2, \ab x_3, \ab x_4)) \rightsquigarrow
\pasevent(e_3(x_1, \ab x_2, \ab x_3, \ab x_4))$ is similar, after swapping the roles
of $A$ and $B$.
\end{example}

\subsection{Injective Correspondences}\label{sec:definjcorresp}

\begin{definition}\label{def:execevent}
We say that the event $\asevent(M)$ is executed at step $\step$ in a
trace $\trace = \env_0, \pset_0 \rightarrow^* \env', \pset'$ if and only if the
$\step$-th reduction of $\trace$ is of the form $\env,\pset \cup
\{\,\asevent(M).P\, \}\rightarrow \env, \pset \cup \{ \, P\,\}$ for
some $\env$, $\pset$, $P$.
\end{definition}

Intuitively, an injective correspondence $\pasevent(M)
\rightsquigarrow \inj\ \pasevent(M')$ requires that each event
$\asevent(\sigma M)$ is enabled by distinct events $\asevent(\sigma
M')$, while a non-injective correspondence $\pasevent(M)
\rightsquigarrow \pasevent(M')$ allows several events $\asevent(\sigma
M)$ to be enabled by the same event $\asevent(\sigma M')$.
We denote by $\injopt$ an optional $\inj$ marker: it can be either $\inj$ or
nothing. When $\injopt = \inj$, an injective correspondence is required.
When $\injopt$ is nothing, the correspondence does not need to
be injective.

\begin{definition}\label{def:injcorresp}
The closed process $P_0$ \emph{satisfies the correspondence} 
\[\pasevent(M) \Rightarrow \sor_{j=1}^{m} \left(\pasevent(N_j) \rightsquigarrow 
\sand_{k = 1}^{l_j} \injopt_{jk}\pasevent(M_{jk})\right)\]
against $\rw$-adversaries if and only if, for any
$\rw$-adversary $Q$, for any $\env_0$ containing $\fn(P_0) \cup \rw
\cup \fn(M) \cup \bigcup_j \fn(N_j) \cup \bigcup_{j,k} \fn(M_{jk})$, 
for any trace
$\trace = \env_0, \{ P_0, Q \} \rightarrow^* \env', \pset'$, there exist 
functions $\phi_{jk}$ from a subset of steps in $\trace$ to
steps in $\trace$ such that
\begin{itemize}

\item For all $\step$, if the event $\asevent(\sigma M)$ 
is executed at step $\step$ in $\trace$ for some $\sigma$,
then there exist $\sigma'$ and $j$ such that $\sigma' N_j = \sigma M$
and, for all $k \in \{ 1, \ldots, l_j \}$, $\phi_{jk}(\step)$ is defined
and $\asevent(\sigma' M_{jk})$ is executed at
step $\phi_{jk}(\step)$ in $\trace$.

\item If $\injopt_{jk} = \inj$, then $\phi_{jk}$ is injective.

\end{itemize}
\end{definition}
The functions $\phi_{jk}$ map execution steps of events $\asevent(\sigma M)$
to the execution steps of the events $\asevent(\sigma' M_{jk})$
that enable $\asevent(\sigma M)$. When $\injopt_{jk} = \inj$,
the injectivity of $\phi_{jk}$ guarantees that distinct executions
of $\asevent(\sigma M)$ correspond to distinct executions of 
$\asevent(\sigma' M_{jk})$.
When $M=N_j$ for all $j$, we abbreviate the correspondence by
$\pasevent(M) \rightsquigarrow \sor_{j=1}^{m} \sand_{k = 1}^{l_j}
\injopt_{jk} \pasevent(M_{jk})$, as in the non-injective case.

Woo and Lam's correspondence assertions~\cite{Woo93} are a particular
case of this definition. Indeed, they consider properties of the
form: if $\gamma_1$ or \ldots or $\gamma_k$ have been executed,
then $\mu_1$ or \ldots or $\mu_m$ must have been executed, denoted by
$\gamma_1 \mid \ldots \mid \gamma_k\hookrightarrow \mu_1 \mid \ldots \mid
\mu_m$. Such a correspondence assertion is formalized in our setting
by for all $i \in \{ 1, \ldots, k\}$, the process
satisfies the correspondence $\pasevent(\gamma_i) \rightsquigarrow \sor_{j = 1}^m \inj\ \pasevent(\mu_j)$.

\begin{Remark}
  Correspondences $\act \Rightarrow \sor_{j=1}^{m} \left(\act_j
    \rightsquigarrow \sand_{k = 1}^{l_j}
    \injopt_{jk}\pasevent(M_{jk})\right)$ with $\act = \attacker(M)$
  and at least one $\inj$ marker would always be wrong: the adversary
  can always repeat the output of $M$ on one of his channels any
  number of times. With $\act = \mess(M,M')$ and at least one $\inj$
  marker, the correspondence may be true only when the adversary
  cannot execute the corresponding output. For simplicity, we focus on
  the case $\act = \pasevent(M)$ only.
\end{Remark}

\begin{definition}
\emph{Injective agreement} is a correspondence of the form
$\pasevent(e(x_1, \ab \ldots, \ab x_n)) \rightsquigarrow \inj\ 
\pasevent(e'(x_1,\ab  \ldots, \ab x_n))$.
\end{definition}
Injective agreement requires that the number of executions of
$\asevent(e(M_1, \ab \ldots, \ab M_n))$ is smaller than the number of
executions of $\asevent(e'(M_1, \ab \ldots, \ab M_n))$: each execution of
$\asevent(e(M_1, \ab \ldots, \ab M_n))$ corresponds to a distinct execution of
$\asevent(e'(M_1, \ab \ldots, \ab M_n))$. This corresponds to Lowe's agreement
specification~\cite{Lowe97}.

\begin{example}
In the example of Section~\ref{sec:example}, the correspondence
$\pasevent(e_A(x_1, \ab x_2, \ab x_3, \ab x_4)) \rightsquigarrow
\inj\ \pasevent(e_2(x_1, \ab x_2, \ab x_3, \ab x_4))$
means that each execution of $\pasevent(e_A(x_1, \ab x_2, \ab x_3, \ab x_4))$
corresponds to a distinct execution of $\pasevent(e_2(x_1, \ab x_2, \ab x_3, 
\ab x_4))$.
So each completed session of $A$ talking to $B$ corresponds
to a distinct session of $B$ talking to $A$, and $A$ and $B$ agree on the
values of the nonces.

The correspondence
$\pasevent(e_B(x_1, \ab x_2, \ab x_3, \ab x_4)) \rightsquigarrow
\inj\ \pasevent(e_3(x_1, \ab x_2, \ab x_3, \ab x_4))$
is similar, after swapping the roles
of $A$ and $B$.
\end{example}

\subsection{General Correspondences}\label{sec:defgencorresp}

Correspondences also give information on the order in
which events are executed. Indeed, if we have the correspondence
\[\pasevent(M) \Rightarrow \sor_{j=1}^{m} 
\left(\pasevent(N_j) \rightsquigarrow 
\sand_{k = 1}^{l_j} \injopt_{jk}\pasevent(M_{jk})\right)\]
then the events $\pasevent(M_{jk})$ for $k \leq l_j$ have been executed before $\pasevent(N_j)$.
Formally, in the definition of injective correspondences, we can
define $\phi_{jk}$ such that $\phi_{jk}(\step) \leq \step$ when
$\phi_{jk}$ is defined. (The inequality $\step' \leq \step$ means that $\step'$ occurs
before $\step$ in the trace.) Indeed, otherwise, by considering the
prefix of the trace that stops just after $\step$, we would contradict
the correspondence. In this section, we exploit this point to define more
general properties involving the ordering of events. 

Let us first consider some examples.  
Using the process of Section~\ref{sec:example}, we will denote by 
\begin{equation}
\begin{split}
&\pasevent(e_B(x_1, \ab x_2, \ab x_3, \ab x_4)) \rightsquigarrow
(\inj\ \pasevent(e_3(x_1, \ab x_2, \ab x_3, \ab x_4)) \rightsquigarrow\\
&\quad (\inj\ \pasevent(e_2(x_1, \ab x_2, \ab x_3, \ab x_4)) \rightsquigarrow
\inj\ \pasevent(e_1(x_1, \ab x_2, \ab x_3))))
\end{split}
\label{corresp:nested1}
\end{equation}
the correspondence that
means that each execution of the event $e_B(x_1, x_2, x_3, x_4)$
corresponds to distinct executions of the events
$e_1(x_1, x_2, x_3)$, $e_2(x_1, x_2, x_3, x_4)$, and $e_3(x_1, x_2, x_3, x_4)$
in this order: each execution of $e_B(x_1, x_2, x_3, x_4)$ is preceded
by a distinct execution of $e_3(x_1, x_2, x_3, x_4)$, which is itself preceded by a distinct execution of $e_2(x_1, x_2, x_3, x_4)$, which is itself preceded by a distinct execution of $e_1(x_1, x_2, x_3)$. 
This correspondence shows that, when $B$ terminates
the protocol talking with $A$, $A$ and $B$ have exchanged
all messages of the protocol in the expected order.
This correspondence is not equivalent to the conjunction of the
correspondences $\pasevent(e_B(x_1, \ab x_2, \ab x_3, \ab x_4)) \rightsquigarrow
\inj\ \pasevent(e_3(x_1, \ab x_2, \ab x_3, \ab x_4))$, 
$\pasevent(e_3(x_1, \ab x_2, \ab x_3, \ab x_4)) \rightsquigarrow
\inj\ \pasevent(e_2(x_1, \ab x_2, \ab x_3, \ab x_4))$, and
$\pasevent(e_2(x_1, \ab x_2, \ab x_3, \ab x_4)) \rightsquigarrow
\inj\ \pasevent(e_1(x_1, \ab x_2, \ab x_3))$, because 
\eqref{corresp:nested1} may be true even when, in order to prove that
$e_2$ is executed, we need to know that $e_B$ has been executed,
and not only that $e_3$ has been executed and, similarly, in
order to prove that $e_1$ has been executed, we need to know that
$e_B$ has been executed, and not only that $e_2$ has been executed.
Using general correspondences such as~\eqref{corresp:nested1} is therefore
strictly more expressive than using injective correspondences.
A correspondence similar to~\eqref{corresp:nested1}
has been used in our study of the Just
Fast Keying protocol, one of the proposed replacements for IKE in
IPSec, in collaboration with Mart{\'\i}n Abadi and
C{\'e}dric Fournet~\cite[Appendix~B.5]{Abadi07}. 

As a more generic example, the correspondence 
$\pasevent(M) \Rightarrow\ab  \sor_{j=1}^{m} 
\big(\pasevent(M_j) \ab \rightsquigarrow\ab \sand_{k = 1}^{l_j} \big(\injopt_{jk} 
\pasevent(M_{jk}) \rightsquigarrow \sor_{j'=1}^{m_{jk}} \sand_{k'=1}^{l_{jkj'}}
\injopt_{jkj'k'} \pasevent(M_{jkj'k'}) \big)\big)$
means that, if an instance of
$\asevent(M)$ has been executed, then there exists $j$ such that this
instance of $\asevent(M)$ is an instance of $\asevent(M_j)$ and for
all $k$, a corresponding instance of $\asevent(M_{jk})$ has been
executed before $\asevent(M_j)$, and there exists $j'_k$ such that
for all $k'$ a corresponding instance of
$\asevent(M_{j k j'_k k'})$ has been executed before
$\asevent(M_{jk})$.

Let us now consider the general definition.
We denote by $\overline{k}$ a sequence of indices $k$.
The empty sequence is denoted $\epsilon$. 
When $\overline{j} = j_1\ldots j_n$ and 
$\overline{k} = k_1\ldots k_n$ are sequences of
the same length, we denote by $\overline{jk}$
the sequence obtained by taking alternatively one index in each
sequence $\overline{j}$ and $\overline{k}$:
$\overline{jk} = j_1 k_1 \ldots j_n k_n$. We sometimes use $\overline{jk}$ as
an identifier that denotes a sequence obtained in this way; for instance,
``for all $\overline{jk}$, $\phi_{\overline{jk}}$ is injective'' abbreviates
``for all $\overline{j}$ and $\overline{k}$ of the same length, $\phi_{\overline{jk}}$ is injective''. We only consider sequences $\overline{jk}$ that occur
in the correspondence. For instance, for the correspondence
$\pasevent(M) \Rightarrow\ab  \sor_{j=1}^{m} 
\big(\pasevent(M_j) \ab \rightsquigarrow\ab \sand_{k = 1}^{l_j} \big(\injopt_{jk} 
\pasevent(M_{jk}) \rightsquigarrow \sor_{j'=1}^{m_{jk}} \sand_{k'=1}^{l_{jkj'}}
\injopt_{jkj'k'} \pasevent(M_{jkj'k'}) \big)\big)$, we consider the
sequences $\overline{jk} = \epsilon$, $\overline{jk} = jk$, and 
$\overline{jk} = jkj'k'$ where $1 \leq j \leq m$, $1 \leq k \leq l_j$,
$1 \leq j' \leq m_{jk}$, and $1 \leq k' \leq l_{jkj'}$.

Given a family of indices $J = (j_{\overline{k}})_{\overline{k}}$ indexed 
by sequences of indices $\overline{k}$, we define
$\indp{\overline{k}}{J}$ by $\indp{\epsilon}{J} = \epsilon$ and 
$\indp{\overline{k}k}{J} = \indp{\overline{k}}{J} j_{\overline{k}} k$.
Less formally, if $\overline{k} = k_1k_2k_3 \ldots$, we have 
$\indp{\overline{k}}{J} = j_{\epsilon} k_1 j_{k_1} k_2 j_{k_1k_2} k_3\ldots$
Intuitively, the correspondence contains disjunctions over indices $j$ and
conjunctions over indices $k$, so we would like to express quantifications
of the form $\exists j_{\epsilon} \forall k_1 \exists j_{k_1} \forall k_2 \exists j_{k_1k_2} \forall k_3 \ldots$
on the sequence $j_{\epsilon} k_1 j_{k_1} k_2 j_{k_1k_2} k_3\ldots$. The notation $\indp{\overline{k}}{J}$
allows us to replace such a quantification with the quantification $\exists J 
\forall \overline{k}$ on the sequence $\indp{\overline{k}}{J}$.
\begin{definition}\label{def:nestedquery}
The closed process $P_0$ \emph{satisfies the correspondence} 
\[\pasevent(M) \Rightarrow \sor_{j=1}^{m} 
\left(\pasevent(M_j) \rightsquigarrow \sand_{k = 1}^{l_j} \injopt_{jk} q_{jk}\right)\]
where 
\[q_{\overline{jk}} = \pasevent(M_{\overline{jk}}) \rightsquigarrow \sor_{j=1}^{m_{\overline{jk}}}
\sand_{k = 1}^{l_{\overline{jk}j}} \injopt_{\overline{jk}jk} q_{\overline{jk}jk}\]
against $\rw$-adversaries if and only if, for any
$\rw$-adversary $Q$, for any $\env_0$ containing $\fn(P_0) \cup \rw
\cup \fn(M) \cup \bigcup_j \fn(M_j) \cup \bigcup_{\overline{jk}} \fn(M_{\overline{jk}})$, for any trace
$\trace = \env_0, \{ P_0, Q \} \rightarrow^* \env', \pset'$, there exists 
a function $\phi_{\overline{jk}}$ for each non-empty $\overline{jk}$,
such that for all non-empty $\overline{jk}$, $\phi_{\overline{jk}}$ maps 
a subset of steps of $\trace$ to steps of $\trace$ and
\begin{itemize}

\item For all $\step$, if the event $\asevent(\sigma M)$ 
is executed at step $\step$ in $\trace$ for some $\sigma$,
then there exist $\sigma'$ and $J = (j_{\overline{k}})_{\overline{k}}$
such that $\sigma' M_{j_\epsilon} = \sigma M$ and, 
for all non-empty $\overline{k}$, $\phi_{\indp{\overline{k}}{J}}(\step)$ is 
defined and
$ \asevent(\sigma' M_{\indp{\overline{k}}{J}})$ is executed at
step $\phi_{\indp{\overline{k}}{J}}(\step)$ in $\trace$.

\item For all non-empty $\overline{jk}$, 
if $\injopt_{\overline{jk}} = \inj$, then $\phi_{\overline{jk}}$ is injective.

\item For all non-empty $\overline{jk}$, for all $j$ and $k$,  
if $\phi_{\overline{jk}jk}(\step)$ is defined, then 
$\phi_{\overline{jk}}(\step)$ is defined and 
$\phi_{\overline{jk}jk}(\step) \leq \phi_{\overline{jk}}(\step)$.
For all $j$ and $k$, if $\phi_{jk}(\step)$ is defined, then 
$\phi_{jk}(\step) \leq \step$.

\end{itemize}
\end{definition}

We abbreviate by $q_{\overline{jk}} = \pasevent(M_{\overline{jk}})$
the correspondence $q_{\overline{jk}} = \pasevent(M_{\overline{jk}}) \rightsquigarrow \sor_{j=1}^{m_{\overline{jk}}}
\sand_{k = 1}^{l_{\overline{jk}j}} \injopt_{\overline{jk}jk} q_{\overline{jk}jk}$ when $m_{\overline{jk}} = 1$ and $l_{\overline{jk}1} = 0$, that is,
the disjunction $\sor_{j=1}^{m_{\overline{jk}}}
\sand_{k = 1}^{l_{\overline{jk}j}} \injopt_{\overline{jk}jk} q_{\overline{jk}jk}$ is true. Injective correspondences are then a particular case
of general correspondences.

The function $\phi_{\overline{jk}}$ maps the execution steps
of instances of $\asevent(M)$ to the execution steps
of the corresponding instances of $\asevent(M_{\overline{jk}})$.
The first item of Definition~\ref{def:nestedquery} guarantees 
that the required events have been executed. The second item means that,
when the $\inj$ marker is present, the correspondence is injective.
Finally, the third item guarantees that the events have been executed
in the expected order.

\begin{example}
Let us consider again the correspondence~\eqref{corresp:nested1}.
Using the notations of Definition~\ref{def:nestedquery}, this correspondence
is written $\pasevent(e_B(x_1, \ab x_2, \ab x_3, \ab x_4)) \rightsquigarrow \inj \ q_{11}$ (or $\pasevent(e_B(x_1, \ab x_2, \ab x_3, \ab x_4)) \Rightarrow \pasevent(e_B(x_1, \ab x_2, \ab x_3, \ab x_4)) \rightsquigarrow \inj \ q_{11}$), where $q_{11} = \pasevent(e_3(x_1, \ab x_2, \ab x_3, \ab x_4)) \rightsquigarrow \inj\ q_{1111}$, $q_{1111} = \pasevent(e_2(x_1, \ab x_2, \ab x_3, \ab x_4)) \rightsquigarrow \inj\ q_{111111}$, and $q_{111111} = \pasevent(e_1(x_1, \ab x_2, \ab x_3))$.
By Definition~\ref{def:nestedquery}, this correspondence
means that there exist functions $\phi_{11}$, $\phi_{1111}$, and $\phi_{111111}$
such that:
\begin{itemize}

\item For all $\step$, if the event $\asevent(\sigma e_B(x_1, x_2, x_3, x_4))$ 
is executed at step $\step$ for some $\sigma$,
then $\phi_{11}(\step)$, $\phi_{1111}(\step)$, and $\phi_{111111}(\step)$ are defined, 
and
$\asevent(\sigma e_3(x_1, \ab x_2, \ab x_3, \ab x_4))$ is executed at step $\phi_{11}(\step)$, 
$\asevent(\sigma e_2(x_1, \ab x_2, \ab x_3, \ab x_4))$ is executed at step $\phi_{1111}(\step)$, 
and $\asevent(\sigma e_1(x_1, \ab x_2, \ab x_3))$ is executed at step $\phi_{111111}(\step)$. (Here, $\sigma' =\sigma$ since all variables of the correspondence occur in $\pasevent(e_B(x_1, \ab x_2, \ab x_3, \ab x_4))$. Moreover, $j_{\overline{k}} = 1$ for all $\overline{k}$ and the non-empty sequences $\overline{k}$ are 1, 11, and 111, since all conjunctions and disjunctions have a single element. The sequences $\indp{\overline{k}}{J}$ are then 11, 1111, and 111111.)

\item The functions $\phi_{11}$, $\phi_{1111}$, and $\phi_{111111}$ are injective,
so distinct executions of $e_B(x_1, \ab x_2, \ab x_3, \ab x_4)$ correspond to 
distinct executions of $e_1(x_1, \ab x_2, \ab x_3)$, $e_2(x_1, \ab x_2, \ab x_3, \ab x_4)$, and 
$e_3(x_1, \ab x_2, \ab x_3, \ab x_4)$.

\item When $\phi_{111111}(\step)$ is defined,
$\phi_{111111}(\step) \leq \phi_{1111}(\step) \leq \phi_{11}(\step) \leq \step$,
so the events $e_1(x_1, \ab x_2, \ab x_3)$, $e_2(x_1, \ab x_2, \ab x_3, \ab x_4)$, and 
$e_3(x_1, \ab x_2, \ab x_3, \ab x_4)$ are executed in this order, before 
$e_B(x_1, \ab x_2, \ab x_3, \ab x_4)$.

\end{itemize}
\end{example}
Similarly, general correspondences allow us to express that, if
a protocol participant successfully terminates with honest interlocutors,
then the expected messages of the protocol have been exchanged
between the protocol participants, in the expected order. This notion is 
the formal counterpart of the notion of matching conversations initially
introduced in the computational model by Bellare and Rogaway~\cite{Bellare93}. 
This notion of authentication is also used in~\cite{Datta05}.

We first focus on non-injective correspondences,
and postpone the treatment of general correspondences to
Section~\ref{sect:injag}.

\section{Automatic Verification: from Secrecy to Correspondences}
\label{sec:veriffirst}

Let us first summarize our analysis for secrecy. The clauses use two
predicates: $\attacker$ and $\mess$, where $\attacker(M)$ means that
the attacker may have the message $M$ and $\mess(M,M')$ means that the
message $M'$ may be sent on channel $M$. The clauses relate atoms that
use these predicates as follows. A clause $\mess(M_1, M'_1) \wedge
\ldots \wedge \mess(M_n, M'_n) \rewrite \mess (M, M')$ is generated
when the process outputs $M'$ on channel $M$ after receiving $M'_1$,
\ldots, $M'_n$ on channels $M_1$, \ldots, $M_n$ respectively.  A
clause $\attacker(M_1) \wedge \ldots \wedge \attacker(M_n)\rewrite
\attacker(M)$ is generated when the attacker can compute $M$ from
$M_1$, \ldots, $M_n$.  The clause $\mess(x,y) \wedge \attacker(x)
\rewrite \attacker(y)$ means that the attacker can listen on channel
$x$ when he has $x$, and the clause $\attacker(x) \wedge \attacker(y)
\rewrite \mess(x,y)$ means that the attacker can send any message $y$
he has on any channel $x$ he has. 
When $\attacker(M)$ is derivable from the clauses the attacker
\emph{may} have $M$, that is, when $\attacker(M)$ is not derivable
from the clauses, we are sure that the attacker cannot have $M$, but
the converse is not true, because the Horn clauses can be applied any
number of times, which is not true in general for all actions of the
process. Similarly, when $\mess(M,M')$ is derivable from the clauses,
the message $M'$ \emph{may} be sent on channel $M$.
Hence our analysis overapproximates the execution of actions.

Let us now consider that we want to prove a correspondence, for instance 
$\pasevent(e_1(x)) \rightsquigarrow \pasevent(e_2(x))$.
In order to prove this correspondence, we can overapproximate the
executions of event $e_1$: if we prove the correspondence with 
this overapproximation, it will also hold in the exact semantics.
So we can easily extend our analysis for secrecy with an additional
predicate $\pasevent$, such that $\pasevent(M)$ means that 
$\asevent(M)$ may have been executed. We generate clauses
$\mess(M_1, M'_1) \wedge
\ldots \wedge \mess(M_n, M'_n) \rewrite \pasevent(M)$ when the process
executes $\asevent(M)$ after receiving $M'_1$,
\ldots, $M'_n$ on channels $M_1$, \ldots, $M_n$ respectively.
However, such an overapproximation cannot be done for the event $e_2$:
if we prove the correspondence after overapproximating the
execution of $e_2$, we are not really sure that $e_2$ will be executed,
so the correspondence may be wrong in the exact semantics.
Therefore, we have to use a different method for treating $e_2$.

We use the following idea: we fix the exact set $\eventset$ of
allowed events $e_2(M)$ and, in order to prove
$\pasevent(e_1(x)) \rightsquigarrow \pasevent(e_2(x))$,
we check that only events $e_1(M)$ for $M$ such that $e_2(M) \in \eventset$
can be executed. If we prove this property for any value
of $\eventset$, we have proved the desired correspondence.
So we introduce a predicate $\pasbegin$, such that $\pasbegin(e_2(M))$ 
is true if and only if $e_2(M) \in \eventset$.
We generate clauses $\mess(M_1, M'_1) \wedge \ldots \wedge \mess(M_n,
M'_n) \wedge \pasbegin(e_2(M_0)) \rewrite \mess(M, M')$ when the
process outputs $M'$ on channel $M$ after executing the event $e_2(M_0)$ and
receiving $M'_1$, \ldots, $M'_n$ on channels $M_1$, \ldots, $M_n$
respectively.  In other words, the output of $M'$ on channel $M$ can
be executed only when $\pasbegin(e_2(M_0))$ is true, that is,
$e_2(M_0) \in \eventset$.  
(When the output of $M'$ on channel $M$ is under several events, the
clause contains several $\pasbegin$ atoms in its hypothesis. We also
have similar clauses with $\pasevent(e_1(M))$ instead of $\mess(M,M')$
when the event $e_1$ is executed after executing $e_2$ and
receiving $M'_1$, \ldots, $M'_n$ on channels $M_1$, \ldots, $M_n$
respectively.)

For instance, if the events $e_2(M_1)$ and $e_2(M_2)$ are executed in
a certain trace of the protocol, we define $\eventset = \{ e_2(M_1), 
e_2(M_2) \}$,
so that $\pasbegin(e_2(M_1))$ and $\pasbegin(e_2(M_2))$ are true and
all other $\pasbegin$ facts are false. Then we show that the only
events $e_1$ that may be executed are $e_1(M_1)$ and $e_1(M_2)$.
We prove a similar result for all values of $\eventset$, which
proves the desired correspondence. 

In order to determine whether an atom is derivable from the clauses,
we use a resolution-based algorithm. The resolution is performed
for an unknown value of $\eventset$. So, basically, 
we keep $\pasbegin$ atoms without trying to evaluate
them (which we cannot do since $\eventset$ is unknown).
In the vocabulary of resolution, we never select $\pasbegin$ atoms.
(We detail this point in Section~\ref{sec:basicsolv}.) 
Thus the obtained result holds for any value of $\eventset$,
which allows us to prove correspondences. 
In order to prove the correspondence $\pasevent(e_1(x))
\rightsquigarrow \pasevent(e_2(x))$, we show that $\pasevent(e_1(M))$
is derivable only when $\pasbegin(e_2(M))$ holds. We transform the
initial set of clauses into a set of clauses that derives the same
atoms. If, in the obtained set of clauses, all clauses that conclude
$\pasevent(e_1(M))$ contain $\pasbegin(e_2(M))$ in their hypotheses,
then $\pasevent(e_1(M))$ is derivable only when $\pasbegin(e_2(M))$
holds, so the desired correspondence holds.

We still have to solve one problem. For simplicity, we have considered
that terms, which represent messages, are directly used in clauses.
However, in order to represent nonces in our analysis for secrecy, we
use a special encoding of names: a name $a$ created by a restriction
$\Res{a}$ is represented by a function $a[M_1, \ldots, M_n]$ of the
messages $M_1, \ldots, M_n$ received above the restriction, so that
names created after receiving different messages are distinguished in
the analysis (which is important for the precision of the analysis). 
However, this encoding still merges names created by the
same restriction after receiving the same messages. For example, in
the process $\Repl{} \cinput{c}{x} \Res{a}$, the names created by
$\Res{a}$ are represented by $a[x]$, so several names created for the same
value of $x$ are merged. This merging is not acceptable for the
verification of correspondences, because when we prove
$\pasevent(e_1(x)) \rightsquigarrow \pasevent(e_2(x))$, we must make
sure that $x$ contains exactly the same names in $e_1(x)$ and in
$e_2(x)$.
In order to solve this problem, we label each replication with a
\emph{session identifier} $i$, which is an integer that takes a
different value for each copy of the process generated by the
replication.  We add session identifiers as arguments
to our encoding of names, which becomes $a[M_1, \ldots, M_n, i_1,
\ldots, i_{n'}]$ where $i_1, \ldots, i_{n'}$ are the session
identifiers of the replications above the restriction $\Res{a}$.
For example, in the process $\Repl{} \cinput{c}{x} \Res{a}$,
the names created by $\Res{a}$ are represented by $a[x,i]$.
Each execution of the restriction is then associated with
a distinct value of the session identifiers $i_1, \ldots, i_{n'}$,
so each name has a distinct encoding.
We detail and formalize this encoding in Section~\ref{sect:instr}.

\section{From Processes to Horn Clauses}\label{sec:Horn}

In this section, we first explain the instrumentation of processes
with session identifiers. Next, we explain the translation
of processes into Horn clauses.

\subsection{Instrumented Processes}\label{sect:instr}

We consider a closed process $P_0$ representing the protocol we wish
to check.  We assume that the bound names of $P_0$ have been renamed
so that they are pairwise distinct and distinct from names in $\rw \cup 
\fn(P_0)$ and in the correspondence to prove. 
We denote by $Q$ a particular adversary; below, we prove the correspondence
properties for any $Q$.
Furthermore, we assume that, in the initial configuration
$E_0, \{ P_0, Q \}$, the names of $E_0$ not in $\rw \cup \fn(P_0)$ or
in the correspondence to prove have been renamed to fresh names, and the
bound names of $Q$ have been renamed so that they are pairwise
distinct and fresh. (These renamings do not change the satisfied
correspondences, since $\Res{a}P$ and the renamed process $\Res{a'}P\{a'/a\}$
reduce to the same configuration by (Red Res).)
After encoding names, the terms are represented by
\emph{patterns} $p$ (or ``terms'', but we prefer the word ``patterns''
in order to avoid confusion), which are generated by the following
grammar:
\begin{defn}
\categ{p}{patterns}\\
\entry{x,y,z,i}{variable}\\
\entry{a[p_1, \ldots, p_n, i_1, \ldots, i_{n'}]}{name}\\
\entry{f(p_1, \ldots, p_n)}{constructor application}
\end{defn}
For each name $a$ in $P_0$ we have a corresponding pattern construct
$a[p_1, \ldots, p_n, i_1, \ab \ldots, i_{n'}]$.  We treat $a$
as a function symbol, and write $a[p_1, \ldots, p_n, i_1, \ldots,
i_{n'}]$ rather than $a(p_1, \ldots, p_n, i_1, \ab \ldots,
\ab i_{n'})$ only to distinguish names from constructors. 
The symbol $a$ in $a[\ldots]$ is called a \emph{name function symbol}.
If $a$ is a free name, then its encoding is simply $a[\,]$.
If $a$ is bound by a restriction $\Res{a} P$ in $P_0$, then
its encoding $a[\ldots]$ takes as argument session identifiers $i_1, \ldots,
i_{n'}$, which can be constant session identifiers $\lambda$ or
variables $i$ (taken in a set $V_s$ disjoint from the set $V_o$ of
ordinary variables).  There is one session identifier for each
replication above the restriction $\Res{a}$. The pattern $a[\ldots]$
may also take as argument patterns $p_1, \ldots, p_n$ containing the
messages received by inputs above the
restriction $\Res{a} P$ in the abstract syntax tree of $P_0$
and the result of destructor applications above the
restriction $\Res{a} P$. (The precise definition is given below.)

In order to define formally the patterns associated with a name,
we use a notion of instrumented processes.
The syntax of instrumented processes is defined as follows:
\begin{itemize}

\item The replication $\Repl{P}$ is labeled with a variable $i$ in $V_s$:
  $\ReplInstr{i}{P}$. The process $\ReplInstr{i}{P}$ represents copies
  of $P$ for a countable number of values of $i$. The variable $i$ is
  a session identifier. It indicates which copy of $P$, that is, which
  session, is executed.

\item The restriction $\Res{a}P$ is labeled with a restriction label $\rlbl$:
  $\ResInstr{a}{\rlbl} P$, where $\rlbl$ is either $a[M_1,\ab
  \ldots,\ab M_n,\ab i_1, \ab \ldots, \ab i_{n'}]$ for restrictions in honest
  processes or $\advnfs[a[i_1, \ab \ldots, \ab i_{n'}]]$ for restrictions in
  the adversary.  The symbol $\advnfs$ is a special name function
  symbol, distinct from all other such symbols. Using a specific
  instrumentation for the adversary is helpful so that all names
  generated by the adversary are encoded by instances of $\advnfs[x]$.
  They are therefore easy to generate.  This labeling of restrictions
  is similar to a
  Church-style typing: $\rlbl$ can be considered as the type of $a$.
  (This type is polymorphic since it can contain variables.)

\end{itemize}
The instrumented processes are then generated by the following grammar:
\begin{defn}
\categ{P, Q}{instrumented processes}\\
\entry{\ReplInstr{i}{P}}{replication}\\
\entry{\ResInstr{a}{\rlbl}P}{restriction}\\
\entry{\ldots \text{(as in the standard calculus)}}{}
\end{defn}
For instrumented processes, a semantic configuration $S, \env, \pset$ 
consists of a set
$S$ of session identifiers that have not yet been used by $\pset$, an
environment $\env$ that is a mapping from names to closed patterns of
the form $a[\ldots]$, and a finite multiset of instrumented processes
$\pset$. The first semantic configuration uses any countable set of
session identifiers $S_0$. The domain of $\env$ must always contain all
free names of processes in $\pset$, and the initial environment maps
all names $a$ to the pattern $a[\,]$. The semantic rules
(Red Repl) and (Red Res) become:
\begin{align}
&S,\env,\pset \cup \{ \,\ReplInstr{i}{P}\, \} \rightarrow S \setminus \{ \lambda\},\env, \pset \cup \{ \,P\{\lambda/i\}, \ReplInstr{i}{P}\, \}\text{ where $\lambda \in S$}\tag{Red Repl}\\
\begin{split}
&S,\env,\pset \cup \{ \,\ResInstr{a}{\rlbl}P\, \} \\
&\qquad\rightarrow S,\env[ a' \mapsto \env(\rlbl)\, ], \pset \cup \{ \,P\{a'/a\}\,\}\text{ if $a' \notin \dom(\env)$}
\end{split}\tag{Red Res}
\end{align}
where the mapping $\env$ is extended to all terms 
as a substitution by $\env(f(M_1, \ab \ldots, M_n)) = f(\env(M_1), \ab
\ldots, \ab \env(M_n))$ and to restriction labels
by $\env(a[M_1, \ab \ldots,\ab M_n, \ab i_1, \ab \ldots, \ab i_{n'}]) = 
a[\env(M_1), \ab \ldots, \ab \env(M_n), \ab i_1, \ab \ldots, \ab i_{n'}]$
and $\env(\advnfs[a[i_1, \ab \ldots, \ab i_{n'}]]) = \advnfs[a[i_1, \ab 
\ldots, \ab i_{n'}]]$, 
so that it maps terms and restriction labels to patterns.
The rule (Red Repl) takes an unused constant session identifier $\lambda$ in $S$,
and creates a copy of $P$ with session identifier $\lambda$.
The rule (Red Res) creates a fresh name $a'$, substitutes it
for $a$ in $P$, and adds to the
environment $E$ the mapping of $a'$ to its encoding
$\env(\rlbl)$.
Other semantic rules $\env,\pset \rightarrow \env, \pset'$ simply become
$S,\env,\pset \rightarrow S,\env, \pset'$. 

The instrumented process $P'_0 = \instr{P_0}$ associated with the
process $P_0$ is built from $P_0$ as follows:
\begin{itemize}

\item We  label each replication $\Repl{P}$ of $P_0$ with a distinct, fresh 
session identifier $i$, so that it becomes $\ReplInstr{i}{P}$.

\item We label each restriction $\Res{a}$ of $P_0$ with $a[t,s]$, so
  that it becomes $\ResInstr{a}{a[t,s]}$, where $s$ is the sequence of
  session identifiers that label replications above $\Res{a}$ in the
  abstract syntax tree of $P'_0$, in the order from top to bottom; $t$
  is the sequence of variables $x$ that store received messages in
  inputs $\cinput{M}{x}$ above $\Res{a}$ in $P_0$ and results of
  non-deterministic destructor applications
  $\letfun{x}{g(\ldots)}{P}{Q}$ above $\Res{a}$ in $P_0$.
(A destructor is said to be non-deterministic when it may return
several different results for the same arguments.
Adding the result of destructor applications to $t$ is
useful to improve precision, only for non-deterministic destructors.
For deterministic destructors, the result of the destructor can be uniquely
determined from the other elements of $t$, so the addition is useless.
If we add the result of non-deterministic destructors to $t$,
we can show that the relative completeness result of~\cite{Abadi04c}
still holds in the presence of non-deterministic destructors. This result 
shows that, for secrecy, the Horn clause approach is at least as precise as 
a large class of type systems.)

Hence names are represented by functions $a[t,s]$ of the inputs and
results of destructor applications in $t$ and the session identifiers
in $s$. In each trace of the process, at most one name corresponds to
a given $a[t,s]$, since different copies of the restriction have
different values of session identifiers in $s$. Therefore, different
names are not merged by the verifier.

\end{itemize}
For the adversary, we use a slightly different instrumentation.
We build the instrumented process $Q' = \instradv{Q}$ as follows:
\begin{itemize}

\item We  label each replication $\Repl{P}$ of $Q$ with a distinct, fresh 
session identifier $i$, so that it becomes $\ReplInstr{i}{P}$.

\item We label each restriction $\Res{a}$ of $Q$ with $\advnfs[a[s]]$, so
that it becomes $\ResInstr{a}{\advnfs[a[s]]}$, where 
$s$ is the sequence of session identifiers that label
replications above $\Res{a}$ in $Q'$.
(Including the session identifiers as arguments of nonces is necessary
for soundness, as discussed in Section~\ref{sec:veriffirst}.
Including the messages previously received as arguments of nonces is
important for precision in the case of honest processes, in order to
relate the nonces to these messages. It is however useless for the
adversary: since we consider any $\rw$-adversary $Q$, we have no
definite information on the relation between nonces generated by the
adversary and messages previously received by the adversary.)

\end{itemize}

\begin{Remark}
By moving restrictions downwards in the syntax tree of the process
(until the point at which the fresh name is used), one can add more
arguments to the pattern that represents the fresh name, when the
restriction is moved under an input, replication, or destructor
application. Therefore,
this transformation can make our analysis more precise. The tool can
perform this transformation automatically.
\end{Remark}

\begin{example}
The instrumentation of the
process of Section~\ref{sec:example} yields:
{\allowdisplaybreaks\begin{align*}
&P_A'(\skA, \pkA, \pkB) = \ReplInstr{i_A}{} \cinput{c}{x\_\pkB}.
\ResInstr{a}{a[x\_\pkB, i_A]}\ldots\ResInstr{r_1}{r_{1}[x\_\pkB, i_A]}\ldots\\*
&\qquad\cinput{c}{m}\ldots \ResInstr{r_3}{r_{3}[x\_\pkB, m, i_A]]}\\
&P_B'(\skB, \pkB, \pkA) = \ReplInstr{i_B}{}  \cinput{c}{m'}
\ldots
\ResInstr{b}{b[m', i_B]}\ldots\ResInstr{r_2}{r_{2}[m', i_B]}\ldots\\
&P' = \ResInstr{\skA}{\skA[\,]}\ResInstr{\skB}{\skB[\,]}\ldots(P_A'(\skA, \pkA, \pkB) \parpop P_B'(\skB, \pkB, \pkA))
\end{align*}}%
The names created by the restriction $\Res{a}$ will be represented
by the pattern $a[x\_\pkB, \ab i_A]$, so we have a different pattern 
for each copy of the process, indexed by $i_A$, and the pattern
also records the public key $x\_\pkB$ of the interlocutor of $A$.
Similarly, the names created by the restriction $\Res{b}$ will
be represented by the pattern $b[m', i_B]$.
\end{example}

The semantics of instrumented processes allows exactly the same
communications and events as the one of standard processes. More
precisely, let $\pset$ be a multiset of instrumented processes. We
define $\delete(\pset)$ as the multiset of processes of $\pset$
without the instrumentation. Thus we have:

\begin{proposition}\label{prop:equivsem}
If $\env_0, \{ P_0, Q \} \rightarrow^* \env_1, \pset_1$, then there exist
$\env_1'$ and $\pset_1'$ such that for any $S$, countable set of session
identifiers, there exists $S'$ such that $S, \{ a \mapsto a[\,] \mid a \in \env_0 \}, \ab \{
\instr{P_0}, \ab \instradv{Q} \} \ab \rightarrow^* \ab S',
\env_1',\pset_1'$, $\dom(\env_1') = \env_1$, $\delete(\pset_1') =
\pset_1$, and both traces execute the same events at the same steps
and satisfy the same atoms.

Conversely, if 
$S, \{ a \mapsto a[\,] \mid a \in \env_0 \},\{
\instr{P_0}, \instradv{Q}\} \rightarrow^* S',
\env_1',\pset_1'$, then $\env_0, \{ P_0, Q\} \rightarrow^* \dom(\env_1'), \delete(\pset_1')$, and both traces execute the same events at the same steps
and satisfy the same atoms.
\end{proposition}
\begin{proof}
This is an easy proof by induction on the length of the traces. 
The reduction rules applied in both traces are rules with the same name.
\proofcomplete
\end{proof}

We can define correspondences for instrumented processes. These
correspondences and the clauses use \emph{facts} defined by the
following grammar:
\begin{defn}
\categ{F}{facts}\\
\entry{\attacker(p)}{attacker knowledge}\\
\entry{\mess(p,p')}{message on a channel}\\
\entry{\pasbegin(p)}{must-event}\\
\entry{\pasend(p)}{may-event}
\end{defn}
The fact $\attacker(p)$ means that the attacker may have
$p$, and the fact $\mess(p,p')$ means that the message $p'$ may appear
on channel $p$. The fact $\pasbegin(p)$ means that $\asevent(M)$ must have been
executed with $M$ corresponding to $p$, and $\pasend(p)$ that
$\asevent(M)$ may have been executed with $M$ corresponding to
$p$. We use the word ``fact'' to distinguish them from
atoms $\attacker(M)$, $\mess(M,M')$, and $\pasevent(M)$. The correspondences
do not use the fact $\pasbegin(p)$, but the clauses use it.

The mapping $\env$ of a semantic configuration is extended to  
atoms by $\env(\attacker(M)) = \attacker(\env(M))$,
$\env(\mess(M,\ab M')) = \mess(\env(M), \ab \env(M'))$, and $\env(\pasevent(M)) =
\pasevent(\env(M))$, so that it maps atoms to facts.
We define that an instrumented trace $\trace$ satisfies an atom 
$\act$ by naturally adapting Definition~\ref{def:exectrace}.  
When $F$ is not $\pasbegin(p)$, we say that an
instrumented trace $\trace = S_0, \env_0, \pset_0 \rightarrow^* S',
\env', \pset'$ satisfies a fact $F$ when there exists an atom $\act$
such that $\trace$ satisfies $\act$ and $\env'(\act) = F$.
We also define that $\asevent(M)$ is executed at step $\step$ in 
the instrumented trace $\trace$ by naturally adapting 
Definition~\ref{def:execevent}. 
We say that $\asevent(p)$ is executed at step $\step$ in 
the instrumented trace $\trace = S_0, \env_0, \pset_0 \rightarrow^* S',
\env', \pset'$ when there exists a term $M$
such that $\asevent(M)$ is executed at step $\step$ in $\trace$ and $\env'(M) = p$.

\begin{definition}
Let $P_0$ be a closed process and $P'_0 = \instr{P_0}$. 
The instrumented process $P'_0$ satisfies the 
correspondence
\[\acti \Rightarrow \sor_{j=1}^{m} \left(\acti_j \rightsquigarrow 
\sand_{k = 1}^{l_j} \pasevent(p_{jk})\right)\]
against $\rw$-adversaries if and only if, for any
$\rw$-adversary $Q$, for any trace $\trace = S_0, \env_0, \{ P'_0, Q'
\}\rightarrow^* S', \env', \pset'$, with $Q' = \instradv{Q}$, 
$\env_0(a) = a[\,]$ for all $a
\in \dom(\env_0)$, and $\fn(P'_0) \cup \rw \subseteq \dom(\env_0)$, if
$\trace$ satisfies $\sigma \acti$
for some substitution $\sigma$, then there exist $\sigma'$
and $j \in \{ 1, \ldots, m \}$ such that $\sigma' \acti_j = \sigma \acti$ and
for all $k \in \{ 1, \ldots, l_j \}$, $\trace$ satisfies 
$\pasevent(\sigma' p_{jk})$.
\end{definition}

A correspondence for instrumented processes implies
a correspondence for standard processes, as shown
by the following lemma, proved 
in Appendix~\ref{app:instr}.

\begin{lemma}\label{lem:instrcorresp}
Let $P_0$ be a closed process and $P'_0 =
\instr{P_0}$.  Let $M_{jk}$
($j \in \{ 1, \ldots, m\}$, $k \in \{ 1, \ldots, l_j\}$)
be terms; let $\act$ and $\act_j$ ($j \in \{ 1, \ldots, m\}$) be atoms. 
Let $p_{jk}, \acti, \acti_j$
be the patterns and facts obtained by replacing names $a$ with 
patterns $a[\,]$ in
the terms and atoms $M_{jk}, \act, \act_j$ respectively. 
If $P'_0$ satisfies the correspondence
\[\acti \Rightarrow \sor_{j=1}^{m} \left(\acti_j \rightsquigarrow 
\sand_{k = 1}^{l_j} \pasevent(p_{jk})\right)\]
against $\rw$-adversaries then $P_0$ satisfies the correspondence
\[\act \Rightarrow \sor_{j=1}^{m} \left(\act_j \rightsquigarrow 
\sand_{k = 1}^{l_j} \pasevent(M_{jk})\right)\]
against $\rw$-adversaries.
\end{lemma}

For instrumented processes, we can specify
properties referring to bound names of the process, which are
represented by patterns. Such a specification is impossible in
standard processes, because bound names can be renamed, so they cannot
be referenced in terms in correspondences. 

\subsection{Generation of Horn Clauses}\label{sec:clausegen}

Given a closed process $P_0$ and a set of names $\rw$, the
protocol verifier first instruments $P_0$ to obtain $P'_0 = \instr{P_0}$, 
then it builds a set of Horn clauses,
representing the protocol in parallel with any $\rw$-adversary.  The
clauses are of the form $F_1 \wedge \ldots \wedge F_n \rewrite F$,
where $F_1, \ab \ldots, \ab F_n, \ab F$ are facts.
They comprise clauses for the attacker and clauses for the protocol,
defined below. These clauses form the set $\rset{P'_0,
\rw}$. 
The
predicate $\pasbegin$ is defined by a set of closed facts $\beginset$,
such that $\pasbegin(p)$ is true if and only if $\pasbegin(p) \in \beginset$.
The facts in $\beginset$ do not belong to $\rset{P'_0, \rw}$. 
The set $\beginset$ is the set of facts that corresponds to the
set of allowed events $\eventset$, mentioned in 
Section~\ref{sec:veriffirst}.

\subsubsection{Clauses for the Attacker}

The clauses describing the attacker are almost the same as for the
verification of secrecy in~\cite{Abadi04c}. The only difference is
that, here, the attacker is given an infinite set of fresh names
$\advnfs[x]$, instead of only one fresh name $\advnfs[\,]$. Indeed, we cannot merge
all fresh names created by the attacker, since we have to make sure
that different terms are represented by different patterns for the
verification of correspondences to be correctly implemented, as seen in
Section~\ref{sec:veriffirst}. The abilities of the
attacker are then represented by the following clauses:
{\allowdisplaybreaks\begin{align}
&\text{For each $a\in \rw$, }\attacker(a[\,])\tag{Init}\label{ruleInit}\\
&\attacker(\advnfs[x])\tag{Rn}\label{ruleRn}\\
\begin{split}
&\text{For each public constructor $f$ of arity $n$,}\\*
&\quad\attacker(x_1) \wedge \ldots \wedge \attacker(x_n) \rewrite \attacker(f(x_1, \ldots, x_n))
\end{split}\tag{Rf}\label{ruleRf}\\
\begin{split}
&\text{For each public destructor $g$,}\\*
&\quad\text{for each rewrite rule $g(M_1, \ldots, M_n) \rightarrow M$ in $\defg$,}\\*
&\quad\attacker(M_1) \wedge \ldots \wedge \attacker(M_n) \rewrite
\attacker(M)
\end{split}\tag{Rg}\label{ruleRg}\\
&\mess(x,y) \wedge \attacker(x) \rewrite
\attacker(y)\tag{Rl}\label{ruleRl}\\
&\attacker(x) \wedge \attacker(y) \rewrite
\mess(x,y)\tag{Rs}\label{ruleRs}
\end{align}}%
The clause~\eqref{ruleInit} represents the initial knowledge of the attacker.
The clause~\eqref{ruleRn} means that the attacker can generate an
unbounded number of new names. The clauses~\eqref{ruleRf} 
and~\eqref{ruleRg} mean that the attacker can apply all operations to all
terms it has, \eqref{ruleRf} for constructors, \eqref{ruleRg} for
destructors.
For~\eqref{ruleRg}, notice that
the rewrite rules in $\defg$ do not contain names and that terms without
names are also patterns, so the clauses have the required format.
Clause~\eqref{ruleRl} means that the attacker can listen on all
channels it has, and~\eqref{ruleRs} that it can send all messages
it has on all channels it has.

If $c \in \rw$, we can replace all occurrences of $\mess(c[\,],M)$ with
$\attacker(M)$ in the clauses. Indeed, these facts are equivalent by the
clauses~\eqref{ruleRl} and~\eqref{ruleRs}.

\subsubsection{Clauses for the Protocol}\label{sec:protclauses}

When a function $\rho$ associates a pattern with each name and
variable, and $f$ is a constructor, we
extend $\rho$ as a substitution by $\rho( f(M_1, \ldots, M_n) ) =
f(\rho(M_1), \ldots, \rho(M_n))$.

The translation $\lp P\rp \rho H$ of a process $P$ is a set of clauses,
where $\rho$ is a function that associates a pattern with each name
and variable, and $H$ is a sequence of facts of the form
$\mess(p,p')$ or $\pasbegin(p)$. 
The environment $\rho$ maps each variable and name to its associated
pattern representation.
The sequence $H$ keeps track of events that have been executed and of
messages received by the process, since these may trigger other
messages. 
The
empty sequence is denoted by $\emptyseq$; the concatenation of a fact
$F$ to the sequence $H$ is denoted by $H \wedge F$. The pattern $\rho
i$ is always a session identifier variable of $V_s$.
{\allowdisplaybreaks\begin{align*}
&\lp 0 \rp \rho H = \emptyset\\
&\lp P \parpop Q \rp \rho H = \lp P \rp \rho H  \cup \lp Q\rp \rho H\\
&\lp \ReplInstr{i}{P} \rp \rho H = \lp P \rp (\rho[i \mapsto i]) H\\
&\lp \ResInstr{a}{a[M_1, \ldots, M_n, i_1, \ldots, i_{n'}]} P\rp \rho H =\\*
&\quad \lp P \rp (\rho[a \mapsto a[\rho(M_1), \ldots, \rho(M_n), \rho(i_1), \ldots, \rho(i_{n'})]\,]) H\\
&\lp \cinput{M}{x}.P\rp \rho H = 
\lp P \rp (\rho[x \mapsto x]) (H \wedge \mess(\rho(M), x))\\
&\lp \coutput{M}{N}.P \rp \rho H = \lp P \rp \rho H
\cup \{ H  \rewrite \mess(\rho(M), \rho(N)) \}\\
&\lp \letfun{x}{g(M_1, \ldots, M_n)}{P}{Q} \rp \rho H = \bigcup \{ \lp P \rp ((\sigma\rho)[x \mapsto \sigma' p']) (\sigma H)\\*
&\quad \mid g(p'_1, \ldots, p'_n)
\rightarrow p'\text{ is in $\defg$ and $(\sigma, \sigma')$ is a most
general pair of}\\*
&\quad\text{ substitutions such that }\sigma\rho(M_1) = \sigma'
p'_1, \ldots, \sigma \rho(M_n) = \sigma' p'_n \} \cup \lp Q \rp \rho H\\ 
&\lp \guard{M}{N}{P}{Q} \rp \rho H = \lp P \rp (\sigma\rho) (\sigma H) 
\cup \lp Q \rp \rho H\\*
&\quad\text{where $\sigma$ is the most general unifier of
$\rho(M)$ and $\rho(N)$}\\
&\lp \asevent(M).P\rp \rho H = \lp P\rp \rho (H
\wedge \pasbegin(\rho(M))) \cup \{ H \rewrite \pasend(\rho(M)) \}
\end{align*}}%

The translation of a process is a set of Horn clauses that express
that it may send certain messages or execute certain events.
The clauses are similar to those of~\cite{Abadi04c},
except in the cases of replication, restriction, and the addition
of events.
\begin{itemize}

\item The nil process does nothing, so its translation is empty.

\item The clauses for the parallel composition
of processes $P$ and $Q$ are the union of clauses for $P$ and $Q$.

\item
The replication only inserts the new session identifier $i$ in
the environment $\rho$. It is otherwise ignored, because all Horn
clauses are applicable arbitrarily many times.  

\item
For the restriction, we
replace the restricted name $a$ in question with the pattern 
$a[\rho(M_1), \ab \ldots, \ab \rho(M_n), \ab \rho(i_1), \ab 
\ldots, \ab \rho(i_{n'})]$. By definition of the instrumentation, 
this pattern contains the previous inputs, results of non-deterministic
destructor applications, and session identifiers.

\item
The sequence $H$ is
extended in the translation of an input, with the input in
question. 

\item
The translation of an output adds a clause, meaning that the
output is triggered when all conditions in $H$ are true. 

\item 
The
translation of a destructor application is the union of the clauses
for the cases where the destructor succeeds (with an appropriate
substitution) and where the destructor fails. 
For simplicity, we assume that the $\kw{else}$ branch of destructors
may always be executed; this is sufficient in most cases,
since the $\kw{else}$ branch is often empty or just sends an error message. 
We outline a more precise treatment in Section~\ref{sect:elsebranches}.

\item
The conditional
$\guard{M}{N}{P}{Q}$ is in fact equivalent to
$\letfun{x}{\equal(M,N)}{P}{Q}$, where the destructor $\equal$ is
defined by $\equal(x,x) \rightarrow x$, so the translation of the
conditional is a particular case of the destructor application. We
give it explicitly since it is particularly simple. 

\item
The translation of an event adds the hypothesis $\pasbegin(\rho(M))$ to $H$, 
meaning that $P$ can be executed only if the event has been executed
first. 
Furthermore, it adds a
clause, meaning that the event is triggered when all conditions in
$H$ are true. 

\end{itemize}

\begin{Remark}\label{rem:event}
Depending on the form of the correspondences we want to prove, we can
sometimes simplify the clauses generated for events.
Suppose that all arguments of events in the process and in correspondences
are of the form $f(M_1, \ldots, M_n)$ for some function symbol $f$. 

If, for a certain function symbol $f$, events $\pasevent(f(\ldots))$
occur only before $\rightsquigarrow$ in the desired correspondences,
then it is easy to see in the following theorems that hypotheses of
the form $\pasbegin(f(\ldots))$ in clauses can be removed without
changing the result, so the clauses generated by the event
$\asevent(M)$ when $M$ is of the form $f(\ldots)$ can be simplified
into:
\[\lp \asevent(M).P\rp \rho H = \lp P\rp \rho H
\cup \{ H \rewrite \pasend(\rho(M)) \}\] 
(Intuitively, since the events $\pasevent(f(\ldots))$ occur only
before $\rightsquigarrow$ in the desired correspondences, we never
prove that an event $\asevent(f(\ldots))$ has been executed, so the
facts $\pasbegin(f(\ldots))$ are useless.)

Similarly, if $\pasevent(f(\ldots))$ occurs only after
$\rightsquigarrow$ in the desired correspondences, then clauses that
conclude a fact of the form $\pasend(f(\ldots))$ can be removed
without changing the result, so the clauses generated by the event
$\asevent(M)$ when $M$ is of the form $f(\ldots)$ can be simplified
into:
\[\lp \asevent(M).P\rp \rho H = \lp P\rp \rho (H
\wedge \pasbegin(\rho(M)))\]
(Intuitively, since the events $\pasevent(f(\ldots))$ occur only after 
$\rightsquigarrow$ in the desired correspondences, we never prove properties
of the form ``if $\asevent(f(\ldots))$ has been executed, then \ldots'',
so clauses that conclude $\pasend(f(\ldots))$ are useless.)
\end{Remark}

This translation of the protocol into Horn clauses introduces 
approximations. The actions are considered as
implicitly replicated, since the clauses can be applied any number of
times. 
This approximation implies that the tool fails to prove 
protocols that first need to keep some value secret and later reveal it.
For instance, consider the process $\Res{d}(\coutput{d}{s}.\coutput{c}{d} 
\parpop \cinput{d}{x})$. This process preserves the secrecy of $s$, because
$s$ is output on the private channel $d$ and received by the input on $d$,
before the adversary gets to know $d$ by the output of $d$ on the public
channel $c$. However, the Horn clause method cannot prove this property,
because it treats this process like a variant with additional
replications $\Res{d}(\Repl{\coutput{d}{s}.\coutput{c}{d}} 
\parpop \Repl{\cinput{d}{x}})$, which does not preserve the secrecy $s$.
Similarly, the process $\Res{d}(\coutput{d}{M} \parpop
\cinput{d}{x}.\cinput{d}{x}.\asevent(e_1))$ never executes the event
$e_1$, but the Horn clause method cannot prove this property because it
treats this process like $\Res{d}(\Repl{\coutput{d}{M}} \parpop
\cinput{d}{x}.\cinput{d}{x}.\asevent(e_1))$, which may execute
$e_1$.
The only exception to this implicit replication of processes 
is the creation of new names: since
session identifiers appear in patterns, the created name is precisely
related to the session that creates it, so name creation cannot be
unduly repeated inside the same session. Due to these approximations, 
our tool is not complete
(it may produce false attacks) but, as we show below, it is sound
(the security properties that it proves are always true).

\subsubsection{Summary and Correctness}

Let $\rho = \{ a \mapsto a[\,] \mid a \in \fn(P'_0) \}$.
We define the clauses corresponding to the instrumented process $P'_0$ as:
\[\rset{P'_0, \rw} = \lp P'_0 \rp \rho \emptyseq \cup 
\{ \attacker(a[\,]) \mid a \in \rw \} \cup \{ \eqref{ruleRn}, \eqref{ruleRf}, 
\eqref{ruleRg}, \eqref{ruleRl}, \eqref{ruleRs} \}\]

\begin{example}
The clauses for the process $P$ of
Section~\ref{sec:example} are the clauses for the adversary, plus:
{\allowdisplaybreaks\begin{align}
&\attacker(\pk(\skA[\,]))\label{rulePka}\\%17
&\attacker(\pk(\skB[\,]))\label{rulePkb}\\%18
&H_1 \rewrite \attacker(\pencrypt((a[x\_\pkB, i_A], \pk(\skA[\,])), x\_\pkB, r_1[x\_\pkB, i_A]))\label{ruleOutA1}\\%19
\begin{split}
&H_2 \rewrite \attacker(\pencrypt(x\_b, x\_\pkB, r_3[x\_\pkB, p_2, i_A]))
\end{split}\label{ruleOutA2}\\%20
&H_3 \rewrite \pasend(e_A(\pk(\skA[\,]), \pk(\skB[\,]), a[\pk(\skB[\,]), i_A], x\_b))\label{ruleEndA}\\%21
&H_3 \rewrite \attacker(\sencrypt(\sAa[\,], a[\pk(\skB[\,]), i_A]))\label{rulesAa}\\%22
&H_3 \rewrite \attacker(\sencrypt(\sAb[\,], x\_b))\label{rulesAb}\\%23
&\text{where } p_2 = \pencrypt((a[x\_\pkB, i_A], x\_b, x\_\pkB), \pk(\skA[\,]), x\_r_2)\notag\\
&\phantom{\text{where}} H_1 = \attacker(x\_\pkB) \wedge \pasbegin(e_1(\pk(\skA[\,]), x\_\pkB, a[x\_\pkB, i_A]))\notag\\
&\phantom{\text{where}} H_2 = H_1 \wedge \attacker(p_2) \wedge \pasbegin(e_3(\pk(\skA[\,]), x\_\pkB, a[x\_\pkB, i_A], x\_b))\notag\\
&\phantom{\text{where}} H_3 = H_2 \{ \pk(\skB[\,]) / x\_\pkB \}\notag\\[2mm]
\begin{split}
&\attacker(p_1) \wedge \pasbegin(e_2(x\_\pkA, \pk(\skB[\,]), x\_a, b[p_1, i_B]))\\*
&\quad \rewrite \attacker(\pencrypt((x_a, b[p_1, i_B], \pk(\skB[\,])), x\_\pkA, r_2[p_1, i_B]))
\end{split}\label{ruleOutB}\\%24
&\text{where } p_1 = \pencrypt((x\_a, x\_\pkA), \pk(\skB[\,]), x\_r_1)\notag\\[2mm]
&H_4 \rewrite \pasend(e_B(\pk(\skA[\,]), \pk(\skB[\,]), x\_a, b[p'_1, i_B]))\label{ruleEndB}\\%25
&H_4 \rewrite \attacker(\sencrypt(\sBa[\,], x\_a))\label{rulesBa}\\%26
&H_4 \rewrite \attacker(\sencrypt(\sBb[\,], b[p'_1, i_B]))\label{rulesBb}\\%27
&\text{where } p'_1 = \pencrypt((x\_a, \pk(\skA[\,])), \pk(\skB[\,]), x\_r_1)\notag\\
&\phantom{\text{where}} H_4 = \attacker(p'_1) \wedge \pasbegin(e_2(\pk(\skA[\,]), \pk(\skB[\,]), x\_a, b[p'_1, i_B])) \wedge{}\notag\\*
&\phantom{\text{where }}\quad \qquad \attacker(\pencrypt(b[p'_1, i_B], \pk(\skB[\,]), x\_r_3))\notag
\end{align}}%
Clauses~\eqref{rulePka} and~\eqref{rulePkb} correspond to the outputs in
$P$; they mean that the adversary has the public keys of the
participants. Clauses~\eqref{ruleOutA1} and~\eqref{ruleOutA2} correspond to
the first two outputs in $P_A$. For example, \eqref{ruleOutA2} means that, 
if the attacker has $x\_\pkB$ and the second message of the protocol
$p_2$ and the events
$e_1(\pk(\skA[\,]), x\_\pkB, a[x\_\pkB, i_A])$ and
$e_3(\pk(\skA[\,]), x\_\pkB, a[x\_\pkB, i_A], x\_b)$ are allowed, 
then the attacker can get
$\pencrypt(x\_b, x\_\pkB, r_3[x\_\pkB, p_2, i_A])$,
because $P_A$ sends this message after receiving $x\_\pkB$ and $p_2$
and executing the events $e_1$ and $e_3$.
When furthermore $x\_\pkB = \pk(\skB[\,])$, $P_A$ executes event $e_A$
and outputs the encryption of $\sAa[\,]$ under  $a[x\_\pkB, \ab i_A]$
and the encryption of $\sBb[\,]$ under $x\_b$.
These event and outputs are taken into account by Clauses~\eqref{ruleEndA},
\eqref{rulesAa}, and~\eqref{rulesAb} respectively.
Similarly, Clauses~\eqref{ruleOutB}, \eqref{rulesBa}, and~\eqref{rulesBb}
correspond to the outputs in $P_B$ and~\eqref{ruleEndB} to the
event $e_B$.
These clauses have been simplified using Remark~\ref{rem:event},
taking into account that $e_1$, $e_2$, and $e_3$ appear only on
the right-hand side of $\rightsquigarrow$, and $e_A$ and $e_B$
only on the left-hand side of $\rightsquigarrow$ in the queries
of Examples~\ref{exa:q1}, \ref{exa:q2}, and~\ref{exa:q3}.
\end{example}

\begin{theorem}[Correctness of the clauses]\label{th:main}
Let $P_0$ be a closed process and $Q$ be an $\rw$-adversary. Let $P'_0
= \instr{P_0}$ and $Q' = \instradv{Q}$.
Consider a trace $\trace = S_0, \env_0, \{ P'_0, Q'\} \rightarrow^*
S', \env', \pset'$, with $\fn(P'_0) \cup \rw
\subseteq \dom(\env_0)$ and $\env_0(a) = a[\,]$ for all $a \in \dom(\env_0)$. 
Assume that, if $\trace$ satisfies $\pasevent(p)$, 
then $\pasbegin(p) \in \beginset$.
Finally, assume that $\trace$ satisfies $\acti$.
Then $\acti$ is derivable from $\rset{P'_0, \rw} \cup \beginset$. 
\end{theorem}

This result shows that, if the only
executed events are those allowed in $\beginset$ and a fact $\acti$
is satisfied, then $\acti$ is derivable
from the clauses. It is proved 
in Appendix~\ref{app:main}. 
Using a 
technique similar to that of~\cite{Abadi04c}, its proof
relies on a type system to express the soundness of the clauses on
$P'_0$, and on the subject reduction of this type system to show that
soundness of the clauses is preserved during all executions of the process.

\section{Solving Algorithm}\label{sec:solv}

We first describe a basic solving algorithm without optimizations.
Next, we list the optimizations that we use in our implementation,
and we prove the correctness of the algorithm.
The termination of the algorithm is discussed in 
Section~\ref{sect:termination}.

\subsection{The Basic Algorithm}
\label{sec:basicsolv}

To apply the previous results, we have to determine whether a fact is
derivable from $\rset{P'_0, \rw} \cup \beginset$. This may be undecidable, but
in practice there exist algorithms that terminate on numerous examples
of protocols. In particular, we can use variants of resolution
algorithms, such as the algorithms described
in~\cite{Weidenbach99,Blanchet2001,Blanchet2002,Blanchet04e}. The algorithm 
that we describe here is the one of~\cite{Blanchet2002}, extended with
a second phase to determine derivability of any query. It also
corresponds to the extension to $\pasbegin$ facts of the algorithm
of~\cite{Blanchet04e}.

We first define resolution: when the conclusion of a clause $R$ unifies with an
hypothesis $F_0$ of a clause $R'$, we can infer a new clause $R \circ_{F_0} R'$,
that corresponds to applying $R$ and $R'$ one after the
other. Formally, this is defined as follows:
\begin{definition}
Let $R = H \rewrite C$ and $R' = H' \rewrite
C'$ be two clauses. Assume that there exists $F_0 \in H'$ such that
$C$ and $F_0$ are unifiable, and $\sigma$ is the most general
unifier of $C$ and $F_0$.
In this case, we define 
$R \circ_{F_0} R' = \sigma(H \cup (H'\setminus \{F_0\})) \rewrite \sigma C'$.
\end{definition}
An important idea to obtain an efficient solving algorithm
is to specify conditions that limit the application of resolution,
while keeping completeness. The conditions that we use
correspond to resolution with free 
selection~\cite{Nivelle95,Lynch97,Bachmair01}: a selection function
chooses selected facts in each clause, and resolution is performed
only on selected facts, that is, the clause
$R \circ_{F_0} R'$ is generated only when the conclusion is selected in $R$
and $F_0$ is selected in $R'$.

\begin{definition}
We denote by $\sel$ a selection function, that is, a function from
clauses to sets of facts, such that $\sel(H \rewrite C) \subseteq H$. 
If $F \in \sel(R)$, we say that $F$ is selected in $R$. If $\sel(R) =
\emptyset$, we say that no hypothesis is selected in $R$, or that the
conclusion of the clause is selected.
\end{definition}  
The choice of the selection function can change dramatically the
speed of the algorithm. Since the algorithm 
combines clauses by resolution only when the facts unified in the
resolution are selected, we will choose the selection
function to reduce the number of possible unifications between
selected facts. Having several selected facts slows down the
algorithm, because it has more choices of resolutions to perform,
therefore we will select at most one fact in each clause. 
In the case of protocols, facts of the
form $\attacker(x)$, with $x$ variable, can be unified will all facts
of the form $\attacker(p)$. Therefore we should avoid selecting
them. The $\pasbegin$ facts must never be selected 
since they are not defined by known clauses.

\begin{definition}
We say that a fact $F$ is \emph{unselectable} when $F = \attacker(x)$
for some variable $x$ or $F = \pasbegin(p)$ for some pattern $p$. 
Otherwise, we say that $F$ is \emph{selectable}.

We require that the selection function 
never selects unselectable hypotheses and  that
$\sel(H \rewrite \attacker(x)) \neq \emptyset$ when 
$H$ contains a selectable fact.
\end{definition}
A basic selection function for security protocols is then
\[\sel_0(H \rewrite C) = 
\begin{cases}
\emptyset&\text{if $\forall F \in H$, $F$ is unselectable}\\
\{ F_0 \}&\text{where $F_0 \in H$ and $F_0$ is selectable, otherwise}
\end{cases}\]
In the implementation, the hypotheses are represented by a list, and
the selected fact is the first selectable element of the list of hypotheses.

The solving algorithm works in two phases, summarized in
Figure~\ref{fig:solv1}. The first
phase, $\saturate$, transforms the set of clauses into an equivalent but
simpler one. The second phase, $\reach$, uses a depth-first search to determine
whether a fact can be inferred or not from the clauses.

\begin{figure}[t]
\begin{tabbing}
\textbf{First phase: saturation}\\
$\saturate(\satstset)=$\\
\quad 1. \=$\sattmp \leftarrow \emptyset$.\\
\>For each $R \in \satstset$, $\sattmp \leftarrow \elim(\simplify(R) \cup \sattmp)$.\\
\quad 2. Repeat until a fixpoint is reached\\
\>\quad for each $R \in \sattmp$ such that $\sel(R) = \emptyset$,\\
\>\qquad for each $R' \in \sattmp$, for each $F_0 \in \sel(R')$ such that $R
\circ_{F_0} R'$ is defined,\\
\>\qquad\quad $\sattmp
\leftarrow \elim(\simplify(R \circ_{F_0} R') \cup \sattmp)$.\\
\quad 3. Return $\{ R \in \sattmp \mid \sel(R) = \emptyset \}$.\\[2mm]
\textbf{Second phase: backwards depth-first search}\\
$\deriv(R, \derivtmp, \derivstset) =
\begin{cases}
\emptyset&\text{ if }\exists R' \in \derivtmp, R' \impl R\\
\{ R \}\hspace*{4.2cm}&\text{ otherwise, if }\sel(R) = \emptyset\\
\multicolumn{2}{@{}l}{\bigcup \{ \deriv(\simplify'(R' \circ_{F_0} R), 
\{ R \} \cup \derivtmp, \derivstset) \mid R' \in \derivstset, }\\
\multicolumn{2}{@{}l}{\qquad F_0 \in \sel(R)\text{ such that }R' \circ_{F_0} R\text{ is defined}\,\}\quad\text{ otherwise}}
\end{cases}$\\
$\reach(F,\derivstset) = \deriv(F \rewrite F, \emptyset, \derivstset)$
\end{tabbing}
\caption{Solving algorithm}\label{fig:solv1}
\end{figure}

The first phase contains 3 steps. 
\begin{itemize}

\item The first
step inserts in $\sattmp$ the initial clauses representing the
protocol and the attacker (clauses that are in $\satstset$), after
simplification by $\simplify$ (defined below in Section~\ref{sec:simplif}) and 
elimination of subsumed clauses by $\elim$. 
We say that $H_1 \rewrite C_1$ subsumes $H_2 \rewrite C_2$, and we
write $(H_1 \rewrite C_1) \impl (H_2 \rewrite C_2)$, when there exists
a substitution $\sigma$ such that $\sigma C_1 = C_2$ and $\sigma H_1
\subseteq H_2$. ($H_1$ and $H_2$ are multisets, and we use here
multiset inclusion.)
If $R'$ subsumes $R$, and $R$ and $R'$ are in $\sattmp$, then $R$ is removed
by $\elim(\sattmp)$.

\item
The second step is a fixpoint iteration that adds clauses created by
resolution. The composition of clauses $R$ and $R'$ is added only if
no hypothesis is selected in $R$, and the hypothesis $F_0$ of $R'$
that we unify is selected.
When a clause is created by resolution, it is added to the set of clauses
$\sattmp$ after simplification. Subsumed clauses are eliminated from $\sattmp$.

\item
At last, the third step returns the set of clauses of $\sattmp$ with no selected 
hypothesis.

\end{itemize}
Basically, $\saturate$ preserves derivability: $F$ is derivable from
$\satstset \cup \beginset$ if and only if it is derivable from
$\saturate(\satstset) \cup \beginset$. A formal statement of this result is
given in Lemma~\ref{lem:phase1corr} below.

The second phase searches the facts that can be inferred from
$\derivstset = \saturate(\satstset)$. This is simply a backward
depth-first search. 
The call $\reach(F, \derivstset)$ returns a set of clauses $R = H \rewrite
C$ with empty selection, such that $R$ can be obtained by resolution from
$\derivstset$, $C$ is an instance of $F$, and all instances of $F$
derivable from $\derivstset$ can be derived by using as last clause
a clause of $\reach(F, \derivstset)$. 
(Formally, if $F'$ is an instance of $F$ derivable from $\derivstset$,
then there are a clause $H \rewrite C \in \reach(F, \derivstset)$ 
and a substitution $\sigma$ such
that $F' = \sigma C$ and $\sigma H$ is derivable from $\derivstset$.)

The search itself is performed by $\deriv(R, \derivtmp, \derivstset)$.
The function $\deriv$ starts with $R = F \rewrite F$ and transforms
the hypothesis of $R$ by using a clause $R'$ of $\derivstset$ to
derive an element $F_0$ of the hypothesis of $R$.  So $R$ is replaced
with $R' \circ_{F_0} R$ (third case of the definition of $\deriv$).
The fact $F_0$ is chosen using the selection function $\sel$.
The obtained clause $R' \circ_{F_0} R$ is
then simplified by the function $\simplify'$ defined in
Section~\ref{sec:simplif}.
(Hence $\deriv$ derives the hypothesis of $R$ using a backward depth-first
search.
At each step, the clause $R$ can be obtained by resolution from
clauses of $\derivstset$, and $R$ concludes an instance of $F$.)
The set $\derivtmp$ is the set of clauses that we have already seen
during the search. Initially, $\derivtmp$ is empty, and the clause $R$
is added to $\derivtmp$ in the third case of the definition of
$\deriv$.

The transformation of $R$ described above is repeated until one of
the following two conditions is satisfied:
\begin{itemize}

\item $R$ is subsumed by a clause in $\derivtmp$: we are in a cycle;
  we are looking for instances of facts that we have already looked
  for (first case of the definition of $\deriv$);

\item $\sel(R)$ is empty: we have obtained a suitable clause $R$ and we
  return it (second case of the definition of $\deriv$).

\end{itemize}

\subsection{Simplification Steps}\label{sec:simplif}

Before adding a clause to the clause base, it is first simplified using
the following functions. Some of them are standard, such as the
elimination of tautologies and of duplicate hypotheses; others
are specific to protocols. The simplification functions 
take as input a clause or a set of clauses and return a set of clauses. 

\paragraph{Decomposition of Data Constructors}

A data constructor is a constructor $f$ of arity $n$ that comes with
associated destructors $g_i$ for $i \in \{ 1, \ldots, n \}$ defined by
$g_i(f(x_1, \ab \ldots, \ab x_n)) \rightarrow x_i$. Data constructors are
typically used for representing data structures. Tuples are examples
of data constructors. For each data constructor $f$, the following
clauses are generated:
\begin{align}
&\attacker(x_1) \wedge \ldots \wedge \attacker(x_n) \rewrite \attacker(f(x_1,
\ldots, x_n))\tag{Rf}\\
&\attacker(f(x_1, \ldots, x_n)) \rewrite \attacker(x_i)\tag{Rg}
\end{align}
Therefore, $\attacker(f(p_1, \ldots, p_n))$ is derivable if
and only if $\forall i \in \interv{1}{n}$, $\attacker(p_i)$ is
derivable. So the function $\decomp$ transforms clauses as follows.
When a fact of the form $\attacker(f(p_1, \ab \ldots, \ab 
p_n))$ is met, it is replaced with $\attacker(p_1) \wedge \ldots \wedge
\attacker(p_n)$. If this replacement is done in the conclusion of a
clause $H \rewrite \attacker(f(p_1, \ab \ldots, \ab p_n))$, $n$ clauses are
created: $H \rewrite \attacker(p_i)$ for each $i \in \interv{1}{n}$.
This replacement is of course done recursively: if $p_i$ itself is a
data constructor application, it is replaced again. 
The function $\decomphyp$ performs
this decomposition only in the hypothesis of clauses.
The functions $\decomp$ and $\decomphyp$ leave the 
clauses~\eqref{ruleRf} and~\eqref{ruleRg} for data constructors unchanged.
(When $\attacker(x)$ cannot be selected, the clauses~\eqref{ruleRf}
and~\eqref{ruleRg} for data constructors are in fact not necessary,
because they generate only tautologies during resolution. However,
when $\attacker(x)$ can be selected, which cannot be excluded in
extensions such as the one presented in Section~\ref{sec:phases},
these clauses may become necessary for soundness.)

\paragraph{Elimination of Tautologies}
The function $\elimtaut$ removes clauses whose conclusion is already 
in the hypotheses, since such clauses do not generate new facts. 

\paragraph{Elimination of Duplicate Hypotheses}
The function $\elimdup$ eliminates duplicate hypotheses of clauses.

\paragraph{Elimination of Useless $\attacker(x)$ Hypotheses}
If a clause $H \rewrite C$ contains in its hypotheses $\attacker(x)$,
where $x$ is a variable that does not appear elsewhere in the clause,
the hypothesis $\attacker(x)$ is removed by the function $\elimattx$. 
Indeed, the attacker
always has at least one message, so $\attacker(x)$ is always
satisfied.

\paragraph{Secrecy Assumptions}

\begin{figure}[t]
\begin{tabbing}
$\solve{P'_0, \rw}(F) =$\\[0.5mm]
\quad 1. Let $\satend = \saturate(\rset{P'_0, \rw})$.\\
\quad 2. For each $F' \in \fnot$, if $\reach(F', \satend) \neq \emptyset$, then
terminate with error.\\
\quad 3. Return $\reach(F, \satend)$.
\end{tabbing}
\caption{Summary of the solving algorithm}\label{fig:solve}
\end{figure}

When the user knows that a fact $F$ will not be derivable, he can tell it
to the verifier. (When this fact is of the form $\attacker(p)$, the
user tells that $p$ remains secret; that is why we use the name 
``secrecy assumptions''.) 
Let $\fnot$ be a set of facts, for which the user claims that no
instance of these facts is derivable.
The function $\elimnot$ removes all clauses
that have an instance of a fact in $\fnot$ in their hypotheses. 
As shown in Figure~\ref{fig:solve}, at the end of the
saturation, the solving algorithm checks that the facts in $\fnot$ are indeed underivable
from the obtained clauses. If this condition is satisfied, 
$\solve{P'_0, \rw}(F)$ returns clauses that conclude instances of $F$.
Otherwise, the user has given erroneous information, so
an error message is displayed. Even when the user gives erroneous
secrecy assumptions, the verifier never
wrongly claims that a protocol is secure.

Mentioning such underivable facts prunes the search space, by removing
useless clauses. This speeds up the search process. In most cases,
the secret keys of the principals cannot be known by the attacker, so
examples of underivable facts are $\attacker(\skA[\,])$
and $\attacker(\skB[\,])$.

\paragraph{Elimination of Redundant Hypotheses}

When a clause is of the form $H \wedge H' \rewrite C$, and there
exists $\sigma$ such that $\sigma H \subseteq H'$ and $\sigma$ does
not change the variables of $H'$ and $C$, then the clause is replaced
with $H' \rewrite C$ by the function $\elimredundanthyp$. These
clauses are semantically equivalent: obviously, $H' \rewrite C$
subsumes $H \wedge H' \rewrite C$; conversely, if
a fact can be derived by an instance $\sigma' H' \rewrite \sigma' C$ 
of $H' \rewrite C$, then it can also be derived by the instance
$\sigma' \sigma H \wedge \sigma' H' \rewrite \sigma' C$ of
$H \wedge H' \rewrite C$, since the elements of $\sigma' \sigma H$
can be derived because they are in $\sigma' H'$.

This replacement is especially useful when $H$ contains $\pasbegin$
facts. Otherwise, the elements of $H$ could be selected and
transformed by resolution, until they are of the form $\attacker(x)$,
in which case they are removed by $\elimattx$ if $\sigma x \neq x$
(because $x$ does not occur in $H'$ and $C$ since $\sigma$ does not
change the variables of $H'$ and $C$) or by $\elimdup$ if $\sigma x =
x$ (because $\attacker(x) = \sigma \attacker(x) \in \sigma H \subseteq
H'$). In contrast, $\pasbegin$ facts remain forever, because they are
unselectable.
Depending on user settings, this replacement can be applied for all
$H$, applied only when $H$ contains a $\pasbegin$ fact, or switched
off, since testing this property takes time and slows down small
examples. On the other hand, on big examples, such as some of those
generated by TulaFale~\cite{Bhargavan04} for verifying Web services,
this technique can yield important speedups.

\paragraph{Putting All Simplifications Together}

The function $\simplify$ groups all these simplifications. We define
$\simplify = \elimattx \circ 
\elimtaut \circ \elimnot \circ \elimredundanthyp \circ \elimdup \circ \decomp$.
In this definition, the simplifications are ordered in such a way
that $\simplify  \circ \simplify = \simplify$, so it is not
necessary to repeat the simplification.

Similarly, $\simplify' = \elimattx \circ \elimnot 
\circ \elimredundanthyp \circ \elimdup \circ \decomphyp$.
In $\simplify'$, we use $\decomphyp$ instead of $\decomp$, because
the conclusion of the considered clause is the fact we want to derive,
so it must not be modified.

\subsection{Soundness}

The following lemmas show the correctness of $\saturate$ and $\reach$ 
(Figure~\ref{fig:solv1}). 
Proofs can be found in Appendix~\ref{app:corr}.
Intuitively, the correctness of $\saturate$ expresses that saturation
preserves derivability, provided the secrecy assumptions are
satisfied.

\begin{lemma}[Correctness of $\saturate$]\label{lem:phase1corr}
Let $F$ be a closed fact. If, for all $F' \in \fnot$, no instance of $F'$
is derivable from $\saturate(\satstset) \cup \beginset$, 
then $F$ is derivable from $\satstset \cup \beginset$ if
and only if $F$ is derivable from $\saturate(\satstset) \cup \beginset$.
\end{lemma}
This result is proved by transforming a derivation of $F$ from
$\satstset \cup \beginset$ into a derivation of $F$ (or a fact in
$\fnot$) from $\saturate(\satstset) \cup \beginset$. Basically, when
the derivation contains a clause $R'$ with $\sel(R') \neq \emptyset$,
we replace in this derivation two clauses $R$, with $\sel(R) =
\emptyset$, and $R'$ that have been combined by resolution during the
execution of $\saturate$ with a single clause $R \circ_{F_0} R'$. This
replacement decreases the number of clauses in the derivation, so it
terminates, and, upon termination, all clauses of the obtained
derivation satisfy $\sel(R') = \emptyset$ so they are in
$\saturate(\satstset) \cup \beginset$.

Intuitively, the correctness of $\reach$ expresses that if $F'$,
instance of $F$, is derivable, then $F'$ is derivable from
$\derivstset$ by a derivation in which the clause that concludes $F'$
is in $\reach(F, \derivstset)$, provided the secrecy assumptions
are satisfied.

\begin{lemma}[Correctness of $\reach$]\label{lem:reach}
Let $F'$ be a closed instance of $F$.  
If, for all $F'' \in \fnot$, $\reach(F'', \derivstset) = \emptyset$, then
$F'$ is derivable from $\derivstset
\cup \beginset$ if and only if there exist a clause $H \rewrite C$ in
$\reach(F, \derivstset)$ and a substitution $\sigma$ such that $\sigma C = F'$
and all elements of $\sigma H$ are derivable from $\derivstset \cup \beginset$.
\end{lemma}
Basically, this result is proved by transforming a derivation of
$F'$ from $\derivstset \cup \beginset$ into a derivation of $F'$ (or a
fact in $\fnot$) whose last clause (the one that concludes $F'$) is $H
\rewrite C$ and whose other clauses are still in $\derivstset \cup
\beginset$. The transformation relies on the replacement of
clauses combined by resolution during the execution of $\reach$.

It is important to apply $\saturate$ before $\reach$, so that all
clauses in $\derivstset$ have no selected hypothesis. 
Then the conclusion of these
clauses is in general not $\attacker(x)$ (with the simplifications
of Section~\ref{sec:simplif} and the selection function $\sel_0$, it is never
$\attacker(x)$), so that we avoid unifying with
$\attacker(x)$.

Finally, the following theorem shows the correctness of $\solve{P'_0,\rw}$
(Figure~\ref{fig:solve}). 
Below, when we require that $\solve{P'_0,\rw}(F)$ has a certain value,
we also implicitly require that $\solve{P'_0,\rw}(F)$ does not 
terminate with error.
Intuitively, if an instance $\acti'$ of $\acti$ is satisfied by a
trace $\trace$, then $\acti'$ is derivable from $\rset{P'_0, \rw} \cup
\beginset$, so, by the soundness of the solving algorithm, it is
derivable by a derivation whose last clause is in
$\solve{P'_0,\rw}(\acti)$. Then there must exist a clause $H \rewrite
C \in \solve{P'_0,\rw}(\acti)$ that can be used to derive $\acti'$, so
$\acti' = \sigma C$ and the hypothesis $\sigma H$ is derivable from
$\rset{P'_0, \rw} \cup \beginset$. In particular, the events in
$\sigma H$ are satisfied, that is, are in $\beginset$, so these events
have been executed in the trace $\trace$.
Theorem~\ref{thbeginend} below states this result formally.
It is proved by combining
Lemmas~\ref{lem:phase1corr} and~\ref{lem:reach}, and
Theorem~\ref{th:main}.

\begin{theorem}[Main theorem]\label{thbeginend}
Let $P_0$ be a closed process and $P'_0 = \instr{P_0}$. 
Let $Q$ be an $\rw$-adversary and $Q' = \instradv{Q}$. 

Consider a trace $\trace = S_0, \env_0,
\{ P'_0, Q'\} \rightarrow^* S', \env', P'$, with 
$\fn(P'_0) \cup \rw \subseteq
\dom(\env_0)$ and $\env_0(a) = a[\,]$ for
all $a \in \dom(\env_0)$. 

If $\trace$ satisfies an instance $\acti'$ of $\acti$, then there exist a clause $H \rewrite
C \in \solve{P'_0,\rw}(\acti)$ and a substitution $\sigma$
such that $\acti' = \sigma C$ and, for all $\pasbegin(p)$ in 
$\sigma H$, $\trace$ satisfies $\pasevent(p)$.
\end{theorem}
\begin{proof}
Since for all $F'' \in \fnot$, $\reach(F'', \satend) = \emptyset$,
by Lemma~\ref{lem:reach}, no instance of $F''$ is derivable from 
$\satend \cup \beginset = \saturate(\rset{P'_0, \rw}) \cup \beginset$. 
This allows us to apply Lemma~\ref{lem:phase1corr}.

Let $\beginset = \{ \pasbegin(p') \mid \trace\text{ satisfies }\pasevent(p')\}$.
By Theorem~\ref{th:main}, since $\trace$ satisfies $\acti'$, $\acti'$ is
derivable from $\rset{P'_0, \rw} \cup \beginset$.  By
Lemma~\ref{lem:phase1corr}, $\acti'$ is derivable from
$\saturate(\rset{P'_0, \rw}) \cup \beginset = \satend \cup \beginset$. 
By Lemma~\ref{lem:reach}, there
exist a clause $R = H \rewrite C$ in $\solve{P'_0, \rw}(\acti) =
\reach(\acti, \satend)$ and a substitution
$\sigma$ such that $\sigma C = \acti'$ and all elements of
$\sigma H$ are derivable from $\satend \cup \beginset$.  For all
$\pasbegin(p)$ in $\sigma H$, $\pasbegin(p)$ is derivable from
$\satend \cup \beginset$. Since no clause in
$\satend$ has a conclusion of the form
$\pasbegin(p')$, $\pasbegin(p) \in \beginset$. Given the choice of
$\beginset$, this means that $\trace$ satisfies $\pasevent(p)$.
\proofcomplete
\end{proof}

Theorem~\ref{thbeginend} is our main correctness result: it allows
one to show that some events must have been executed. The
correctness of the analysis for correspondences follows from
this theorem.

\begin{example}\label{exa:noninj1}
For the process $P$ of Section~\ref{sec:example}, $\rw = \{ c \}$,
and $P' = \instr{P}$, our tool shows that 
\begin{align*}
\begin{split}
&\solve{P', \rw}(\pasend(e_B(x_1, x_2, x_3, x_4))) = \{ \pasbegin(e_1(\pkA, \ab\pkB, \ab p_a)) \wedge {}\\
&\phantom{\solve{P', \rw}(\pasend(e_B(x_1, x_2, x_3, x_4))) = \{} \pasbegin(e_2(\pkA, \ab \pkB, \ab p_a, \ab p_b)) \wedge{}\\
&\phantom{\solve{P', \rw}(\pasend(e_B(x_1, x_2, x_3, x_4))) = \{} \pasbegin(e_3(\pkA, \ab \pkB, \ab p_a, \ab p_b))\\
&\phantom{\solve{P', \rw}(\pasend(e_B(x_1, x_2, x_3, x_4))) = \{} \rewrite \pasend(e_B(\pkA, \ab \pkB, \ab p_a, \ab p_b))\}
\end{split}\\
&\text{where }\pkA = \pk(\skA[\,]),\ \pkB = \pk(\skB[\,]),\ p_a = a[\pkB, i_A]\\
&\phantom{\text{where }}p_b = b[\pencrypt((p_a, \ab \pkA), \ab \pkB, \ab r_1[\pkB, i_A]), \ab i_B]
\end{align*}
By Theorem~\ref{thbeginend}, if $\trace$ satisfies 
$\pasevent(e_B(p_1,\ab p_2, \ab p_3, \ab p_4))$, 
this event is an instance of $\pasend(e_B(x_1, \ab x_2, \ab x_3, \ab x_4))$, so,
given the value of $\solve{P', \rw}(\pasend(e_B(x_1, \ab x_2, \ab x_3, \ab x_4)))$,
there exists $\sigma$ such that $\pasevent(e_B(p_1,\ab p_2, \ab p_3, \ab p_4)) = 
\sigma \pasend(e_B(\pkA, \ab \pkB, \ab p_a, \ab p_b))$ and $\trace$ satisfies
\begin{align*}
&\pasevent(\sigma e_1(\pkA, \ab\pkB, \ab p_a)) = \pasevent(e_1(p_1, \ab p_2, \ab p_3))\\
&\pasevent(\sigma e_2(\pkA, \ab \pkB, \ab p_a, \ab p_b))
= \pasevent(e_2(p_1,\ab p_2, \ab p_3, \ab p_4))\\
&\pasevent(\sigma e_3(\pkA, \ab \pkB, \ab p_a, \ab p_b))
= \pasevent(e_3(p_1,\ab p_2, \ab p_3, \ab p_4))
\end{align*}
Therefore, if $\asevent(e_B(M_1,\ab M_2, \ab M_3, \ab M_4))$ has been
executed, then $\asevent(e_1(M_1, \ab M_2, \ab M_3))$,
$\asevent(e_2(M_1,\ab M_2, \ab M_3, \ab M_4))$,
and $\asevent(e_3(M_1,\ab M_2, \ab M_3, \ab M_4))$ have been executed.
\end{example}

\section{Application to Correspondences}\label{sec:verif}

\subsection{Non-injective Correspondences}

Correspondences for instrumented processes can be checked as
shown by the following theorem:

\begin{theorem}\label{thbeginend3}
Let $P_0$ be a closed process and $P'_0 = \instr{P_0}$.  Let $p_{jk}$
($j \in \{ 1, \ldots, m\}$, $k \in \{ 1, \ldots, l_j\}$)
be patterns; let $\acti$ and $\acti_j$ ($j \in \{ 1, \ldots, m\}$) be facts. 
Assume that for all
$R \in \solve{P'_0, \rw}(\acti)$, there exist $j \in \{ 1,
\ldots, m \}$, $\sigma'$, and $H$ such that $R = H \wedge
\pasbegin(\sigma' p_{j1}) \wedge \ldots \wedge \pasbegin(\sigma'
p_{jl_j}) \rewrite \sigma' \acti_j$.

Then $P'_0$ satisfies the correspondence
$\acti \Rightarrow \sor_{j=1}^{m} \left(\acti_j \rightsquigarrow 
\sand_{k = 1}^{l_j} \pasevent(p_{jk})\right)$
against $\rw$-adversaries.
\end{theorem}
\begin{proof}
Let $Q$ be an $\rw$-adversary and $Q' = \instradv{Q}$. 
Consider a trace $\trace = S_0, \env_0, \{
P'_0, Q' \}\rightarrow^* S', \env', \pset'$, with $\fn(P'_0) \cup \rw
\subseteq \dom(\env_0)$ and $\env_0(a) = a[\,]$ for all $a \in
\dom(\env_0)$.  Assume that $\trace$ satisfies $\sigma \acti$.
By Theorem~\ref{thbeginend}, there exist $R = H' \rewrite C' \in
\solve{P'_0, \rw}(\acti)$ and $\sigma''$ such that $\sigma \acti =
\sigma'' C'$ and for all $\pasbegin(p)$ in $\sigma'' H'$, $\trace$
satisfies $\pasevent(p)$.
All clauses $R$ in $\solve{P'_0, \rw}(\acti)$ are of the form $H
\wedge \pasbegin(\sigma' p_{j1}) \wedge \ldots \wedge
\pasbegin(\sigma' p_{jl_j}) \rewrite \sigma' \acti_j$ for some $j$ and
$\sigma'$. So, there exist $j$ and $\sigma'$ such that for all $k \in
\{ 1, \ldots, l_j\}$, $\pasbegin(\sigma' p_{jk}) \in H'$ and $C' =
\sigma' \acti_j$. Hence $\sigma \acti = \sigma'' C' = \sigma'' \sigma'
\acti_j$ and for all $k \in \{ 1, \ldots, l_j \}$, $\pasbegin(\sigma''
\sigma' p_{jk}) \in \sigma'' H'$, so $\trace$ satisfies
$\pasevent(\sigma'' \sigma' p_{jk})$, so we have the result.
\proofcomplete
\end{proof}

From this theorem and Lemma~\ref{lem:instrcorresp}, we obtain 
correspondences for standard processes.

\begin{theorem}\label{thbeginend2}
Let $P_0$ be a closed process and $P'_0 = \instr{P_0}$. 
Let $M_{jk}$ ($j \in \{ 1, \ab \ldots, \ab m\}$, $k \in \{ 1,
\ldots, l_j\}$) be terms; let $\act$ and $\act_j$ ($j \in \{ 1, \ldots, m\}$)
be atoms. 
Let $p_{jk}, \acti, \acti_j$ be the patterns and facts
obtained by replacing names $a$ with patterns $a[\,]$ in the terms
and atoms
$M_{jk}, \act, \act_j$ respectively.  Assume that, for all clauses $R$ in
$\solve{P'_0, \rw}(\acti)$, there exist $j \in \{ 1, \ldots,
m\}$, $\sigma'$, and $H$ such that $R = H \wedge \pasbegin(\sigma'
p_{j1}) \wedge \ldots \wedge \pasbegin(\sigma' p_{jl_j}) \rewrite
\sigma' \acti_j$.  

Then $P_0$ satisfies the correspondence $\act
\Rightarrow \sor_{j=1}^{m} \left(\act_j \rightsquigarrow 
\sand_{k = 1}^{l_j} \pasevent(M_{jk})\right)$ against $\rw$-adversaries.
\end{theorem}

\begin{example}
For the process $P$ of Section~\ref{sec:example}, $\rw = \{ c \}$,
and $P' = \instr{P}$, the value of 
$\solve{P', \rw}(\pasend(e_B(x_1, x_2, x_3, x_4)))$
given in Example~\ref{exa:noninj1} shows  that
$P$ satisfies the correspondence
$\pasevent(e_B(x_1, \ab x_2, \ab x_3, \ab x_4)) \rightsquigarrow
\pasevent(e_1(x_1, \ab x_2, \ab x_3)) \wedge 
\pasevent(e_2(x_1, \ab x_2, \ab x_3, \ab x_4)) \wedge 
\pasevent(e_3(x_1, \ab x_2, \ab x_3, \ab x_4))$ against $\rw$-adversaries.
\end{example}

As particular cases of correspondences, we can show secrecy and non-injective agreement:

\begin{corollary}[Secrecy]
Let $P_0$ be a closed process and $P'_0 = \instr{P_0}$. 
Let $N$ be a term. Let $p$ be
the pattern obtained by replacing names $a$ with patterns $a[\,]$ in the
term $N$. Assume that
$\solve{P'_0, \rw}(\attacker(p)) = \emptyset$. Then $P_0$ preserves
the secrecy of all instances of $N$ from $\rw$.
\end{corollary}
Intuitively, if no instance of $\attacker(p)$ is derivable
from the clauses representing the protocol, then the adversary
cannot have an instance of the term $N$ corresponding to $p$.

\begin{example}
For the process $P$ of Section~\ref{sec:example}, $\rw = \{ c \}$,
and $P' = \instr{P}$,
our tool shows that $\solve{P', \rw}(\attacker(\sAa[\,])) =
\emptyset$. So $P$ preserves
the secrecy of $\sAa$ from $\rw$. The situation is similar
for $\sAb$, $\sBa$, and $\sBb$.
\end{example}

\begin{corollary}[Non-injective agreement]\label{th:noninjsolv}
Let $P_0$ be a closed process and $P'_0 = \instr{P_0}$.  
Assume that, for each $R \in
\solve{P'_0,\rw}(\pasend(e(x_1, \ldots, x_n)))$ such that $R = H
\rewrite \pasend(e(p_1, \ldots, \ab p_n))$, we have $\pasbegin(e'(p_1,
\ab \ldots, \ab p_n)) \in H$. Then $P_0$ satisfies the correspondence
$\pasevent(e(x_1, \ab \ldots, \ab x_n)) \rightsquigarrow \pasevent(e'(x_1,
\ab \ldots, \ab x_n))$ against $\rw$-adversaries.
\end{corollary}
Intuitively, the condition means that, if $\pasend(e(p_1, \ab \ldots,
\ab p_n))$ can be derived, $\pasbegin(e'(p_1, \ab \ldots, \ab p_n))$
occurs in the hypotheses. Then the theorem says that, if
$\asevent(e(M_1, \ab \ldots, \ab M_n))$ has been executed, then
$\asevent(e'(M_1, \ab \ldots, \ab M_n))$ has been executed.

\begin{example}
For the process $P$ of Section~\ref{sec:example}, $\rw = \{ c \}$,
and $P' = \instr{P}$, the value of 
$\solve{P', \rw}(\pasend(e_B(x_1, x_2, x_3, x_4)))$
given in Example~\ref{exa:noninj1} also shows that $P$
satisfies the correspondence $\pasevent(e_B(x_1, x_2, x_3, x_4)) 
\rightsquigarrow
\pasevent(e_3(x_1, x_2, x_3, x_4))$
against $\rw$-adversaries.
The tool shows in a similar way that $P$ satisfies the correspondence 
$\pasevent(e_A(x_1, x_2, x_3, x_4)) \rightsquigarrow
\pasevent(e_2(x_1, x_2, x_3, x_4))$ against $\rw$-adversaries.
\end{example}

\subsection{General Correspondences}\label{sect:injag}

\newcommand{\aseventInstr}[2]{\asevent({#2},{#1})}

In this section, we explain how to prove general correspondences.
Moreover, we also show that, when our verifier proves injectivity, it
proves recentness as well. For example, when it proves a correspondence
$\pasevent(M) \rightsquigarrow \inj\ \pasevent(M')$, it shows that,
when the event $\asevent(M)$ has been executed, not only the event
$\asevent(M')$ has been executed, but also this event has been
executed recently.
As explained by Lowe~\cite{Lowe97}, the precise meaning of ``recent''
depends on the circumstances: it can be that $\asevent(M)$ is executed
within the duration of the part of the process after $\asevent(M')$,
or it can be within a certain number of time units. Here, we define
recentness as follows: the runtime of the session that executes
$\asevent(M)$ overlaps with the runtime of the session that executes
the corresponding $\asevent(M')$ event.

We can formally define recent correspondences for instrumented
processes as follows.
We assume that, in $P_0$, the events are 
under at least one replication. We define an
instrumented process $P'_0 = \instrinj{P_0}$, where
$\instrinj{P_0}$ is defined like $\instr{P_0}$, except that
the events $\asevent(M)$ in $P_0$ are replaced with 
$\aseventInstr{i}{M}$, where $i$ is the session identifier
that labels the down-most replication above $\asevent(M)$ in $P_0$.
The session identifier $i$ indicates the session in which the 
considered event is executed.

When $\overline{k} = k_1\ldots k_n$ is a non-empty sequence of
indices, we denote by $\cutlast{\overline{k}}$ the sequence obtained
by removing the last index from $\overline{k}$:
$\cutlast{\overline{k}} = k_1\ldots k_{n-1}$.

\begin{definition}
Let $P_0$ be a closed process and $P'_0 =\instrinj{P_0}$.
We say that $P'_0$ \emph{satisfies the recent correspondence}
\[\pasevent(p) \Rightarrow \sor_{j=1}^{m} 
\left(\pasevent(p'_j) \rightsquigarrow \sand_{k = 1}^{l_j} 
\injopt_{jk}q_{jk}\right)\]
where 
\[q_{\overline{jk}} = \pasevent(p_{\overline{jk}}) \rightsquigarrow \sor_{j=1}^{m_{\overline{jk}}}
\sand_{k = 1}^{l_{\overline{jk}j}} \injopt_{\overline{jk}jk} q_{\overline{jk}jk}\]
against $\rw$-adversaries if and only if for any
$\rw$-adversary $Q$, for any trace $\trace = S_0, \env_0, \{ P'_0, Q'
\}\rightarrow^* S', \env', \pset'$, with $Q' =
\instradv{Q}$, $\env_0(a) = a[\,]$ for all $a \in
\dom(\env_0)$, and $\fn(P'_0) \cup \rw \subseteq \dom(\env_0)$, there
exists a function $\phi_{\overline{jk}}$ for each non-empty $\overline{jk}$,
such that for all non-empty $\overline{jk}$, $\phi_{\overline{jk}}$ maps 
a subset of steps of $\trace$ to steps of $\trace$ and
\begin{itemize}

\item For all $\step$, if the event
  $\aseventInstr{\lambda_{\epsilon}}{\sigma p}$ is
  executed at step $\step$ in $\trace$ for some $\sigma$ and 
  $\lambda_{\epsilon}$, then there exist $\sigma'$ and
  $J = (j_{\overline{k}})_{\overline{k}}$ such that $\sigma'
  p'_{j_{\epsilon}} = \sigma p$ and, for all non-empty $\overline{k}$,
  $\phi_{\indp{\overline{k}}{J}}(\step)$ is defined,
  $\aseventInstr{\lambda_{\overline{k}}}{\sigma' p_{\indp{\overline{k}}{J}}}$ is
  executed at step $\phi_{\indp{\overline{k}}{J}}(\step)$ in $\trace$, and if
  $\injopt_{\indp{\overline{k}}{J}} = \inj$, then the runtimes of
  $\session(\lambda_{\cutlast{\overline{k}}})$ and
  $\session(\lambda_{\overline{k}})$ overlap (recentness).

The runtime of $\session(\lambda)$ begins when the rule $S,\ab \env,\ab \pset
\cup \{ \,\ReplInstr{i}{P}\, \} \rightarrow S \setminus \{ \lambda\},\ab \env,\ab
\pset \cup \{ \,P\{\lambda/i\}, \ReplInstr{i}{P}\, \}$ is applied and
ends when $P\{\lambda/i\}$ has disappeared.

\item For all non-empty $\overline{jk}$, 
if $\injopt_{\overline{jk}} = \inj$, then $\phi_{\overline{jk}}$ is injective.

\item For all non-empty $\overline{jk}$, for all $j$ and $k$,
if $\phi_{\overline{jk}jk}(\step)$ is defined, then 
$\phi_{\overline{jk}}(\step)$ is defined and 
$\phi_{\overline{jk}jk}(\step) \leq \phi_{\overline{jk}}(\step)$.
For all $j$ and $k$, if $\phi_{jk}(\step)$ is defined, then 
$\phi_{jk}(\step) \leq \step$.

\end{itemize}
\end{definition}
We do not define recentness for standard processes, since it is
difficult to track formally the runtime of a session in these processes.
Instrumented processes make that very easy thanks to session
identifiers. It is easy to infer correspondences for
standard processes from recent correspondences for instrumented
processes, with a proof similar to that of Lemma~\ref{lem:instrcorresp}.

\begin{lemma}\label{lem:instrinjcorresp}
Let $P_0$ be a closed process and $P'_0 =
\instrinj{P_0}$.  Let $M_{\overline{jk}}$,  $M$, and $M'_j$ 
be terms. Let $p_{\overline{jk}}, p, p'_j$
be the patterns obtained by replacing names $a$ with patterns $a[\,]$ in
the terms $M_{\overline{jk}}, M, M'_j$ respectively.  
If $P'_0$ satisfies the recent correspondence 
\[\pasevent(p) \Rightarrow \sor_{j=1}^{m}  \left(
\pasevent(p'_j) \rightsquigarrow 
\sand_{k = 1}^{l_j} \injopt_{jk}q_{jk}\right) \]
where 
\[q_{\overline{jk}} = \pasevent(p_{\overline{jk}}) \rightsquigarrow \sor_{j=1}^{m_{\overline{jk}}}
\sand_{k = 1}^{l_{\overline{jk}j}} \injopt_{\overline{jk}jk} q_{\overline{jk}jk}\]
against $\rw$-adversaries then $P_0$ satisfies the  
correspondence 
\[\pasevent(M) \Rightarrow \sor_{j=1}^{m} \left(
\pasevent(M'_j) \rightsquigarrow  \sand_{k = 1}^{l_j} 
\injopt_{jk} q'_{jk}\right)\]
where 
\[q'_{\overline{jk}} = \pasevent(M_{\overline{jk}}) \rightsquigarrow \sor_{j=1}^{m_{\overline{jk}}}
\sand_{k = 1}^{l_{\overline{jk}j}} \injopt_{\overline{jk}jk} q'_{\overline{jk}jk}\]
against $\rw$-adversaries.
\end{lemma}

Let $P_0$ be a closed process and $P'_0 = \instrinj{P_0}$.
We adapt the generation of clauses as follows:
the set of clauses $\rsetp{P'_0,\rw}$ is defined as $\rset{P'_0, \rw}$ 
except that
{\allowdisplaybreaks
\begin{align*}
&\lp \coutput{M}{N}.P \rp \rho H = \lp P \rp \rho H
\cup \{ H\{ \rho_{|V_o \cup V_s} / \square \} \rewrite \mess(\rho(M), \rho(N)) \}\\
&\lp \ReplInstr{i}{P} \rp \rho H = \lp P \rp (\rho[i \mapsto i]) 
(H\{ \rho_{|V_o \cup V_s} / \square \})\\
&\lp \aseventInstr{i}{M}.P\rp \rho H = \lp P\rp \rho (H
\wedge \pasbegin(\rho(M),\square)) \cup \{ H \rewrite \pasend(\rho(M),i) \}
\end{align*}}%
where $\square$ is a special variable.  The predicate $\pasend$
has as additional argument the session identifier in which the event
is executed. The predicate $\pasbegin$ has as additional argument an
environment $\rho$ that gives values that variables will contain at
the first output or replication
that follows the event; $\square$ is a placeholder
for this environment. We define $\solvep{P'_0, \rw}$ as
$\solve{P'_0, \rw}$ except that it applies to $\rsetp{P'_0,\rw}$
instead of $\rset{P'_0, \rw}$.

Let us first consider the particular case of injective correspondences.
We consider general correspondences in Theorem~\ref{th:recentcorresp} below.

\begin{proposition}[Injective correspondences]\label{prop:recinjprop}
  Let $P_0$ be a closed process and $P'_0 =
  \instrinj{P_0}$. We assume that, in $P_0$, all events
  are of the form $\asevent(f(M_1, \ldots, M_n))$ and that different
  occurrences of $\asevent$ have different root function symbols.

We also assume that the patterns $p, p'_j, p_{jk}$ 
satisfy the following conditions: $p$ and $p'_j$ for $j \in \{1,
\ldots, m\}$ are of the form $f(\ldots)$ for some function symbol $f$ 
and for all $j$, $k$ such that $\injopt_{jk} = \inj$,
$p_{jk} = f_{jk}(\ldots)$ for some function symbol $f_{jk}$. 

Let $\solvep{P'_0, \rw}(\pasend(p,i)) 
= \{ R_{jr} \mid j \in \{ 1, \ldots, m \}, r \in
\{ 1, \ldots, n_j \} \}$. Assume that there exist $x_{jk}$, $i_{jr}$, and $\rho_{jrk}$ ($j \in \{1, \ldots, m \}$, $r \in \{1, \ldots, n_j\}$, $k
\in \{ 1, \ldots, l_j \}$) such that 
\begin{itemize}

\item For all $j \in \{1, \ldots, m \}$, for all $r \in \{1, \ldots,
n_j\}$, there exist $H$ and $\sigma$ such that $R_{jr} = H \wedge
\pasbegin(\sigma p_{j1}, \rho_{jr1}) \wedge \ldots
\wedge \pasbegin(\sigma p_{j l_j}, \ab \rho_{jrl_j})
\rewrite \pasend(\sigma p'_j, \ab i_{jr})$.

\item For all $j \in \{1, \ldots, m \}$, for all $r$ and $r'$ in $\{
1, \ldots, n_j \}$, for all $k \in \{ 1, \ldots, l_j \}$ such that
$\injopt_{jk} = \inj$,
$\rho_{jrk}(x_{jk})\{\lambda/i_{jr}\}$ does not unify with
$\rho_{jr'k}(x_{jk})\{\lambda'/i_{jr'}\}$ when $\lambda \neq \lambda'$.

\end{itemize}
Then $P'_0$ satisfies the recent correspondence 
\[\pasevent(p) \Rightarrow \sor_{j=1}^{m} \left(
\pasevent(p'_j) \rightsquigarrow 
\sand_{k = 1}^{l_j} \injopt_{jk} \pasevent(p_{jk})\right)\]
against $\rw$-adversaries.
\end{proposition}
This proposition is a particular case of Theorem~\ref{th:recentcorresp} below.
It is proved in Appendix~\ref{app:recentness}.
By Theorem~\ref{thbeginend3}, after deleting session identifiers
and environments, the first item shows that $P'_0$ satisfies the
correspondence
\begin{equation}
\pasevent(p) \Rightarrow \sor_{j=1..m, r} \left(\pasevent(p'_j) \rightsquigarrow 
\sand_{k = 1}^{l_j} \pasevent(p_{jk})\right)
\end{equation}
The environments and session identifiers as well as the second item
serve in proving injectivity. Suppose that $\injopt_{jk} = \inj$,
and denote by $\_$ an unknown term.
If two instances of $\asevent(p, i)$ are executed in $P'_0$ for
the branch $j$ of the correspondence, by the first item, they are instances of
$\asevent(\sigma_{jr} p'_j, i_{jr})$ for some $r$, so they are
$\asevent(\sigma'_1 \sigma_{jr_1} p'_j, \sigma'_1 i_{jr_1})$ and
$\asevent(\sigma'_2 \sigma_{jr_2} p'_j, \sigma'_2 i_{jr_2})$ for some
$\sigma'_1$ and $\sigma'_2$. Furthermore, there is only one occurrence
of $\asevent(f(\ldots), i)$ in $P'_0$, so the event
$\asevent(f(\ldots), i)$ can be executed at most once for each value
of the session identifier $i$, so $\sigma'_1 i_{jr_1} \neq \sigma'_2
i_{jr_2}$. Then, by the first item, corresponding events 
$\asevent(\sigma'_1 \sigma_{jr_1} p_{jk}, \_)$ and $\asevent(\sigma'_2 \sigma_{jr_2}
p_{jk}, \_)$ have been executed, with associated
environments $\sigma'_1 \rho_{jr_1k}$ and $\sigma'_2 \rho_{jr_2k}$. By
the second item, $\rho_{jr_1k}(x_{jk}) \{ \lambda_1/i_{jr_1}\}$ does not
unify with $\rho_{jr_2k}(x_{jk}) \{ \lambda_2/i_{jr_2}\}$ for
different values $\lambda_1 = \sigma'_1 i_{jr_1}$ and $\lambda_2 =
\sigma'_2 i_{jr_2}$ of the session identifier.  
(In this condition, $r_1$ can be equal to $r_2$, and
when $r_1 = r_2 = r$, the condition simply means that $i_{jr}$ occurs
in $\rho_{jrk}$.)
So $\sigma'_1
\rho_{jr_1k}(x_{jk}) \neq \sigma'_2 \rho_{jr_2k}(x_{jk})$, so the
events $\asevent(\sigma'_1 \sigma_{jr_1} p_{jk}), \_)$ and
$\asevent(\sigma'_2 \sigma_{jr_2} p_{jk}), \_)$ are distinct,
which shows injectivity.
This point is very similar to the fact that injective agreement is
implied by non-injective agreement when the parameters of events
contain nonces generated by the agent to whom authentication is being made, 
because the event can be executed at most once
for each value of the nonce. (The session identifier $i_{jr}$ in our theorem
plays the role of the nonce.) [Andrew Gordon, personal communication].

\begin{corollary}[Recent injective agreement]\label{cor:recinjag}
  Let $P_0$ be a closed process and $P'_0 =
  \instrinj{P_0}$.  We assume that, in $P_0$, all events
  are of the form $\asevent(f(M_1, \ldots, M_k))$ and that different
  occurrences of $\asevent$ have different root function symbols.
Let $\{ R_1, \ldots, R_n \} = \solvep{P'_0, \rw}(\pasend(e(x_1, \ab \ldots, \ab x_m),i))$.
Assume that there exist $x$, $i_r$, and $\rho_r$ ($r \in \{ 1, \ldots, n\}$)
such that 
\begin{itemize}

\item For all $r \in \{1, \ldots, n\}$, $R_r = H \wedge
\pasbegin(e'(p_1, \ldots, p_m), \rho_r)\rewrite \pasend(e(p_1, \ab \ldots, \ab p_m), \ab i_r)$ for some $p_1, \ldots, p_m$, and $H$.

\item For all $r$ and $r'$ in $\{ 1, \ldots, n \}$,
$\rho_r(x)\{\lambda/i_r\}$ does not unify with
$\rho_{r'}(x)\{\lambda'/i_{r'}\}$ when $\lambda \neq \lambda'$.

\end{itemize}
Then $P'_0$ satisfies the recent correspondence
$\pasevent(e(x_1, \ab \ldots, \ab x_m)) \rightsquigarrow \inj$ $\pasevent(e'(x_1, \ab \ldots, \ab x_m))$ against
$\rw$-adversaries.
\end{corollary}
\begin{proof}
This result is an immediate consequence of Proposition~\ref{prop:recinjprop}.
\proofcomplete
\end{proof}

\begin{example}\label{ex:injag}
For the process $P$ of Section~\ref{sec:example}, 
$P' = \instrinj{P}$, and $\rw = \{ c \}$, we have
\begin{align*}
&\solvep{P',\rw}(\pasend(e_B(x_1, \ab x_2, \ab x_3, \ab x_4), i)) =\\*
&\quad \{ H \wedge \pasbegin(e_3(\pkA,\pkB,a[\pkB,i_{A0}],b[p_1,i_{B0}]), \rho)\\*
&\qquad \rewrite \pasend(e_B(\pkA,\pkB,a[\pkB,i_{A0}],b[p_1,i_{B0}]),i_{B0}) \}\\
&\text{where }\pkA = \pk(\skA[\,]),\ \pkB = \pk(\skB[\,])\\*
&\phantom{\text{where }}p_1 = \pencrypt((a[\pkB,i_{A0}],\pkA),\pkB,r_1[\pkB,i_{A0}])\\*
&\phantom{\text{where }}p_2 = \pencrypt((a[\pkB,i_{A0}],b[p_1,i_{B0}],\pkB),\pkA,r_2[p_1,i_{B0}])\\*
&\phantom{\text{where }}\rho = \{ i_A \mapsto i_{A0}, x\_\pkB \mapsto \pkB, m \mapsto p_2\}
\end{align*}
Intuitively, this result shows that each event $e_B(\pkA,\ab \pkB,\ab a[\pkB,i_{A0}],\ab b[p_1,i_{B0}])$, executed
in the session of index $i_B = i_{B0}$ is preceded by an event $e_3(\pkA,\ab \pkB,\ab a[\pkB,i_{A0}],\ab b[p_1,i_{B0}])$
executed in the session of index $i_A = i_{A0}$ with $x\_\pkB = \pkB$ and $m = p_2$.
Since $i_{B0}$ occurs in this event (or in its environment\footnote{In general, the environment may contain more variables than the event itself, so looking for the session identifiers in the environment instead of the event is more powerful.}), different executions of $e_B$,
which have different values of $i_{B0}$, cannot correspond to the same execution of $e_3$,
so we have injectivity.
More formally, the second hypothesis of Corollary~\ref{cor:recinjag} is satisfied because
$\rho(m) \{ \lambda / i_{B0} \}$ does not unify with
$\rho(m) \{ \lambda' / i_{B0} \}$ when $\lambda \neq \lambda'$,
since $i_{B0}$ occurs in $\rho(m) = p_2$.
Then, $P'$ satisfies the recent correspondence
$\pasevent(e_B(x_1, \ab x_2, \ab x_3, \ab x_4)) \rightsquigarrow
\inj\ \pasevent(e_3(x_1, \ab x_2, \ab x_3, \ab x_4))$
against $\rw$-adversaries.

The tool shows in a similar way that $P'$ satisfies the recent 
correspondence
$\pasevent(e_A(x_1, \ab x_2, \ab x_3, \ab x_4)) \rightsquigarrow
\inj\ \pasevent(e_2(x_1, \ab x_2, \ab x_3, \ab x_4))$
against $\rw$-adversaries.
\end{example}

Let us now consider the case of general correspondences.
The basic idea is to decompose the general correspondence to prove
into several correspondences. For instance, the correspondence
$\pasevent(e_B(x_1, \ab x_2, \ab x_3, \ab x_4)) \rightsquigarrow
(\pasevent(e_3(x_1, \ab x_2, \ab x_3, \ab x_4)) \rightsquigarrow
\pasevent(e_2(x_1, \ab x_2, \ab x_3, \ab x_4)))$ is implied by the conjunction
of the correspondences 
$\pasevent(e_B(x_1, \ab x_2, \ab x_3, \ab x_4)) \rightsquigarrow
\pasevent(e_3(x_1, \ab x_2, \ab x_3, \ab x_4))$ and 
$\pasevent(e_3(x_1, \ab x_2, \ab x_3, \ab x_4)) \rightsquigarrow
\pasevent(e_2(x_1, \ab x_2, \ab x_3, \ab x_4))$. However, as noted in 
Section~\ref{sec:defgencorresp}, this proof technique would often
fail because, in order to prove that $e_2(x_1, \ab x_2, \ab x_3, \ab x_4)$
has been executed, we may need to know that $e_B(x_1, \ab x_2, \ab x_3, \ab x_4)$
has been executed, and not only that $e_3(x_1, \ab x_2, \ab x_3, \ab x_4)$
has been executed. To solve this problem, we use the following idea:
when we know that $e_B(x_1, \ab x_2, \ab x_3, \ab x_4)$ has been executed,
we may be able to show that certain particular instances of 
$e_3(x_1, \ab x_2, \ab x_3, \ab x_4)$ have been executed, and we can exploit
this information in order to prove that $e_2(x_1, \ab x_2, \ab x_3, \ab x_4)$
has been executed. In other words, we rather prove the correspondences
$\pasevent(e_B(x_1, \ab x_2, \ab x_3, \ab x_4)) \Rightarrow
\sor_{r=1}^m \sigma_r \pasevent(e_B(x_1, \ab x_2, \ab x_3, \ab x_4)) 
\rightsquigarrow
\sigma_r \pasevent(e_3(x_1, \ab x_2, \ab x_3, \ab x_4))$ and
for all $r \leq m$, $\sigma_r \pasevent(e_3(x_1, \ab x_2, \ab x_3, \ab x_4)) \rightsquigarrow
\sigma_r \pasevent(e_2(x_1, \ab x_2, \ab x_3, \ab x_4))$.
When the considered general correspondence has several nesting levels,
we perform such a decomposition recursively.
The next theorem generalizes and formalizes these ideas.

\newcommand{\checkrc}{\mathrm{verify}}
\newcommand{\envset}{\mathit{Env}}

Below, the notation $(\envset_{\overline{jk}})_{\overline{jk}}$
represents a family $\envset_{\overline{jk}}$ of sets of pairs $(\rho,
i)$ where $\rho$ is an environment and $i$ is a session identifier,
one for each non-empty $\overline{jk}$.  The notation
$(\envset_{jk\overline{jk}})_{\overline{jk}}$ represents a subfamily
of $(\envset_{\overline{jk}})_{\overline{jk}}$ in which the first
two indices are $jk$, and this family is reindexed by omitting the
fixed indices $jk$.

\begin{theorem}\label{th:recentcorresp}
Let $P_0$ be a closed process and $P'_0 =\instrinj{P_0}$.
We assume that, in $P_0$, all events are of the form 
$\asevent(f(M_1, \ldots, M_n))$ and
that different occurrences of $\asevent$ have different root function
symbols. 

Let us define $\checkrc(q', (\envset_{\overline{jk}})_{\overline{jk}})$, 
where $\overline{jk}$ is non-empty, by:
\begin{enumerate}

\item[V1.] If $q' = \pasevent(p)$ for some $p$, then $\checkrc(q',
  (\envset_{\overline{jk}})_{\overline{jk}})$ is true.

\item[V2.] If $q' = \pasevent(p) \Rightarrow \sor_{j=1}^{m}
  \left(\pasevent(p'_j) \rightsquigarrow \sand_{k = 1}^{l_j}
    \injopt_{jk}q'_{jk}\right)$ and $q'_{jk} = \pasevent(p_{jk})
  \rightsquigarrow \ldots$ for some $p$, $p'_j$, and $p_{jk}$, where
  $m \neq 1$, $l_j \neq 0$, or $p \neq p'_1$, then $\checkrc(q',
  (\envset_{\overline{jk}})_{\overline{jk}})$ is true if and only if
  there exists $(\sigma_{jr})_{jr}$ such that the following three
  conditions hold:
\begin{enumerate}

\item[V2.1.] 
We have $\solvep{P'_0, \rw}(\pasend(p,i)) \subseteq \{ H \wedge
\bigwedge_{k=1}^{l_j} \pasbegin(\sigma_{jr} p_{jk}, \rho_{jrk}) \rewrite
\pasend(\sigma_{jr} p'_j, i_{jr})$ for some $H$, $j\in \{1, \ldots,
m\}$, $r$, and $(\rho_{jrk},i_{jr}) \in \envset_{jk}$ for all $k \}$.

\item[V2.2.]
For all $j, r, k_0$, the common variables between 
$\sigma_{jr} q'_{jk_0}$ on the one hand and
$\sigma_{jr} p'_j$ and
$\sigma_{jr} q'_{jk}$ for all $k \neq k_0$ on the
other hand occur in $\sigma_{jr} p_{jk_0}$.

\item[V2.3.]
For all $j, r,k$, $\checkrc(\sigma_{jr} q'_{jk}, 
(\envset_{jk\overline{jk}})_{\overline{jk}})$ is true.

\end{enumerate}
\end{enumerate}
Consider the following recent correspondence:
\[q = \pasevent(p) \Rightarrow \sor_{j=1}^{m} 
\left(\pasevent(p'_j) \rightsquigarrow \sand_{k = 1}^{l_j} 
\injopt_{jk}q_{jk}\right)\]
where 
\[q_{\overline{jk}} = \pasevent(p_{\overline{jk}}) \rightsquigarrow \sor_{j=1}^{m_{\overline{jk}}}
\sand_{k = 1}^{l_{\overline{jk}j}} \injopt_{\overline{jk}jk} q_{\overline{jk}jk}\]
We assume that the patterns in the correspondence  
satisfy the following conditions: $p$ and $p'_j$ for $j \in \{1,
\ldots, m\}$ are of the form $f(\ldots)$ for some function symbol $f$
and, for all non-empty $\overline{jk}$ such that $\injopt_{\overline{jk}} = \inj$, 
$p_{\overline{jk}} = f_{\overline{jk}}(\ldots)$ for some function
symbol $f_{\overline{jk}}$.
We also assume that if $\inj$ occurs in $q_{\overline{jk}}$, then
$\injopt_{\overline{jk}} = \inj$.

Assume that there exist 
$(\envset_{\overline{jk}})_{\overline{jk}}$ and $(x_{\overline{jk}})_{\overline{jk}}$, where $\overline{jk}$ is non-empty, such that
\begin{enumerate}

\item[H1.] $\checkrc(q, (\envset_{\overline{jk}})_{\overline{jk}})$ is true.

\item[H2.] For all non-empty $\overline{jk}$, if $\injopt_{\overline{jk}} = \inj$, then for all $(\rho,i), (\rho',i') \in \envset_{\overline{jk}}$,
$\rho(x_{\overline{jk}}) \{ \lambda / i \}$ does not unify with $\rho'(x_{\overline{jk}}) \{ \lambda' / i' \}$ when $\lambda \neq \lambda'$.

\end{enumerate}
Then $P'_0$ satisfies the recent correspondence $q$
against $\rw$-adversaries.
\end{theorem}
This theorem is rather complex, so  
we give some intuition here. Its proof can be found in Appendix~\ref{app:recentness}.

Point~V2.1 allows us to infer correspondences
by Theorem~\ref{thbeginend3}: after deleting session identifiers and 
environments, $P'_0$ satisfies the correspondences:
\begin{equation}
\pasevent(p) \Rightarrow \sor_{j=1..m, r} \left(\pasevent(\sigma_{jr} p'_j) \rightsquigarrow 
\sand_{k = 1}^{l_j} \pasevent(\sigma_{jr} p_{jk})\right)
\label{eq:hyp11}
\end{equation}
and, using the recursive calls of Point~V2.3,
\begin{equation}
\pasevent(\sigma'_{\cutlast{\overline{jrk}}} p_{\overline{jk}}) \Rightarrow \sor_{j=1..m_{\overline{jk}}, r} \left(\pasevent(\sigma'_{\overline{jrk}jr} p_{\overline{jk}}) \rightsquigarrow 
\sand_{k = 1}^{l_{\overline{jk}j}} \pasevent(\sigma'_{\overline{jrk}jr} p_{\overline{jk}jk})\right)
\label{eq:hyp12}
\end{equation}
against $\rw$-adversaries, where 
$\sigma'_{\overline{jrk}jr} = \sigma_{\overline{jrk}jr} \sigma_{\cutlast{\overline{jrk}}} \ldots \sigma_{jr}$ 
and we denote by 
$\sigma_{\overline{jrk}jr}$ the substitution $\sigma_{jr}$ obtained in 
recursive calls to $\checkrc$ indexed by $\overline{jrk}$. In order to infer the desired
correspondence, we need to show injectivity properties and to
combine the correspondences~\eqref{eq:hyp11} and~\eqref{eq:hyp12} 
into a single correspondence.
Injectivity comes from Hypothesis~H2: this hypothesis generalizes the
second item of Proposition~\ref{prop:recinjprop} to the case of general
correspondences.

The correspondences~\eqref{eq:hyp11} and~\eqref{eq:hyp12} are combined
into a single correspondence using Point~V2.2. We illustrate
this point on the simple example of the correspondence
$\pasevent(p) \ab \Rightarrow\ab (\pasevent(p'_1) \ab \rightsquigarrow \ab
(\pasevent(p_{11}) \ab \rightsquigarrow \ab \pasevent(p_{1111})))$.
By V2.1 and the recursive call of V2.3, we have correspondences of the form:
\begin{align}
&\pasevent(p) \Rightarrow \sor_r \left(\pasevent(\sigma_{1r} p'_1) \rightsquigarrow \pasevent(\sigma_{1r} p_{11})\right)
\label{nest1}\\
&\pasevent(\sigma_{1r} p_{11}) \Rightarrow \sor_{r'} \left(\pasevent(\sigma_{1r11r'} \sigma_{1r} p_{11}) \rightsquigarrow 
\pasevent(\sigma_{1r11r'} \sigma_{1r} p_{1111})\right)
\label{nest2}
\end{align}
for some $\sigma_{1r}$ and $\sigma_{1r11r'}$.
The correspondence~\eqref{nest2} implies the simpler correspondence
\begin{equation}
\pasevent(\sigma_{1r} p_{11}) \rightsquigarrow 
\pasevent(\sigma_{1r} p_{1111}).
\label{nest2simp}
\end{equation}
Furthermore, if an instance of $\pasevent(p)$ is executed, 
$e_1 = \pasevent(\sigma p)$, then by~\eqref{nest1},
for some $r$ and $\sigma'_1$ such that $\sigma p
= \sigma'_1 \sigma_{1r} p'_{1}$, the event $e_2 = \pasevent(\sigma'_1 \sigma_{1r} p_{11})$ has been executed before $e_1$.
By~\eqref{nest2simp}, for some $\sigma'_2$ such that
$\sigma'_1 \sigma_{1r} p_{11} = \sigma'_2 \sigma_{1r} p_{11}$,
the event $e_3 = \pasevent(\sigma'_2 \sigma_{1r} p_{1111})$ has been executed before $e_2$.
We now need to reconcile the substitutions $\sigma'_1$ and $\sigma'_2$;
this can be done thanks to V2.2.
Let us define $\sigma''$ such that $\sigma'' x = \sigma'_1 x$ for $x \in \fv(\sigma_{1r} p_{11}) \cup \fv(\sigma_{1r} p'_{1})$ and $\sigma'' x = \sigma'_2 x$ for $x \in \fv(\sigma_{1r} p_{1111}) \cup \fv(\sigma_{1r} p_{11})$.
Such a substitution $\sigma''$ exists because the common variables between
$\fv(\sigma_{1r} p_{11}) \cup \fv(\sigma_{1r} p'_{1})$ and 
$\fv(\sigma_{1r} p_{1111}) \cup \fv(\sigma_{1r} p_{11})$
occur in $\sigma_{1r} p_{11}$ by V2.2, and for the
variables $x \in \fv(\sigma_{1r} p_{11})$, $\sigma'_1 x = \sigma'_2 x$
since $\sigma'_1 \sigma_{1r} p_{11} = \sigma'_2 \sigma_{1r} p_{11}$.
So, for some $r$ and $\sigma''$ such that $\sigma p = \sigma''
\sigma_{1r} p'_{1}$, the event $e_2 = \pasevent(\sigma'' \sigma_{1r} p_{11})$
has been executed before $e_1$ and $e_3 = \pasevent(\sigma''
\sigma_{1r} p_{1111})$ has been executed before $e_2$.  This result
proves the desired correspondence $\pasevent(p) \Rightarrow
\left(\pasevent(p'_1) \rightsquigarrow (\pasevent(p_{11})
\rightsquigarrow \pasevent(p_{1111})\right)$.
Point~V2.2 generalizes this technique to any correspondence.

In the implementation, the hypotheses of this theorem are checked as follows.
In order to check $\checkrc(q', (\envset_{\overline{jk}})_{\overline{jk}})$,
we first compute $\solvep{P'_0, \rw}(\pasend(p,i))$.
By matching, we check V2.1 and obtain the values of $\sigma_{jr}$, $\rho_{jrk}$, and
$i_{jr}$ for all $j$, $r$, and $k$. We add $(\rho_{jrk}, i_{jr})$ to
$\envset_{jk}$. We compute $\sigma_{jr} p'_j$ and 
$\sigma_{jr}q'_{jk}$ for each $j$, $r$, and $k$, and check V2.2
and V2.3. 

After checking $\checkrc(q', (\envset_{\overline{jk}})_{\overline{jk}})$, 
we finally check Hypothesis~H2 for each $\overline{jk}$. We start with
a set that contains the whole domain of $\rho$ for some $(\rho, i) \in
\envset_{\overline{jk}}$.  For each $(\rho, i)$ and $(\rho', i')$ in
$\envset_{\overline{jk}}$, we remove from this set the variables $x$
such that $\rho(x) \{ \lambda / i \}$ unifies with $\rho'(x) \{
\lambda' / i' \}$ for $\lambda \neq \lambda'$.  When the obtained set
is non-empty, Hypothesis~H2 is satisfied by taking for
$x_{\overline{jk}}$ any element of the obtained set. Otherwise,
Hypothesis~H2 is not satisfied.

\begin{example}\label{exa:nestedcorresp}
For the example $P$ of Section~\ref{sec:example}, 
the previous theorem does not enable us to prove the correspondence
$\pasevent(e_B(x_1, \ab x_2, \ab x_3, \ab x_4)) \rightsquigarrow
(\inj\ \pasevent(e_3(x_1, \ab x_2, \ab x_3, \ab x_4)) \rightsquigarrow
(\inj\ \pasevent(e_2(x_1, \ab x_2, \ab x_3, \ab x_4)) \rightsquigarrow
\inj\ \pasevent(e_1(x_1, \ab x_2, \ab x_3))))$
directly. Indeed, Theorem~\ref{th:recentcorresp} would require 
that we show a correspondence of the form
$\pasevent(\sigma e_2(x_1, \ab x_2, \ab x_3, \ab x_4)) \rightsquigarrow
\inj\ \pasevent(\sigma e_1(x_1, \ab x_2, \ab x_3))$. However, such a 
correspondence does not hold, because after executing a single
event $e_1$, the adversary can replay the first message of the protocol,
so that $B$ executes several events $e_2$.

It is still possible to prove this correspondence by combining
the automatic proof of the slightly weaker correspondence
$q = \pasevent(e_B(x_1, \ab x_2, \ab x_3, \ab x_4)) \rightsquigarrow
(\inj\ \pasevent(e_3(x_1, \ab x_2, \ab x_3, \ab x_4)) \rightsquigarrow
(\inj\ \pasevent(e_1(x_1, \ab x_2, \ab x_3)) \wedge 
\inj\ \pasevent(e_2(x_1, \ab x_2, \ab x_3, \ab x_4))))$, which does not order
the events $e_1$ and $e_2$, with a simple manual argument. (This technique
applies to many other examples.) Let us first prove the latter 
correspondence.

Let $P' = \instrinj{P}$ and $\rw = \{ c \}$.
We have
{\allowdisplaybreaks\begin{align*}
&\solvep{P',\rw}(\pasend(e_B(x_1, \ab x_2, \ab x_3, \ab x_4), i)) =\\*
&\quad \{ H \wedge \pasbegin(e_3(\pkA,\pkB,a[\pkB,i_{A0}],b[p_1,i_{B0}]), \rho_{111})\\*
&\qquad \rewrite \pasend(e_B(\pkA,\pkB,a[\pkB,i_{A0}],b[p_1,i_{B0}]),i_{B0}) \}\\
&\solvep{P',\rw}(\pasend(e_3(\pkA,\pkB,a[\pkB,i_{A0}],b[p_1,i_{B0}]), i)) = \\*
&\quad\{ 
\pasbegin(e_1(\pkA,\pkB,a[\pkB,i_{A0}]), \rho_{111111})\\*
&\quad \wedge
\pasbegin(e_2(\pkA,\pkB,a[\pkB,i_{A0}],b[p_1,i_{B0}]), \rho_{111112}) \\*
&\qquad\rewrite \pasend(e_3(\pkA,\pkB,a[\pkB,i_{A0}],b[p_1,i_{B0}]), i_{A0}) \}\\
&\text{where }\pkA = \pk(\skA[\,]),\ \pkB = \pk(\skB[\,])\\*
&\phantom{\text{where }}p_1 = \pencrypt((a[\pkB,i_{A0}],\pkA),\pkB,r_1[\pkB,i_{A0}])\\*
&\phantom{\text{where }}p_2 = \pencrypt((a[\pkB,i_{A0}],b[p_1,i_{B0}],\pkB),\pkA,r_2[p_1,i_{B0}])\\*
&\phantom{\text{where }}\rho_{111} = \rho_{111111} = \{ i_A \mapsto i_{A0}, x\_\pkB \mapsto \pkB, m \mapsto p_2\}\\*
&\phantom{\text{where }}\rho_{111112} = \{ i_B \mapsto i_{B0}, m' \mapsto p_1 \}
\end{align*}}%
Intuitively, as in Example~\ref{ex:injag}, the value of $\solvep{P',\rw}(\pasend(e_B(x_1, \ab x_2, \ab x_3, \ab x_4), i))$
guarantees that each event $e_B(\pkA,\ab \pkB,\ab a[\pkB,i_{A0}],\ab b[p_1,i_{B0}])$, executed
in the session of index $i_B = i_{B0}$ is preceded by an event $e_3(\pkA,\ab \pkB,\ab a[\pkB,i_{A0}],\ab b[p_1,i_{B0}])$
executed in the session of index $i_A = i_{A0}$ with $x\_\pkB = \pkB$ and $m = p_2$.
Since $i_{B0}$ occurs in this event (or in its environment), we have injectivity.
The value of $\solvep{P',\rw}(\pasend(e_3(\pkA,\ab \pkB,\ab a[\pkB,i_{A0}],\ab b[p_1,i_{B0}]), \ab i))$
guarantees that each event $e_3(\pkA,\ab \pkB,\ab a[\pkB,i_{A0}],\ab b[p_1,i_{B0}])$
executed in the session of index $i_A = i_{A0}$ is preceded by events 
$e_1(\pkA,\ab \pkB,\ab a[\pkB,\ab i_{A0}])$ executed in the session of index
$i_A = i_{A0}$ with $x\_\pkB = \pkB$ and $m = p_2$,
and $e_2(\pkA,\ab \pkB,\ab a[\pkB,i_{A0}],\ab b[p_1,i_{B0}])$ executed in the
session of index $i_B = i_{B0}$ with $m' = p_1$.
Since $i_{A0}$ occurs in these events (or in their environments), we have injectivity.
So we obtain the desired correspondence
$\pasevent(e_B(x_1, \ab x_2, \ab x_3, \ab x_4)) \rightsquigarrow
(\inj\ \pasevent(e_3(x_1, \ab x_2, \ab x_3, \ab x_4)) \rightsquigarrow
(\inj\ \pasevent(e_1(x_1, \ab x_2, \ab x_3)) \wedge
\inj\ \pasevent(e_2(x_1, \ab x_2, \ab x_3, \ab x_4))))$.

More formally, let us show that we can apply Theorem~\ref{th:recentcorresp}.
We have $p = p'_1 = e_B(x_1, \ab x_2, \ab x_3, \ab x_4)$,
$p_{11} = e_3(x_1, \ab x_2, \ab x_3, \ab x_4)$,
$p_{1111} = e_1(x_1, \ab x_2, \ab x_3)$,
$p_{1112} = e_2(x_1, \ab x_2, \ab x_3, \ab x_4)$.
We show $\checkrc(q, (\envset_{\overline{jk}})_{\overline{jk}})$.
Given the first value of $\solvep{P',\rw}$ shown above, we satisfy V2.1
by letting
$\sigma_{11} = \{ x_1 \mapsto \pkA, \ab x_2 \mapsto \pkB, \ab x_3 \mapsto a[\pkB,i_{A0}], \ab x_4 \mapsto b[p_1,i_{B0}] \}$
and $i_{11} = i_{B0}$, with
$(\rho_{111},  i_{11}) \in \envset_{11}$.
The common variables between $\sigma_{11} q_{11} = 
\pasevent(e_3(\pkA, \ab \pkB, \ab a[\pkB,\ab i_{A0}], \ab b[p_1,i_{B0}])) 
\rightsquigarrow
(\inj$ $\pasevent(e_1(\pkA, \ab \pkB, \ab a[\pkB,i_{A0}])) \wedge
\inj$ $\pasevent(e_2(\pkA, \ab \pkB, \ab a[\pkB,i_{A0}], \ab b[p_1,i_{B0}])))$ 
and $\sigma_{11} p'_1 = e_B(\pkA,\ab \pkB,\ab a[\pkB,i_{A0}],\ab b[p_1,i_{B0}])$
are $i_{A0}$ and $i_{B0}$, and they occur in 
$\sigma_{11} p_{11} = e_3(\pkA,\ab \pkB,\ab a[\pkB,i_{A0}],\ab b[p_1,i_{B0}])$.
So we have V2.2.
Recursively, in order to obtain V2.3, we have to show 
$\checkrc(\sigma_{11} q_{11}, \ab (\envset_{11\overline{jk}})_{\overline{jk}})$.
Given the second value of $\solvep{P',\rw}$ shown above, we satisfy V2.1
by letting $\sigma_{11111} = \Id$
and $i_{11111} = i_{A0}$,
with $(\rho_{111111}, i_{11111}) \in \envset_{1111}$
and $(\rho_{111112}, i_{11111}) \in \envset_{1112}$.
(We prefix the indices with $111$ in order to represent that these values concern the
recursive call with $j=1$, $r=1$, and $k=1$.)
V2.2 holds trivially, because $\sigma_{11111} \sigma_{11} q_{111k_0} = 
\sigma_{11111} \sigma_{11} \pasevent(p_{111k_0})$,
since the considered correspondence has one nesting level only.
V2.3 holds because $q_{1111}$ reduces to $\pasevent(p_{1111})$, so 
$\checkrc(\sigma_{11111} \sigma_{11} q_{1111}, 
(\envset_{1111\overline{jk}})_{\overline{jk}})$
holds by V1, and the situation is similar for $q_{1112}$.
Therefore, we obtain H1.
In order to show H2,
we have to find $x_{11}$ such that $\rho_{111}(x_{11})\{\lambda/i_{11}\}$
does not unify with
$\rho_{111}(x_{11})\{\lambda'/i_{11}\}$ when $\lambda \neq \lambda'$.
This property holds with $x_{11} = m$, because $i_{11} = i_{B0}$ occurs in $\rho_{111}(m) = p_2$.
Similarly, 
$\rho_{111111}(x_{1111})\{\lambda/i_{11111}\}$
does not unify with
$\rho_{111111}(x_{1111})\{\lambda'/i_{11111}\}$ when $\lambda \neq \lambda'$,
for $x_{1111} = i_A$, since $i_{11111} = i_{A0}$ occurs in $\rho_{111111}(i_A)$.
Finally,
$\rho_{111112}(x_{1112})\{\lambda/i_{11111}\}$
does not unify with
$\rho_{111112}(x_{1112})\{\lambda'/i_{11111}\}$ when $\lambda \neq \lambda'$
for $x_{1112} = m'$, since $i_{11111} = i_{A0}$ occurs in $\rho_{111112}(m') = p_1$. 
So, by Theorem~\ref{th:recentcorresp}, the process $P'$ satisfies the
recent correspondence $\pasevent(e_B(x_1, \ab x_2, \ab x_3, \ab x_4)) \rightsquigarrow
(\inj$ $\pasevent(e_3(x_1, \ab x_2, \ab x_3, \ab x_4)) \rightsquigarrow
(\inj$ $\pasevent(e_1(x_1, \ab x_2, \ab x_3)) \wedge
\inj$ $\pasevent(e_2(x_1, \ab x_2, \ab x_3, \ab x_4))))$ 
against $\rw$-adversaries.

We can then show that $P'$ satisfies the recent correspondence
$\pasevent(e_B(x_1, \ab x_2, \ab
x_3, \ab x_4)) \rightsquigarrow (\inj$ $\pasevent(e_3(x_1, \ab x_2, \ab
x_3, \ab x_4)) \rightsquigarrow (\inj$ $\pasevent(e_2(x_1, \ab x_2, \ab
x_3, \ab x_4)) \rightsquigarrow \inj$ $\pasevent(e_1(x_1, \ab x_2, \ab
x_3))))$.  We just have to show that the event $e_2(x_1, \ab x_2, \ab
x_3, \ab x_4)$ is executed after $e_1(x_1, \ab x_2, \ab x_3)$. The
nonce $a$ is created just before executing $e_1(x_1, \ab x_2, \ab x_3)
= e_1(\pkA, x\_\pkB, a)$, and the event $e_2(x_1, \ab x_2, \ab x_3,
\ab x_4) = e_2(x\_\pkA, \pkB, x\_a, b)$ contains $a$ in the variable
$x_3 = x\_a$. So $e_2$ has been executed after receiving a message
that contains $a$, so after $a$ has been sent in some message, so
after executing event $e_1$.
\end{example}

\section{Termination}\label{sect:termination}

In this section, we study termination properties of our algorithm.
We first show that it terminates on a restricted class of 
protocols, named \emph{tagged protocols}. Then, we study how to
improve the choice of the selection function in order to obtain
termination in other cases.

\subsection{Termination for Tagged Protocols}
\label{sect:taggedterm}

Intuitively, a tagged protocol is a protocol in which each application
of a constructor can be immediately distinguished from others in the
protocol, for example by a tag: for instance, when we want to encrypt
$m$ under $k$, we add the constant tag $\ct{0}$ to $m$, so that the encryption
becomes $\sencrypt((\ct{0}, m), k)$ where the tag $\ct{0}$ is a different
constant for each encryption in the protocol. The tags
are checked when destructors are applied. This condition is easy
to realize by adding tags, and it is also a good protocol design: the
participants use the tags to identify the messages unambiguously, thus
avoiding type flaw attacks~\cite{Heather00b}.

In~\cite{Blanchet04e}, in collaboration with Andreas Podelski, 
we have given conditions on the clauses
that intuitively correspond to tagged protocols, and
we have shown that, for tagged protocols using only public
channels, public-key cryptography with atomic keys, shared-key
cryptography and hash functions, and for secrecy properties, 
the solving algorithm using the selection function $\sel_0$ terminates.

Here, we extend this result by giving a definition of tagged protocols
for processes and showing that the clause generation algorithm yields
clauses that satisfy the conditions of~\cite{Blanchet04e},
so that the solving algorithm terminates.
(A similar result has been proved for strong secrecy in the technical
report~\cite{Blanchet04c}.)

\begin{definition}[Tagged protocol]\label{def:tagged}
A tagged protocol is a process $P_0$ together with a signature
of constructors and destructors such that:
\begin{enumerate}

\pname{C1} The only constructors and destructors are those of
Figure~\ref{fig:cryptoop}, plus $\equal$.\label{HYP1}%

\pname{C2}In every occurrence of $\cinput{M}{x}$ and $\coutput{M}{N}$ in $P_0$,
$M$ is a name free in $P_0$.\label{condCpubchannel}

\pname{C3}In every occurrence of $f(\ldots)$ with $f \in \{ \sencrypt,
\ab \sencryptp, \ab \pencrypt, \ab \sign, \ab \nmrsign, \ab \hash, \ab
\mac \}$ in $P_0$, the first argument of $f$ is a tuple $(\cto, M_1,
\ldots, M_n)$, where the tag $\cto$ is a constant. 
Different occurrences of $f$ have different values of the
tag $\cto$.\label{condC3}
%The tags are put

\pname{C4}In every occurrence of $\letfun{x}{g(\ldots)}{P}{Q}$,
for $g \in \{ \sdecrypt, \ab \sdecryptp, \ab \pdecrypt, \ab
\checksign, \ab \getmess \}$  in $P_0$,
$P = \aletfun{y}{\nth{1}{n}(x)}\aguard{y}{\cto}{P'}$ for some $\cto$
and $P'$.

In every occurrence of $\nmrchecksign$ in $P_0$, its third argument is 
$(\cto, \ab M_1, \ab \ldots, \ab M_n)$ for some $\cto, M_1, \ldots, M_n$.\label{condC4}
%The tags are checked

\pname{C5}The destructor applications (including equality tests) 
have no $\kw{else}$ branches.
There exists a trace of $P_0$ (without adversary) in
which all program points are executed exactly once.\label{condC5}

\pname{C6} The second argument of  $\pencrypt$ in the trace of 
Condition~\ref{condC5} is of the form $\pk(M)$ for some $M$.\label{HYP2}

\pname{C7} The arguments of $\pk$ and $\host$ in the trace of
Condition~\ref{condC5} are atomic constants (free names or names
created by restrictions not under inputs, non-deterministic
destructor applications, or replications) and they are not tags.
\label{HYP5}

\end{enumerate}
\end{definition}

Condition~\ref{HYP1} limits the set of allowed constructors and destructors.
We could give conditions on the form of allowed destructor rules,
but these conditions are complex, so it is simpler and more intuitive 
to give an explicit list. Condition~\ref{condCpubchannel} states that
all channels must be public. This condition avoids the need for the
predicate $\mess$. Condition~\ref{condC3} guarantees that tags are added
in all messages, and Condition~\ref{condC4} guarantees that tags are always 
checked. 

In most cases, the trace of Condition~\ref{condC5} is simply the
intended execution of the protocol. All terms that occur in the trace
of Condition~\ref{condC5} have pairwise distinct tags (since
each program point is executed at most once, and tags at different
program points are different by Condition~\ref{condC3}).  We can prove
that it also guarantees that the terms of all clauses generated for
the process $P_0$ have instances in the set of terms that occur in the
trace of Condition~\ref{condC5} (using the fact that all program
points are executed at least once).  These properties are key in the
termination proof. More concretely, Condition~\ref{condC5} means that,
after removing replications of $P_0$, the resulting process has a
trace that executes each program point (at least) once. In
this trace, all destructor applications succeed and the process
reduces to a configuration with an empty set of processes. Since,
after removing replications, the number of traces of a process is
always finite, Condition~\ref{condC5} is decidable.

Condition~\ref{HYP2} means that, in its intended execution, the
protocol uses public-key encryption only with public keys, and
Condition~\ref{HYP5} means that long-term secret (symmetric and
asymmetric) keys are atomic constants.

\begin{example}
A tagged protocol can easily be obtained by tagging the Needham-Schroeder-Lowe
protocol. The tagged protocol consists of the following messages:
\begin{center}
\begin{tabular}{l l l}
Message 1. &$A \rightarrow B :$&$\{ \cto_0, a,\pkA\}_{\pkB}$\\
Message 2. &$B \rightarrow A :$&$\{ \cto_1, a, b, \pkB\}_{\pkA}$\\
Message 3. &$A \rightarrow B :$&$\{ \cto_2, b \}_{\pkB}$\\
\end{tabular}
\end{center}
Each encryption is tagged with a different tag $\cto_0$, $\cto_1$, and
$\cto_2$.
This protocol can be represented in our calculus by the following process 
$P$:
{\allowdisplaybreaks\begin{align*}
&P_A(\skA, \pkA, \pkB) = \Repl{} \cinput{c}{x\_\pkB}.\Res{a}
\asevent(e_1(\pkA, x\_\pkB, a)).\\*
&\qquad \Res{r_1} \coutput{c}{\pencrypt((\cto_0, a, \pkA), x\_\pkB, r_1)}.\\
&\qquad \cinput{c}{m}.
\aletfun{(=\cto_1, =a, x\_b, =x\_\pkB)}{\pdecrypt(m, \skA)}\\
&\qquad \asevent(e_3(\pkA, x\_\pkB, a, x\_b)).\Res{r_3} \coutput{c}{\pencrypt((\cto_2, x\_b), x\_\pkB, r_3)}\\
&\qquad \aguard{x\_\pkB}{\pkB}
\asevent(e_A(\pkA, x\_\pkB, a, x\_b)).\\
&\qquad \coutput{c}{\sencrypt((\cto_3, \sAa), a)}.
\coutput{c}{\sencrypt((\cto_4, \sAb), x\_b)}\\
&P_B(\skB, \pkB, \pkA) = \Repl{} \cinput{c}{m'}.
\aletfun{(=\cto_1, x\_a, x\_\pkA)}{\pdecrypt(m, \skB)}
\\*
&\qquad\Res{b}\asevent(e_2(x\_\pkA, \pkB, x\_a, b)).\\*
&\qquad \Res{r_2} \coutput{c}{\pencrypt((\cto_2, x\_a, b, \pkB), x\_\pkA, r_2)}.\\*
&\qquad \cinput{c}{m''}.
\aletfun{(=\cto_3, =b)}{\pdecrypt(m'', \skB)}\\*
&\qquad \aguard{x\_\pkA}{\pkA}
\asevent(e_B(x\_\pkA, \pkB, x\_a, b)).\\*
&\qquad \coutput{c}{\sencrypt((\cto_5, \sBa), x\_a)}.
\coutput{c}{\sencrypt((\cto_6, \sBb), b)}\\
&P_T = \Repl{} \cinput{c}{x_1}.\cinput{c}{x_2}.\coutput{c}{x_2}.(\cinput{c}{x_3}.\cinput{c}{x_4} \parpop \cinput{c}{x_5}.\cinput{c}{x_6})\\
&P = \Res{\skA}\Res{\skB}\aletdef{\pkA}{\pk(\skA)}\aletdef{\pkB}{\pk(\skB)}\\*
&\qquad\coutput{c}{\pkA}\coutput{c}{\pkB}.(P_A(\skA, \pkA, \pkB) \parpop P_B(\skB, \pkB, \pkA) \parpop P_T)
\end{align*}}%
The encryptions that are used for testing the secrecy of nonces are 
also tagged, with tags $\cto_3$ to $\cto_6$.
Furthermore, a process $P_T$ is added in order to satisfy 
Condition~\ref{condC5}, because, without $P_T$, in the absence of adversary,
the process would block when it tries to send the public keys $\pkA$ and 
$\pkB$. The execution of Condition~\ref{condC5} is the intended
execution of the protocol. In this execution, 
the process $P_T$ receives the public keys
$\pkA$ and $\pkB$; it forwards $\pkB$ on channel $c$ to $P_A$, so that
a session between $A$ and $B$ starts. Then $A$ and $B$ run this session 
normally, and finally output the encryptions of $\sAa$, $\sAb$, $\sBa$, 
and $\sBb$; these encryptions are received by $P_T$.
The other conditions of Definition~\ref{def:tagged} are easy to check, 
so $P$ is tagged.

Proposition~\ref{prop:termination} below applies to $P$, and also to the 
process
without $P_T$, because the addition of $P_T$ in fact does not change
the clauses. (The only clause generated from $P_T$ is a tautology, 
immediately removed by $\elimtaut$.)
\end{example}

We prove the following termination result in Appendix~\ref{app:termination}:

\begin{proposition}\label{prop:termination}
For $\sel = \sel_0$, the algorithm terminates on tagged protocols for
queries of the form $\act \rightsquigarrow \cfalse$ when $\act$ is
closed and all facts in $\fnot$ are closed.
\end{proposition}
The proof first considers the particular case in which $\pk$
and $\host$ have a single argument in the execution of Condition~\ref{condC5},
and then generalizes by mapping all arguments of $\pk$ and $\host$
(which are atomic constants by Condition~\ref{HYP5}) to a single constant.
The proof of the particular case proceeds in two steps.
The first step shows that the clauses generated from a tagged protocol
satisfy the conditions of~\cite{Blanchet04e}. Basically, these conditions 
require that the clauses for the protocol satisfy the following
properties:
\begin{enumerate}

\pname{T1}\label{term1} 
  The patterns in the clauses are \emph{tagged}, that is,
  the first argument of all occurrences of constructors except tuples, $\pk$,
  and $\host$ is of the form $(\cto, \ab M_1, \ab \ldots, \ab M_n)$.  The proof of
  this property relies on Conditions~\ref{condC3} and~\ref{condC4}.

\pname{T2}\label{term2} 
  Let $S_1$ be the set of subterms of patterns that correspond to the
  terms that occur in the execution of Condition~\ref{condC5}.  Every
  clause has an instance in which all patterns are in
  $S_1$.  The proof of this property relies on Condition~\ref{condC5}.

\pname{T3}\label{term3} Each non-variable, non-data tagged pattern has at most one instance in $S_1$.
(A pattern is said to be \emph{non-data} when it is not of the form
$f(\ldots)$ with $f$ a data constructor, that is, here, a tuple.)
This property comes from Condition~\ref{condC3} which guarantees that the tags
at distinct occurrences are distinct and, for $\pk(p)$ and $\host(p)$, from
the hypothesis that $\pk$ and $\host$ have a single argument in the execution of Condition~\ref{condC5}.

\end{enumerate}
  Note that the patterns in the clauses~\eqref{ruleRf}
  and~\eqref{ruleRg} that come from constructors and destructors are
  not tagged, so we need to handle them specially;
  Conditions~\ref{HYP1} and~\ref{HYP2} are useful for that.

The second step of the proof uses the result of~\cite{Blanchet04e} in
order to conclude termination. Basically, this result shows that 
Properties~\ref{term1} and~\ref{term2} are preserved by resolution.
The proof of this result relies on the fact that, if two non-variable 
non-data tagged
patterns unify and have instances in $S_1$, then their instances in
$S_1$ are equal (by~\ref{term3}). So, when unifying two such patterns,
their unification still has an instance in $S_1$.
Furthermore, we show that the size of the instance in $S_1$ of a
clause obtained by resolution is not greater than the size of the
instance in $S_1$ of one of the initial clauses.  Hence, we can bound
the size of the instance in $S_1$ of generated clauses, which shows
that only finitely many clauses are generated.

The hypothesis that all facts in $\fnot$ are closed is not really a 
restriction, since we can always remove facts from $\fnot$ without
changing the result. (It may just slow down the resolution.)
The restriction to queries $\act \rightsquigarrow \cfalse$ allows us to remove
$\pasbegin$ facts from clauses (by Remark~\ref{rem:event}).
For more general queries, $\pasbegin$ facts may occur in clauses,
and one can find examples on which the algorithm does not terminate.
Here is such an example:
{\allowdisplaybreaks\begin{align*}
P_S = {} &\cinput{c'_1}{y}; \aletdef{z}{\sencrypt((\cto_0, y), k_{SB})}\\*
&\coutput{c'_2}{\sencrypt((\cto_2, \sencrypt((\cto_1, z), k_{SA})), k_{SB})}; \asevent(h((\cto_3, y))); \coutput{c'_3}{z}\\
P_B = {} &\cinput{c'_2}{z'}; \cinput{c'_3}{z}; 
\aletdef{(=\cto_0, y)}{\sdecrypt(z, k_{SB})}\\*
&\aletdef{(=\cto_2, y')}{\sdecrypt(z', k_{SB})}
\asevent(h((\cto_4, y, y'))); \coutput{c'_4}{y'}\\
P_0 = {} &\Res{k_{SB}}; (\coutput{c'_1}{C_0} \parpop \Repl{P_S} \parpop \Repl{P_B} \parpop \cinput{c'_4}{y'})
\end{align*}}%
This example has been built on purpose for exhibiting non-termination, 
since we did not meet such non-termination cases in our experiments
with real protocols. 
One can interpret this example as follows. 
The participant $A$ shares a key $k_{SA}$ with a server $S$.
Similarly, $B$ shares a key $k_{SB}$ with $S$.
The code of $S$ is represented
by $P_S$, the code of $B$ by $P_B$, and $A$ is assumed to be dishonest,
so it is represented by the adversary.
The process $P_S$ builds two tickets $\sencrypt((\cto_0, y), k_{SB})$
and $\sencrypt((\cto_2, \ab \sencrypt((\cto_1, \ab 
\sencrypt((\cto_0, \ab y), \ab k_{SB})), \ab 
k_{SA})), \ab k_{SB})$. The first ticket is for $B$, the second 
ticket should 
first be decrypted by $B$, then sent to $A$, which is going to decrypt it
again and sent it back to $B$. In the example, $P_B$ just decrypts the
two tickets and forwards the second one to $A$.
It is easy to check that this process is a tagged protocol.
This process generates the following clauses:
{\allowdisplaybreaks\begin{align}
\begin{split}
&\attacker(y) \rewrite\\*
&\quad \attacker(\sencrypt((\cto_2, \sencrypt((\cto_1, \sencrypt((\cto_0, y), k_{SB})), k_{SA})), k_{SB}))
\end{split}\label{C:PS1}\\
&\attacker(y) \wedge \pasbegin(h((\cto_3, y))) \rewrite \attacker(\sencrypt((\cto_0, y), k_{SB}))\label{C:PS2}\\
\begin{split}
&\attacker(\sencrypt((\cto_0, y), k_{SB})) \wedge \attacker(\sencrypt((\cto_2, y'), k_{SB})) \\*
&\qquad \wedge \pasbegin(h((\cto_4, y, y'))) \rewrite \attacker(y')
\end{split}\label{C:PB}\\
&\attacker(C_0)
\end{align}}%
The first two clauses come from $P_S$, the third one from $P_B$, and the last
one from the output in $P_0$. Obviously, clauses \eqref{ruleInit}
(in particular $\attacker(k_{SA})$ since $k_{SA} \in \fn(P_0)$),
\eqref{ruleRf} for $\sencrypt$ and $h$, and \eqref{ruleRg}
for $\sdecrypt$ are also generated. Assuming the first hypothesis is selected
in~\eqref{C:PB}, the solving algorithm performs a resolution step 
between \eqref{C:PS2} 
and \eqref{C:PB}, which yields:
\[\begin{split}
& \attacker(y) \wedge \attacker(\sencrypt((\cto_2, y'), k_{SB})) \wedge {}\\
&\qquad \pasbegin(h((\cto_3, y)))\wedge \pasbegin(h((\cto_4, y, y'))) \rewrite \attacker(y')
\end{split}\]
The second hypothesis is selected in this clause. 
By resolving with \eqref{C:PS1}, we obtain
\[\begin{split}
&\attacker(y) \wedge \attacker(y') \wedge \pasbegin(h((\cto_3, y))) \wedge {} \\*
&\quad \pasbegin(h((\cto_4, y, \sencrypt((\cto_1, \sencrypt((\cto_0, y'), k_{SB})), k_{SA}))))\\
&\qquad \rewrite \attacker(\sencrypt((\cto_1, \sencrypt((\cto_0, y'), k_{SB})), k_{SA}))
\end{split}\]
By applying \eqref{ruleRg} for $\sdecrypt$ and resolving with
$\attacker(\cto_1)$ and $\attacker(k_{SA})$, we obtain:
\[\begin{split}
&\attacker(y) \wedge \attacker(y') \wedge \pasbegin(h((\cto_3, y))) \wedge {}\\
&\quad \pasbegin(h((\cto_4, y, \sencrypt((\cto_1, \sencrypt((\cto_0, y'), k_{SB})), k_{SA}))))\\
&\qquad \rewrite \attacker(\sencrypt((\cto_0, y'), k_{SB}))
\end{split}\]
This clause is similar to \eqref{C:PS2}, so we can repeat this 
resolution process, resolving with \eqref{C:PB}, \eqref{C:PS1}, 
and decrypting the conclusion. Hence we obtain 
\[\begin{split}
&\bigwedge_{j=1}^n \attacker(y_j)  \wedge \pasbegin(h((\cto_3, y_1))) \wedge {}\\[-3mm]
&\qquad  \bigwedge_{j=1}^{n-1} \pasbegin(h((\cto_4, y_j, \sencrypt((\cto_1, \sencrypt((\cto_0, y_{j+1}), k_{SB})), k_{SA}))))\\[-3mm]
&\qquad \qquad \rewrite \attacker(\sencrypt((\cto_0, y_n), k_{SB}))
\end{split}\]
for all $n>0$, so the algorithm does not terminate. 

As noticed in~\cite{Blanchet04e}, termination could be obtained in the
presence of $\pasbegin$ facts with an additional simplification:
\newcommand{\elimbegin}{\kw{elim\text{-}m\text{-}event}}
\begin{quote}
Elimination of useless $\pasbegin$ facts: $\elimbegin$ eliminates
$\pasbegin$ facts in which a variable $x$ occurs, and $x$ only occurs
in $\pasbegin$ facts and in $\attacker(x)$ hypotheses.
\end{quote}
This simplification is always sound, because it creates a stronger
clause.  It does not lead to a loss of precision when all variables of
events after $\rightsquigarrow$ also occur in the event before
$\rightsquigarrow$. (This happens in particular for non-injective
agreement.)  Indeed, assume that $\pasbegin(p)$ contains a variable
which does not occur in the conclusion. This is preserved by
resolution, so when we obtain a clause $\pasbegin(p') \wedge H
\rewrite \pasend(p'')$, where $\pasbegin(p')$ comes from
$\pasbegin(p)$, $p'$ contains a variable that does not occur in $p''$,
so this occurrence of $\pasbegin(p')$ cannot be used to prove the
desired correspondence.  However, in the general case, this
simplification leads to a loss of precision. (It may miss some
$\pasbegin$ facts.)  That is why this optimization was present in
early implementations which verified only authentication, and was
later abandoned. We could reintroduce it when all variables
of events after $\rightsquigarrow$ also occur in the event before
$\rightsquigarrow$, if we had termination problems coming from $\pasbegin$ 
facts for practical examples. No such problems have occurred 
up to now.

\subsection{Choice of the Selection Function}
\label{sect:termselfun}

Unfortunately, not all protocols are tagged. In particular, protocols
using a Diffie-Hellman key agreement (see Section~\ref{sec:DiffieHellman}) 
are not tagged in the sense
of Definition~\ref{def:tagged}. The algorithm still terminates for some of them
(Skeme~\cite{Krawczyk96} for secrecy, SSH) with the previous selection function $\sel_0$. However, it does not terminate with the selection function $\sel_0$ for some other examples (Skeme~\cite{Krawczyk96} for one authentication property, the Needham-Schroeder shared-key protocol~\cite{Needham78}, some versions of the Woo-Lam shared-key protocol~\cite{Woo92} and \cite[Example 6.2]{Abadi96}.)
In this section, we present heuristics to improve the choice of the 
selection function, in order to avoid most simple non-termination cases. 
As reported in more detail in Section~\ref{sec:results}, these heuristics 
provide termination for Skeme~\cite{Krawczyk96} and the Needham-Schroeder shared-key protocol~\cite{Needham78}.

Let us determine which constraints the selection function should
satisfy to avoid loops in the algorithm.
First, assume that there is a clause $H \wedge F \rewrite \sigma F$, where
$\sigma$ is a substitution such that all $\sigma^n F$ are distinct for
$n \in \mathbb{N}$. 

\begin{itemize}

\item
Assume that $F$ is selected in this clause, and there is a clause $H'
\rewrite F'$, where $F'$ unifies with $F$, and the conclusion is selected in $H'
\rewrite F'$. Let $\sigma'$ be the most general unifier of $F$ and
$F'$. So the algorithm generates:
\begin{align*}
&\sigma' H' \wedge \sigma' H \rewrite \sigma' \sigma F
\quad \ldots \quad
\sigma' H' \wedge \sand_{i=0}^{n-1} \sigma' \sigma^i H
\rewrite \sigma' \sigma^n F
\end{align*}
assuming that the conclusion is selected in all these clauses, and that
no clause is removed because it is subsumed by another clause. So the
algorithm would not terminate. Therefore, in order to avoid this situation,
we should avoid selecting 
$F$ in the clause $H \wedge F \rewrite \sigma F$.

\item 
Assume that the conclusion is selected in the clause $H \wedge F
\rewrite \sigma F$, and there is a clause $H'
\wedge \sigma' F \rewrite C$ (up to renaming of variables), where
$\sigma'$ commutes with $\sigma$ (in particular, when $\sigma$ and
$\sigma'$ have disjoint supports), and that $\sigma' F$ is selected in
this clause. So the algorithm generates:
\begin{align*}
&\sigma' H \wedge \sigma H' \wedge \sigma' F \rewrite\sigma C
\quad \ldots \quad
\sand_{i=0}^{n-1} \sigma' \sigma^i H \wedge \sigma^n
H'\wedge \sigma' F \rewrite\sigma^n C
\end{align*}
assuming that $\sigma' F$ is selected in all these clauses, and that
no clause is removed because it is subsumed by another clause. So
the algorithm would not terminate. Therefore, in order to avoid
this situation, if the conclusion is
selected in the clause $H \wedge F \rewrite \sigma F$, we should avoid
selecting facts of the form $\sigma' F$, where $\sigma'$ and $\sigma$ have
disjoint supports, in other clauses.

\end{itemize}
In particular, since there are clauses of the form $\attacker(x_1)
\wedge \ldots \wedge \attacker(x_n) \rewrite \attacker(f(x_1, \ab 
\ldots, \ab 
x_n))$, by the first remark, the facts $\attacker(x_i)$ should not be
selected in this clause. So the conclusion will be selected in this clause
and, by the second remark, facts of the form $\attacker(x)$ with $x$
variable should not be selected in other clauses. We find again the
constraint used in the definition of $\sel_0$.

We also have the following similar remarks after swapping conclusion and hypothesis. 
Assume that there is a clause $H \wedge \sigma F \rewrite F$, where
$\sigma$ is a substitution such that all $\sigma^n F$ are distinct for
$n \in \mathbb{N}$. 
We should avoid selecting the conclusion in this clause and, if we
select $\sigma F$ in this clause, we should avoid selecting conclusions of
the form $\sigma' F$, where $\sigma'$ and $\sigma$ have disjoint
supports, in other clauses.

We define a
selection function that takes into account all these remarks.
For a clause $H \rewrite C$, we define the weight $w_{\mathrm{hyp}}(F)$
of a fact $F \in H$ by:
\[w_{\mathrm{hyp}}(F) = \begin{cases}
-\infty&\text{if $F$ is an unselectable fact}\\
-2&\text{if $\exists \sigma, \sigma F = C$}\\
-1&\text{otherwise, if $F \in S_{\mathrm{hyp}}$}\\
0&\text{otherwise.}
\end{cases}\]
The set $S_{\mathrm{hyp}}$ is defined as follows: at the beginning,
$S_{\mathrm{hyp}} = \emptyset$; if we generate a
clause $H \wedge F \rewrite \sigma F$ where $\sigma$ is a substitution
that maps variables of $F$ to terms that are not all variables
and, in this clause, we select the
conclusion, then we add to $S_{\mathrm{hyp}}$ all facts $\sigma' F$
with $\sigma$ and $\sigma'$ of disjoint support (and renamings of
these facts). For simplicity, we have replaced the
condition ``all $\sigma^n F$ are distinct for $n \in \mathbb{N}$''
with ``$\sigma$ maps variables of $F$ to terms that are not all variables''.
(The former implies the latter but the converse is wrong.)
Our aim is only to obtain good heuristics, since there exists no
perfect selection function that would provide termination in all
cases. The set $S_{\mathrm{hyp}}$ can easily be represented
finitely: just store the facts $F$ with, for each variable, a flag
indicating whether this variable can be substituted by any term by
$\sigma'$, or only by a variable.

Similarly, we define the weight of the conclusion:
\[w_{\mathrm{concl}} = \begin{cases} 
-2&\text{if $\exists \sigma, \exists F\in H, \sigma C = F$}\\
-1&\text{otherwise, if $C \in S_{\mathrm{concl}}$}\\
0&\text{otherwise.}
\end{cases}\]
The set $S_{\mathrm{concl}}$ is defined as follows: at the beginning,
$S_{\mathrm{concl}} = \emptyset$; if we generate a
clause $H \wedge \sigma F \rewrite F$ where $\sigma$ is a substitution
that maps variables of $F$ to terms that are not all variables
and, in this clause, we select $\sigma F$,
then we add to $S_{\mathrm{concl}}$ all facts $\sigma' F$ with
$\sigma$ and $\sigma'$ of disjoint support (and renamings of these
facts).

Finally, we define
\[\sel_1(H \rewrite C) = \begin{cases}
\emptyset &\text{if $\forall F \in H, w_{\mathrm{hyp}}(F) <
w_{\mathrm{concl}}$,}\\
\{F_0\} &\text{where $F_0 \in H$ of maximum weight, otherwise.}
\end{cases}\]
Therefore, we avoid unifying facts of smallest weight when that is
possible. The selected fact $F_0$ can be any element
of $H$ of maximum weight. In the implementation, the hypotheses are
represented by a list, and the selected fact is the first element of
the list of hypotheses of maximum weight. 

We can also notice that the bigger the fact is, the stronger are
constraints to unify it with another fact. So selecting a
bigger fact should reduce the possible unifications. Therefore, we
consider $\sel_2$, defined as $\sel_1$ except that
$w_{\mathrm{hyp}}(F) = \mathit{size}(F)$ instead of $0$ in the last
case.

When selecting a fact that has a negative weight, we are in one of the
cases when termination will probably not be achieved. We therefore
emit a warning in this case, so that the user can stop the program.

\section{Extensions}\label{sect:extensions}

In this section, we briefly sketch a few extensions to the framework
presented previously. The extensions of Sections~\ref{sec:DiffieHellman},
\ref{sect:elsebranches}, and~\ref{sec:phases} were presented
in~\cite{Blanchet07b} for the proof of process equivalences. We sketch
here how to adapt them to the proof of correspondences.

\subsection{Equational Theories and Diffie-Hellman Key Agreements}\label{sec:DiffieHellman}

Up to now, we have defined cryptographic primitives by associating
rewrite rules to destructors. Another way of defining primitives
is by equational theories, as in the applied pi calculus~\cite{Abadi2001}.
This allows us to model, for instance,  variants of encryption for which the 
failure of decryption cannot be detected or more complex primitives such
as Diffie-Hellman key agreements. 
The Diffie-Hellman key agreement~\cite{Diffie76} enables two
principals to build a shared secret. It is used as an elementary step
in more complex protocols, such as Skeme~\cite{Krawczyk96}, SSH, SSL, 
and IPsec.

As shown in~\cite{Blanchet07b}, our verifier can be extended to handle
some equational theories. Basically, one shows that each trace in a model
with an equational theory corresponds to a trace in a model in which
function symbols are equipped with additional rewrite rules, and conversely.
(We could adapt~\cite[Lemma~1]{Blanchet07b} to show that this result also
applies to correspondences.)
Therefore, we can show that a correspondence proved in the model
with rewrite rules implies the same correspondence in the model
with an equational theory.
Moreover, we have implemented algorithms that compute the rewrite
rules from an equational theory.

In the experiments reported in this paper, we use equational
theories only for the Diffie-Hellman key agreement, which can be modeled by using two
functions $f$ and $f'$ that satisfy the equation
\begin{equation}
f(y, f'(x)) = f(x, f'(y)).\label{eqDH}
\end{equation}
In practice, the functions are $f(x,y) = y^x \mod p$ and $f'(x)
= \gdh^x \mod p$, where $p$ is prime and $\gdh$ is a generator of
${\mathbb Z}_p^*$.
The equation $f(y, f'(x)) = (\gdh^x)^y \mod p = (\gdh^y)^x \mod
p = f(x, f'(y))$ is satisfied. In our verifier, following the ideas
used in the applied pi calculus~\cite{Abadi2001}, we do not consider the
underlying number theory; we work abstractly with the
equation~\eqref{eqDH}. 
The Diffie-Hellman key agreement involves two principals $A$ and
$B$. $A$ chooses a random name $x_0$, and sends $f'(x_0)$ to
$B$. Similarly, $B$ chooses a random name $x_1$, and sends $f'(x_1)$ to 
$A$. Then $A$ computes $f(x_0, f'(x_1))$ and $B$ computes
$f(x_1, f'(x_0))$. Both values are equal by~\eqref{eqDH}, and they are
secret: assuming that the attacker cannot have $x_0$ or $x_1$, it
can compute neither $f(x_0, f'(x_1))$ nor $f(x_1, f'(x_0))$.

In our verifier, the equation~\eqref{eqDH} is translated into the rewrite rules
\begin{align*}
&f(y,f'(x)) \rightarrow f(x,f'(y)) \quad f(x,y) \rightarrow f(x,y).
\end{align*}
Notice that this definition of $f$ is non-deterministic: a term such
as $f(a, f'(b))$ can be reduced to $f(b, f'(a))$ and $f(a, f'(b))$,
so that $f(a, f'(b))$ reduces to its two forms modulo the equational
theory. The fact that these rewrite rules model 
the equation~\eqref{eqDH} correctly follows from~\cite[Section~5]{Blanchet07b}.

When using this model, we have to adapt the verification of
correspondences. Indeed, the conditions on the
clauses must be checked \emph{modulo the equational theory}. 
(Using the rewrite rules, we can implement unification modulo
the equational theory, basically by rewriting the terms by
the rewrite rules before performing syntactic unification.)
For example, in the case of non-injective agreement, even if the
process $P_0$ satisfies non-injective agreement against
$\rw$-adversaries, it may happen that a clause $\pasbegin(e'(p_1,\ab
\ldots, \ab p_n)\{f(p_2,\ab f'(p_1))/z\}) \rewrite \pasend(e(p_1, \ab \ldots,\ab
p_n)\{f(p_1,\ab f'(p_2))/z\})$ is in $\solve{P'_0, \rw}(\pasend(e(x_1, \ab
\ldots, \ab x_n)))$.  The specification is still satisfied in this case,
because $(p_1,\ab \ldots, \ab p_n)\{f(p_1,\ab f'(p_2))/z\} = (p_1, \ab \ldots, \ab
p_n)\{f(p_2,\ab f'(p_1))/z\}$ modulo the equational theory. So we have to
test that, if $H \rewrite \pasend(e(p_1, \ab \ldots, \ab p_n))$ is in
$\solve{P'_0, \rw}(\pasend(e(x_1, \ab \ldots, \ab x_n)))$, then there exist
$p'_1, \ab \ldots, \ab p'_n$ equal to $p_1, \ab \ldots, \ab p_n$ modulo 
the equational theory such that $\pasbegin(e'(p'_1, \ab \ldots, \ab p'_n)) \in H$.
More generally, the equality $R = H \wedge
\pasbegin(\sigma' p_{j1}) \wedge \ldots \wedge \pasbegin(\sigma'
p_{jl_j}) \rewrite \pasend(\sigma' p'_j)$ in the hypothesis of
Theorem~\ref{thbeginend3} is checked modulo the equational theory
(using matching modulo the equational theory to find $\sigma'$).
Point~V2.1 of the definition of $\checkrc$
and Hypothesis~H2 of Theorem~\ref{th:recentcorresp} are also
checked modulo the equational theory. Furthermore, the following condition
is added to Point~V2.2 of the definition of $\checkrc$:
\begin{quote}
For all $j$, $r$, and $k$, we let $\oneq = \sigma_{jr} q_{jk}$ and $\onep
= \sigma_{jr} p_{jk}$, and we require that, 
for all substitutions $\sigma$ and $\sigma'$, if
$\sigma \onep = \sigma' \onep$ and for all $x \in \fv(\oneq) \setminus
\fv(\onep)$, $\sigma x = \sigma' x$, then $\sigma \oneq = \sigma'
\oneq$ (where equalities are considered modulo the equational theory).
\end{quote}
This property is useful in the proof of
Theorem~\ref{th:recentcorresp} (see Appendix~\ref{app:recentness}).  
It always holds when
the equational theory is empty, because $\sigma \onep = \sigma' \onep$ implies
that for all $x \in \fv(\onep)$, $\sigma x = \sigma' x$, so for all $x \in
\fv(\oneq)$, $\sigma x = \sigma' x$. However, it does not hold in
general for any equational theory, so we need to check it explicitly
when the equational theory is non-empty.
In the implementation, this condition is checked as follows.  Let
$\rename$ be a renaming of variables of $\onep$ to fresh variables.  We
check that, for every $\sigma_u$ most general unifier of $\onep$ and
$\rename \onep$ modulo the equational theory, $\sigma_u \oneq = \sigma_u \rename
\oneq$ modulo the equational theory.
When this check succeeds, we can prove the condition above as follows.
Let $\sigma_0$ be defined by, for all $x \in \fv(\oneq)$, 
$\sigma_0 x = \sigma x$ and, for all $x \in \fv(\rename \onep)$,
$\sigma_0 x = \sigma' \rename^{-1} x$.
If $\sigma \onep = \sigma' \onep$, then $\sigma_0 \onep  = \sigma \onep = \sigma' \onep = \sigma_0 \rename \onep$,
so $\sigma_0$ unifies $\onep$ and $\rename \onep$, hence there exist $\sigma_1$
and a most general unifier $\sigma_u$ of $\onep$ and $\rename \onep$ such that
$\sigma_0 = \sigma_1 \sigma_u$.
We have $\sigma_u \oneq = \sigma_u \rename \oneq$, so
$\sigma \oneq = \sigma_0 \oneq = \sigma_1 \sigma_u \oneq = \sigma_1 \sigma_u \rename \oneq
= \sigma_0 \rename \oneq = \sigma' \oneq$.

This treatment of equations has the advantage that resolution
can still use syntactic unification, so it remains efficient.
However, it also has limitations; for example, it cannot
handle associative functions, such as XOR, because it would generate 
an infinite number of rewrite rules for the destructors. 
We refer to~\cite{Comon03b,Chevalier05}
for treatments of XOR
and to~\cite{Meadows02,Chevalier03b,Goubault04c,Millen05} for treatments
of Diffie-Hellman key agreements with more detailed algebraic
relations. The NRL protocol analyzer handles a limited version of
associativity for strings of bounded length~\cite{Escobar07}, which
we could handle.

\subsection{Precise Treatment of $\kw{else}$ Branches}
\label{sect:elsebranches}

\newcommand{\nounif}{\mathrm{nounif}}
\newcommand{\GVar}{\mathit{GVar}}

In the generation of clauses described in Section~\ref{sec:clausegen},
we consider that the $\kw{else}$ branch of destructor applications
may always be executed. Our implementation takes into account
these $\kw{else}$ branches more precisely. In order to do that,
it uses a set of special variables $\GVar$ and a predicate $\nounif$, 
also used in~\cite{Blanchet07b},
such that, for all closed patterns $p$ and $p'$, 
$\nounif(p,p')$ holds if and only if there is no closed substitution
$\sigma$ with domain $\GVar$ such that $\sigma p = \sigma p'$.
The fact $\nounif(p,p')$ means that $p \neq p'$ for all values of the special 
variables in $\GVar$. 

One can then check the failure of an equality 
test $M = M'$ by $\nounif(\rho(M), \ab \rho(M'))$
and the failure of a destructor application $g(M_1, \ab \ldots, \ab M_n)$
by $\bigwedge_{g(p_1, \ldots, p_n) \rightarrow p \in \defg}
\nounif((\rho(M_1), \ab \ldots, \ab \rho(M_n)), 
\GVar(p_1, \ab \ldots, \ab p_n))$, where
$\GVar(p)$ is the pattern $p$ after renaming all its variables 
to elements of $\GVar$ and $\rho$ is the environment that maps variables
to their corresponding patterns. Intuitively, the rewrite rule 
$g(p_1, \ldots, p_n) \rightarrow p$ can be applied if and only
if $(\rho(M_1), \ldots, \rho(M_n))$ is an instance of 
$(p_1, \ldots, p_n)$. So the rewrite rule 
$g(p_1, \ldots, p_n) \rightarrow p$ cannot be applied if and only
if $\nounif((\rho(M_1), \ldots, \rho(M_n)), \ab \GVar(p_1, \ldots, p_n))$.

The predicate $\nounif$ is handled by specific simplification
steps in the solver, described and proved correct in~\cite{Blanchet07b}.

\subsection{Scenarios with Several Stages}\label{sec:phases}

Some protocols can be broken into several parts, or stages, numbered 0,
1, \ldots, such that when the protocol starts, stage~0 is executed; at
some point in time, stage~0 stops and stage~1 starts; later,
stage~1 stops and stage~2 starts, and so on. Therefore, stages
allow us to model a global clock.
Our verifier can be extended to such scenarios with several stages, as
summarized in~\cite{Blanchet07b}. We add a construct $t : P$ to the syntax
of processes, which means that process $P$ runs only in stage $t$, where $t$
is an integer. 

The generation of clauses can easily be extended to processes with stages.
We use predicates $\attacker_t$ and $\mess_t$ for each stage $t$, 
generate the clauses for the attacker for each stage, 
and the clauses for the protocol with predicates $\attacker_t$ and $\mess_t$
for each process that runs in stage $t$. Furthermore, we add clauses
\begin{equation}
\attacker_t(x) \rewrite \attacker_{t+1}(x)\tag{Rt}
\end{equation} 
in order to transmit attacker knowledge from each stage $t$ to the next stage $t+1$.

Scenarios with several stages allow us to model properties related to
the compromise of keys. For example, we can model forward secrecy
properties as follows. Consider a public-key protocol $P$ (without
stage prefix) and the process $P' = 0:P \parpop
1:\coutput{c}{sk_A};\coutput{c}{sk_B}$, which runs $P$ in stage 0 and
later outputs the secret keys of $A$ and $B$ on the public channel $c$ 
in stage 1.  If we prove
that $P'$ preserves the secrecy of the session keys of $P$,
then the attacker cannot obtain these session keys even if it
later compromises the private keys of $A$ and $B$, which is
forward secrecy.

\subsection{Compromise of Session Keys}\label{sec:sessionkeycomp}

We consider the situation in which the attacker compromises some
session keys of the protocol. Our goal is then to show that the other
session keys of the protocol are still safe.  For example, this
property does not hold for the Needham-Schroeder shared-key protocol~\cite{Needham78}:
in this protocol, when an attacker manages to get some session keys,
then it can also get the secrets of other sessions.

If we assume that the compromised sessions are all run before the
standard sessions (to model that the adversary needs time to break the
session keys before being able to use the obtained information against
standard sessions), then this can be modeled as a scenario with two
stages: in stage 0, the process runs a modified version of the
protocol that outputs its session keys; in stage 1, the standard
sessions runs; we prove the security of the sessions of stage 1.

However, we can also consider a stronger model, in which the
compromised sessions may run in parallel with the non-compromised
ones. In this case, we have a single stage.

Let $P_0$ be the process representing the whole protocol.  We consider
that the part of $P_0$ not under replications corresponds to the
creation of long-term secrets, and the part of $P_0$ under at least
one replication corresponds to the sessions.  We say that the names
generated under at least one replication in $P_0$ are \emph{session
names}.  We add one argument $i_c$ to the function symbols $a[\ldots]$ that encode
session names in the instrumented process $P'_0$; this additional argument is named \emph{compromise
identifier} and can take two values, $s_0$ or $s_1$. We consider that,
during the execution of the protocol, each replicated subprocess
$\Repl{Q_X}$ of $P_0$ generates two sets of copies of $Q_X$, one with
compromise identifier $s_0$, one with $s_1$. The attacker compromises
sessions that involve only copies of processes $Q_X$ with the
compromise identifier $s_0$. It does not compromise sessions that
involve at least one copy of some process $Q_X$ with compromise
identifier $s_1$.

The clauses for the process $P_0$ are generated as in 
Section~\ref{sec:clausegen}
(except for the addition of a variable compromise identifier
as argument of session names).
The following clauses are added:
{\allowdisplaybreaks\begin{align*}
&\text{For each constructor $f$, }\comp(x_1) \wedge \ldots \wedge \comp(x_k) \rewrite \comp(f(x_1, \ldots, x_k))\\
&\text{For each $\ResInstr{a}{a[\ldots]}$ under 
$n$ replications and $k$ inputs and non-deterministic}\\*
&\text{destructor applications in $P'_0$, }\\*
&\quad \comp(x_1) \wedge \ldots \wedge \comp(x_k)\rewrite 
\comp(a[x_1, \ldots, x_k])\quad \phantom{, i_1, \ldots, i_n, s_0 a[(}\text{ if $n=0$}\\*
&\quad \comp(x_1) \wedge \ldots \wedge \comp(x_k)\rewrite 
\comp(a[x_1, \ldots, x_k, i_1, \ldots, i_n, s_0])\quad \phantom{a[(}\text{ if $n>0$}\\*
&\quad \comp(x_1) \wedge \ldots \wedge \comp(x_k) \rewrite
\attacker(a[x_1, \ldots, x_k, i_1, \ldots, i_n, s_0])\quad \text{ if $n>0$}
\end{align*}}%
The predicate $\comp$ is such that $\comp(p)$ is true when all session
names in $p$ have compromise identifier $s_0$.  These clauses express
that the attacker has the session names that contain only the
compromise identifier $s_0$.

In order to prove the secrecy of a session name $s$, we query the fact
$\attacker(s[x_1, \ab \ldots, \ab x_k, \ab i_1, \ab \ldots, \ab i_n,
\ab s_1])$. If this fact is underivable, then the protocol does not
have the weakness of the Needham-Schroeder shared-key protocol
mentioned above: the attacker cannot have the secret $s$ of a session
that it has not compromised. In contrast, $\attacker(s[x_1, \ab
\ldots, \ab x_k, \ab i_1, \ab \ldots, \ab i_n, \ab s_0])$ is always
derivable, since the attacker has compromised the sessions with
identifier $s_0$.

We can also prove correspondences in the presence of key compromise.
We want to prove
that the non-compromised sessions are secure, so we prove that, if an
event $\asevent(M)$ has been executed in a copy of some $Q_X$ with
compromise identifier $s_1$, then the required events
$\asevent(M_{\overline{jk}})$ have been executed in any process.
(A copy of $Q_X$ with compromise identifier $s_1$ may interact with a
copy of $Q_Y$ with compromise identifier $s_0$ and, in this case, the
events $\asevent(M_{\overline{jk}})$ may be executed in the copy of $Q_Y$ with
compromise identifier $s_0$.)
We obtain this result by adding the compromise identifier $i_c$ 
as argument of the predicates $\pasbegin$ and $\pasend$ in clauses,
and correspondingly adding $s_1$ as argument of $\asevent(M)$
and $\asevent(M_j)$, and a fresh variable as argument of the other
events $\asevent(M_{\overline{jk}})$ in queries.
We can then prove the correspondence in the same way as
in the absence of key compromise. 
The treatment of correspondences $\attacker(M) \rightsquigarrow
\ldots$ and $\mess(M,M') \rightsquigarrow \ldots$ in which $M$ and
$M'$ do not contain bound names remains unchanged.

\section{Experimental Results}\label{sec:results}

We have implemented our verifier in Ocaml and have performed tests on
various protocols of the literature. The tests reported here concern
secrecy and authentication properties for simple examples of protocols. 
More complex examples have been studied, using
our technique for proving correspondences. We do not detail
them in this paper, because they have been the subject of specific
papers~\cite{Abadi04f,Abadi07,Blanchet08b}.

Our results are
summarized in Figure~\ref{fig:results}, with references to the papers
that describe the protocols and the attacks. In these tests, the
protocols are fully modeled, including interaction with the server for
all versions of the Needham-Schroeder, Woo-Lam shared key,
Denning-Sacco, Otway-Rees, and Yahalom protocols.  
The first column indicates the name of the protocol; we use the following
abbreviations: NS for Needham-Schroeder, PK for public-key, SK for shared-key,
corr.\ for corrected, tag.\ for tagged, unid.\ for unidirectional,
and bid.\ for bidirectional.
We have tested the Needham-Schroeder shared key protocol with the modeling 
of key compromise mentioned in Section~\ref{sec:sessionkeycomp}, in which  
the compromised sessions can be executed
in parallel with the non-compromised ones (version marked ``comp.''
in Figure~\ref{fig:results}).
The second column indicates the number of Horn clauses that represent
the protocol. 
The third column indicates the total number of resolution steps performed
for analyzing the protocol.

The fourth column gives the execution time of our
analyzer, in ms, on a Pentium M 1.8~GHz. Several secrecy and agreement
specifications are checked for each protocol. The time given is the
total time needed to check all specifications.
The following factors influence the speed of the system:
\begin{itemize}

\item We use secrecy assumptions to speed up the search. 
These assumptions say that the secret keys of the
principals, and the random values of the Diffie-Hellman key agreement
in the Skeme protocol, remain secret. 
On average, the verifier is two times slower without secrecy assumptions,
in our tests.

\item We mentioned several selection functions, and the speed
of the system can vary substantially depending on the selection
function. In the tests of Figure~\ref{fig:results}, we used the 
selection function $\sel_2$. With $\sel_1$, the system is two times slower
on average on 
Needham-Schroeder shared-key, Otway-Rees, the variant of~\cite{Paulson98} 
of Otway-Rees, and Skeme but faster on the
bidirectional simplified Yahalom (59~ms instead of 91~ms).  The speed
is almost unchanged for our other tests.  
On average, the verifier is 1.8 times slower with $\sel_1$ than with $\sel_2$,
in our tests.

The selection function $\sel_0$ gives approximately the same speed
as $\sel_1$, except for Skeme, for which the analysis
does not terminate with $\sel_0$. (We comment further on termination 
below.)

\item The tests of Figure~\ref{fig:results} have been performed
without elimination of redundant hypotheses. With elimination
of redundant hypotheses that contain $\pasbegin$ facts, we obtain
approximately the same speed. With elimination of
all redundant hypotheses, the verifier is 1.3 times slower on average
in these tests, because of the time spent testing whether hypotheses
are redundant.

\end{itemize}

\begin{figure}[t]
\hbox{}\hfill\begin{tabular}{@{}l@{\hspace{1mm}}|@{\hspace{1.8mm}}r@{\hspace{1.8mm}}|@{\hspace{1.8mm}}r@{\hspace{1.8mm}}|@{\hspace{0.5mm}}r@{\hspace{1.8mm}}|@{\hspace{1.8mm}}l@{\hspace{1mm}}|@{\hspace{1.8mm}}l@{\hspace{1.8mm}}|@{\hspace{1.8mm}}l@{\hspace{1mm}}|@{\hspace{1.8mm}}l@{}}
\hline
Protocol&\#&\# res.&Time&\multicolumn{4}{c}{Cases with attacks}\\
&cl.&steps&(ms)&Secrecy&$A$&$B$&Ref.\\
\hline
NS PK~\cite{Needham78}&32 &1988&95&Nonces $B$&None&All&\cite{Lowe96}\\%8 separate queries; one test has 31 clauses 
NS PK corr.~\cite{Lowe96}&36 &1481&51&None&None&None&\\%one test has 35 clauses
Woo-Lam PK~\cite{Woo92}&23 &104&7&&&All&\cite{Durante01}\\%4 queries
Woo-Lam PK corr.~\cite{Woo97}&27 &156&6&&&None&\\
Woo-Lam SK~\cite{Gordon2003}&25 &184&8&&&All&\cite{Anderson95}\\
Woo-Lam SK corr.~\cite{Gordon2003}&21 &244&4&&&None&\\
\hline
%These protocols establish a session key. The important point is not 
%full agreement but agreement on the session key !
Denning-Sacco~\cite{Denning81}&30 &440&18&Key $B$& &All&\cite{Abadi96}\\%one test has 29 clauses
Denning-Sacco corr.~\cite{Abadi96}& 30&438&16& None&& Inj&\\%one test has 29 clauses
NS SK~\cite{Needham78}, tag.& 31&2721&41&None&None&None&\\%one test has 30 clauses
NS SK corr.~\cite{Needham87}, tag.& 32&2102&57&None&None&None&\\%one test has 31 clauses
NS SK~\cite{Needham78}, tag., comp. & 50&25241&167&Key $B$&None&Inj&\cite{Denning81}\\%one test has 49 clauses
NS SK corr.~\cite{Needham87}, tag., comp. & 53&23956&225&None&None&None&\\%one test has 52 clauses
Yahalom~\cite{Burrows89}& 26&1515&34&None&Key&None&\\%one test has 25 clauses
Simpler Yahalom~\cite{Burrows89}, unid.&21 &1479&30&None&Key&None&\\%7 queries, one test has 20 clauses
Simpler Yahalom~\cite{Burrows89}, bid.&24 &3685&91&None&All&None&\cite{Syverson94}\\%one test has 23 clauses
Otway-Rees~\cite{Otway87}&34 &1878&59&None&Key&Inj,Key&\cite{Burrows89}\\
%one test has 33 clauses
%9 queries, attack mentioned page 17
Simpler Otway-Rees~\cite{Abadi96}&28 &1934&31&None&All&All&\cite{Paulson98}\\%one test has 27 clauses
Otway-Rees, variant of~\cite{Paulson98}&35 &3349&87&Key $B$& All&All&\cite{Paulson98}\\%one test has 34 clauses
Main mode of Skeme~\cite{Krawczyk96}&39 &4139&154&None&None&None&\\%one test has 37 clauses
\hline
\end{tabular}\hfill\hbox{}
\caption{Experimental results}\label{fig:results}
\end{figure}

When our tool successfully proves that a protocol satisfies
a certain specification, we are sure that this specification
indeed holds, by our soundness theorems.
When our tool does not manage to prove that a protocol satisfies
a certain specification, it finds at least one
clause and a derivation of this clause that contradicts the
specification. The existence of such a clause does not prove that
there is an attack: it may correspond to a false attack, due to the
approximations introduced by the Horn clause model.  However,
using an extension of the technique of~\cite{Allamigeon05} to events,
in most cases, our tool reconstructs a trace of the protocol, and thus
proves that there is actually an attack against the considered
specification. In the tests of Figure~\ref{fig:results}, this
reconstruction succeeds in all cases for secrecy and non-injective
correspondences, in the absence of key compromise. The trace
reconstruction is not implemented yet in the presence of key
compromise (Section~\ref{sec:sessionkeycomp}) or for injective
correspondences. (It presents additional difficulties in the latter case,
since the trace should execute some event twice and others once in
order to contradict injectivity, while the derivation corresponds to
the execution of events once, with badly related session identifiers.)
In the cases in which trace reconstruction is not implemented, we have
checked manually that the protocol is indeed subject to an attack, so
our tool found no false attack in the tests of Figure~\ref{fig:results}:
for all specifications that hold, it has proved them.

The last four columns give the results of the analysis.  The column
``Secrecy'' concerns secrecy properties, the column $A$ concerns
agreement specifications $\pasevent(e(x_1, \ab \ldots, \ab x_n))
\rightsquigarrow \injopt\ \pasevent(e'(x_1, \ab \ldots, \ab x_n))$ in which $A$
executes the event $\asevent(e(M_1, \ab \ldots, \ab M_n))$, the column $B$
agreement specifications $\pasevent(e(x_1, \ab \ldots, \ab x_n))
\rightsquigarrow \injopt\ \pasevent(e'(x_1, \ab \ldots, \ab x_n))$ in which $B$
executes the event $\asevent(e(M_1, \ab \ldots, \ab M_n))$.
The last column gives the reference of the attacks when attacks
are found.  
The first six protocols of Figure~\ref{fig:results}
(Needham-Schroeder public key and
Woo-Lam one-way authentication protocols) are authentication
protocols. For them, we have tested non-injective and recent injective
agreement on the name of the participants, and non-injective and
injective full agreement (agreement on all atomic data). For
the Needham-Schroeder public key protocol, we have also tested the secrecy of
nonces. ``Nonces $B$'' means that the nonces $N_a$ and $N_b$
manipulated by $B$ may not be secret, ``None'' means all tested
specifications are satisfied (there is no attack), ``All'' that our
tool finds an attack against all tested specifications.  The
Woo and Lam protocols are \emph{one-way} authentication protocols: they
are intended to authenticate $A$ to $B$, but not $B$ to $A$, so we
have only tested them with $B$ containing $\asevent(e(M_1, \ldots, M_n))$.

Numerous versions of the Woo and Lam shared-key protocol have been
published in the literature~\cite{Woo92}, \cite{Anderson95},
\cite[end of Example 3.2]{Abadi96}, \cite[Example 6.2]{Abadi96}, \cite{Woo97},
\cite{Gordon2003} (flawed and corrected versions). 
Our tool terminates
and proves the correctness of the corrected versions
of~\cite{Anderson95} and of~\cite{Gordon2003}; it terminates and finds
an attack on the flawed version
of~\cite{Gordon2003}. (The messages received or sent by $A$ do not
depend on the host $A$ wants to talk to, so $A$ may start a session
with the adversary $C$, and the adversary can reuse the messages of
this session to talk to $B$ in $A$'s name.) We can easily see that the
versions of~\cite{Woo92} and \cite[Example 6.2]{Abadi96} are also subject
to this attack, even if our tool does not terminate on them. The only
difference between the protocol of~\cite{Gordon2003} and that
of~\cite{Woo92} is that \cite{Gordon2003} adds tags to distinguish
different encryption sites. As shown in Section~\ref{sect:taggedterm},
adding tags enforces termination. Our tool
finds the attack of \cite[bottom of page 52]{Clark97} on the versions
of \cite[end of Example 3.2]{Abadi96} and \cite{Woo97}.
For example, the version of \cite{Woo97} is 
\begin{center}
\begin{tabular}{l l l}
Message 1.& $A \rightarrow B$:&$A$\\
Message 2.& $B \rightarrow A$:&$N_B$\\
Message 3.& $A \rightarrow B$:&$\{A,B,N_B\}_{K_{AS}}$\\
Message 4.& $B \rightarrow S$:&$\{A,B,\{A,B,N_B\}_{K_{AS}} \}_{K_{BS}}$\\
Message 5.& $S \rightarrow B$:&$\{ A,B,N_B\}_{K_{BS}}$\\
\end{tabular}
\end{center}
and the attack is
\begin{center}
\begin{tabular}{l l l}
Message 1.& $I(A) \rightarrow B$:&$A$\\
Message 2.& $B \rightarrow I(A)$:&$N_B$\\
Message 3.& $I(A) \rightarrow B$:&$N_B$\\
Message 4.& $B \rightarrow I(A)$:&$\{A,B,N_B \}_{K_{BS}}$\\
Message 5.& $I(A) \rightarrow B$:&$\{ A,B,N_B\}_{K_{BS}}$\\
\end{tabular}
\end{center}
In message 3, the adversary sends $N_B$ instead of
$\{A,B,N_B\}_{K_{AS}}$. $B$ cannot see the difference and, acting as
defined in the protocol, $B$ unfortunately sends exactly the message
needed by the adversary as message 5. So $B$ thinks he talks to $A$,
while $A$ and $S$ can perfectly be dead. The attack found against the
version of~\cite[end of Example 3.2]{Abadi96} is very similar. 

The last five protocols exchange a session key, so we have tested
agreement on the names of the participants, and agreement on both the
participants and the session key (instead of full agreement, since
agreement on the session key is more important than agreement on other
values). In Figure~\ref{fig:results}, ``Key $B$'' means that the 
key obtained by $B$ may not be secret, ``Key'' means that agreement on
the session key is wrong, ``Inj'' means that injective agreement is
wrong, ``All'' and ``None'' are as before.

In the Needham-Schroeder shared key protocol~\cite{Needham78}, the last messages are
\begin{center}
\begin{tabular}{l l l}
Message 4.&$B \rightarrow A$:& $\{ N_B \}_K$\\
Message 5.&$A \rightarrow B$:& $\{ N_B - 1 \}_K$
\end{tabular}
\end{center}
where $N_B$ is a nonce. Representing $N_B-1$ with a function
$\mathsf{minusone}(x)= x-1$, with associated destructor
$\mathsf{plusone}$ defined by $\mathsf{plusone}(\mathsf{minusone}(x))
\rightarrow x$, the algorithm does not terminate with the selection
function $\sel_0$.
The selection functions $\sel_1$ or $\sel_2$ given
in Section~\ref{sect:termselfun} however yield
termination. We can also notice that the purpose of the subtraction is
to distinguish the reply of $A$ from $B$'s message. As mentioned
in~\cite{Abadi96}, it would be clearer to have:
\begin{center}
\begin{tabular}{l l l}
Message 4.&$B \rightarrow A$: &$\{ \text{Message }4: N_B \}_K$\\
Message 5.&$A \rightarrow B$: &$\{ \text{Message }5: N_B \}_K$
\end{tabular}
\end{center}
We have used this encoding in the tests shown in Figure~\ref{fig:results}. 
Our tool then terminates with selection functions $\sel_0$, $\sel_1$,
and $\sel_2$. \cite{Blanchet04e} explains in more detail why
these two messages encoded with $\mathsf{minusone}$ 
prevent termination with $\sel_0$, and why the addition of tags
``Message 4'', ``Message 5'' yields termination. 
Adding the tags may strengthen the protocol (for instance, in the
Needham-Schroeder shared key protocol, it prevents replaying Message 5
as a Message 4), so the security of the tagged version does not imply
the security of the original version. As mentioned in~\cite{Abadi96},
using the tagged version is a better design choice because it prevents
confusing different messages, so this version should be implemented.
Our tool also does not terminate on Skeme with selection function $\sel_0$,
for an authentication query, 
but terminates with selection functions $\sel_1$ or $\sel_2$.
All other examples of Figure~\ref{fig:results} terminate
with the three selection functions $\sel_0$, $\sel_1$,
and $\sel_2$.

Among the examples of Figure~\ref{fig:results}, only the Woo-Lam
shared key protocol, flawed and corrected versions of~\cite{Gordon2003}
and the Needham-Schroeder shared key protocol have explicit tags. 
Our tool terminates on all other protocols, even if they are not
tagged. The termination can partly be explained by the notion
of ``implicitly tagged'' protocols~\cite{Blanchet04e}: the various
messages are not distinguished by explicit tags, but by other
properties of their structure, such as the arity of the tuples that
they contain. In Figure~\ref{fig:results}, the Denning-Sacco protocol and 
the Woo-Lam public key protocol are implicitly tagged.
Still, the tool terminates on many examples that
are not even implicitly tagged.

For the Yahalom protocol, we show that, if $B$ thinks that $k$ is a key
to talk with $A$, then $A$ also thinks that $k$ is a key to talk with
$B$. The converse is clearly wrong, because the session key is sent
from $A$ to $B$ in the last message, so the adversary can intercept
this message, so that $A$ has the key but not $B$.

For the Otway-Rees protocol, we do not have agreement on the session
key, since the adversary can intercept messages in such a way that one
participant has the key and the other one has no key. There is also an
attack in which both participants get a key, but not the same
one~\cite{Thayer99}. The latter attack is not found by our tool, since it
stops with the former attacks.

For the simplified version of the Otway-Rees protocol given
in~\cite{Abadi96}, $B$ can execute its event $\asevent(e(M_1, \ldots, M_n))$ with $A$
dead, and $A$ can execute its event $\asevent(e(M_1, \ldots, M_n))$ with $B$ dead. As
Burrows, Abadi, and Needham already noted in~\cite{Burrows89}, even
the original protocol does not guarantee to $B$ that $A$ is alive
(attack against injective agreement that we also
find). \cite{Gordon2003} said that the protocol satisfied its
authentication specifications, because they showed that neither $A$
nor $B$ can conclude that $k$ is a key for talking between $A$ and $B$
without the server first saying so. (Of course, this property is also
important, and could also be checked with our verifier.)

\section{Conclusion}\label{sec:concl}

We have extended previous work on the verification of security
protocols by logic programming techniques, from secrecy to
a very general class of correspondences, including not only
authentication but also, for instance, correspondences that express
that the messages of the protocol have been sent and received
in the expected order. 
This technique enables us to check correspondences in a fully
automatic way, without bounding the number of sessions of the
protocols. This technique also yields an efficient verifier, as the
experimental results demonstrate.

\subsection*{Acknowledgments}

We would like to thank Mart{\'\i}n Abadi, J{\'e}r{\^o}me Feret, C{\'e}dric Fournet, and Andrew Gordon for helpful discussions on this paper.
This work was partly done at Max-Planck-Institut f{\"u}r Informatik,
Saarbr{\"u}cken, Germany.

%%%%%%%%%%%%%%%%%%%%%%%%%%%%%%%%%%%%%%%%%%%%%%%%%%%%%%%%%%%%%%%%%%%
\bibliography{../allbib/biblio}
\bibliographystyle{abbrv} %alpha
%%%%%%%%%%%%%%%%%%%%%%%%%%%%%%%%%%%%%%%%%%%%%%%%%%%%%%%%%%%%%%%%%%%

\appendix

\section*{Appendices}

\section{Instrumented Processes}\label{app:instr}

Let $\last(s)$ be the 
last element of the sequence of session identifiers $s$, or
$\emptyset$ when $s = \emptyset$.
Let $\rlabel(\rlbl)$ be defined by
$\rlabel(a[t,s]) = (a, \last(s))$ 
and $\rlabel(\advnfs[a[s]]) = (a, \last(s))$.
We define the multiset
$\labelset(P)$ as follows: $\labelset(\ResInstr{a}{\rlbl} P) = \{ \rlabel(\rlbl)) \} \cup \labelset(P)$, $\labelset(\ReplInstr{i}{P}) = \emptyset$,
and in all other cases, $\labelset(P)$ is the union of the
$\labelset(P')$ for all immediate subprocesses $P'$ of $P$. Let
$\labelset(\env) = \{ \rlabel(\env(a)) \mid a \in \dom(\env) \}$ and
$\labelset(S) = \{ (a, \lambda) \mid \lambda \in S, \ab a$ any name 
function symbol$\}$.

\begin{definition}
An \emph{instrumented semantic configuration} is a triple
$S, \env, \pset$ such that $S$ is a countable set of constant session 
identifiers, the environment $\env$ is a mapping from names to closed 
patterns, and $\pset$ is a multiset of closed processes.
The instrumented semantic configuration is $S, \env, \pset$ 
\emph{well-labeled} when the
multiset $\labelset(S) \cup \labelset(\env) \cup \bigcup_{P \in \pset}
\labelset(P)$ contains no duplicates.
\end{definition}

\begin{lemma}\label{lem:wl1}
Let $P_0$ be a closed process and $P'_0 = \instr{P_0}$.
Let $Q$ be an $\rw$-adversary and $Q' = \instradv{Q}$.
Let $E_0$ such that $\fn(P'_0) \cup \rw
\subseteq \dom(E_0)$ and, for all $a\in \dom(E_0)$, $E_0(a) = a[\,]$.
The configuration $S_0, E_0, \{ P'_0, Q' \}$ is a well-labeled
instrumented semantic configuration.
\end{lemma}
\begin{proof}
We have
$\labelset(E_0) = \{ (a, \emptyset) \mid a \in \dom(E_0) \}$,
$\labelset(P'_0) = \{ (a, \emptyset) \mid \ResInstr{a}{a[\ldots]}$ 
occurs in $P'_0$ not under a replication$\}$, and
$\labelset(Q') = \{ (a, \emptyset) \mid 
\ResInstr{a}{\advnfs[a[\,]]}$ occurs
in $Q'$ not under a replication$\}$. These multisets
contain no duplicates since the bound names of $P'_0$ and $Q'$
are pairwise distinct and distinct from names in $\dom(E_0)$.
So the multiset $\labelset(S_0) \cup \labelset(E_0)\cup \labelset(P'_0) \cup \labelset(Q')$ contains no duplicates. 
\proofcomplete
\end{proof}

\begin{lemma}\label{lem:wl2}
If $S,\env,\pset$ is a well-labeled instrumented semantic configuration 
and $S,\env,\pset \rightarrow
S',\env',\pset'$ then $S',\env',\pset'$ is a well-labeled
instrumented semantic configuration.
\end{lemma}
\begin{proof}
We proceed by cases on the reduction $S,\env,\pset \rightarrow
S',\env',\pset'$.
The rule (Red Repl) removes the labels $(a, \lambda)$ for a certain
$\lambda$ from $\labelset(S)$
and adds some of them to $\labelset(\pset)$. The rule (Red Res)
removes a label from $\labelset(\pset)$ and adds it to
$\labelset(\env)$. Other rules can remove labels when they remove a
subprocess, but they do not add labels.
\proofcomplete
\end{proof}

\begin{lemma}\label{lem:wlsubst}
Let $S, \env, \pset$ be an instrumented semantic configuration.
Let $\sigma$ be a substitution and $\sigma'$ be defined by 
$\sigma' x = \env(\sigma x)$ for all $x$.
For all terms $M$, $\env(\sigma M) = \sigma' \env(M)$
and, for all atoms $\act$, $\env(\sigma \act) = \sigma' \env(\act)$.
\end{lemma}
\begin{proof}
We prove the result for terms $M$ by induction on $M$.
\begin{itemize}
\item If $M = x$, $\env(\sigma x) = \sigma' x = \sigma' \env(x)$
by definition of $\sigma'$.
\item If $M = a$, $\env(\sigma a) = \env(a) = \sigma' \env(a)$,
since $\env(a)$ is closed.
\item If $M$ is a composite term $M = f(M_1, \ldots, M_n)$, $\env(\sigma M) =
f(\env( \sigma M_1), \ab \ldots, \ab \env(\sigma M_n)) = 
f(\sigma' \env(M_1), \ab \ldots, \ab \sigma' \env(M_n)) =
\sigma' \env(M)$, by induction hypothesis.
\end{itemize}
The extension to atoms is similar to the case of composite terms.
\proofcomplete
\end{proof}

\begin{lemma}\label{lem:wl3}
If $S,\env,\pset$ is a well-labeled instrumented semantic configuration, 
$M$ and $M'$ are closed terms, and
$\env(M) = \env(M')$, then $M = M'$.
\end{lemma}
\begin{proof}
The multiset 
$\labelset(\env)$ does not contain duplicates, hence different names
in $\env$ have different associated patterns, therefore different
terms have different associated patterns.
\proofcomplete
\end{proof}

\begin{lemma}\label{lem:wl4}
If $S,\env,\pset$ is a well-labeled instrumented semantic configuration, 
$M'$ is a closed term, and
$\env(M') = \sigma \env(M)$, then there exists a substitution
$\sigma'$ such that $M'
= \sigma' M$ and, for all variables $x$ of $M$, $E(\sigma' x) = \sigma x$.
We have a similar result for atoms and for tuples containing terms and 
atoms.
\end{lemma}
\begin{proof}
We prove the result for terms by induction on $M$.
\begin{itemize}

\item If $M = x$, $\env(M') = \sigma \env(M) = \sigma x$. We define
$\sigma'$ by $\sigma' x = M'$.

\item If $M$ is a name, $\env(M)$ is closed, so $\env(M') = \sigma
\env(M) = \env(M)$. By Lemma~\ref{lem:wl3}, $M' = M = \sigma' M$
for any substitution $\sigma'$.

\item If $M$ is a composite term $M = f(M_1, \ldots, M_n)$, $\env(M')
= f(\sigma\env(M_1), \ab \ldots, \ab \sigma\env(M_n))$.  Therefore,
$M' = f(M'_1, \ab \ldots, \ab M'_n)$ with
$\env(M'_i) = \sigma \env(M_i)$ for all $i \in \{1, \ldots, n\}$. 
By induction hypothesis, for all $i \in \{1, \ldots, n\}$, there exists
$\sigma'_i$ such that $M'_i = \sigma'_i M_i$ and, for all variables $x$
of $M_i$, $E(\sigma'_i x) = \sigma x$. For all $i,j$, if $x$ occurs in
$M_i$ and $M_j$, $E(\sigma'_i x) = \sigma x = E(\sigma'_j x)$, so by
Lemma~\ref{lem:wl3}, $\sigma'_i x = \sigma'_j x$. Thus we can merge
all substitutions $\sigma'_i$ into a substitution $\sigma'$ defined by 
$\sigma' x = \sigma'_i x$ when
$x$ occurs in $M_i$. So we have $M' = \sigma' M$ and, for all
variables $x$ of $M$, $E(\sigma' x) = \sigma x$.

\end{itemize}
The extension to atoms and to tuples of terms and atoms is similar
to the case of composite terms.
\proofcomplete
\end{proof}

\begin{proof}[of Lemma~\ref{lem:instrcorresp}]
  Let $Q$ be an $\rw$-adversary and $Q' = \instradv{Q}$.  
  Let $\env_0$ containing $\fn(P_0) \cup
  \rw \cup \fn(\act) \cup \bigcup_j \fn(\act_j) \cup \bigcup_{j,k}
  \fn(M_{jk})$.  Consider a trace $\trace = \env_0, \{ P_0, Q \}
  \rightarrow \env_1, \pset_1$.  Let $\sigma$ such that $\trace$
  satisfies $\sigma \act$.  By
  Proposition~\ref{prop:equivsem}, letting $\env'_0 = \{ a \mapsto
  a[\,] \mid a \in \env_0 \}$, there is a trace $\trace' = S_0,
  \env'_0, \{ P'_0, Q'\} \rightarrow^* S', \env'_1, \pset'_1$,
  $\delete(\pset'_1) = \pset_1$, and both traces satisfy the same
  atoms, so $\trace'$ also satisfies $\sigma \act$.  Since $\env'_0$
  contains the names of $\act$, $\act_j$, and $M_{jk}$, and $\env'_1$
  is an extension of $\env'_0$, $\env'_1(\act) = \env'_0(\act) =
  \acti$, $\env'_1(\act_j) = \env'_0(\act_j) = \acti_j$, and
  $\env'_1(M_{jk}) = \env'_0(M_{jk}) = p_{jk}$.  Let $\sigma''$ be
  defined by $\sigma'' x = \env_1(\sigma x)$ for all $x$.  By
  Lemma~\ref{lem:wlsubst}, $\env'_1(\sigma \act) = \sigma''
  \env'_1(\act)$, so $\env'_1(\sigma \act) = \sigma'' \acti$.  Hence
  $\trace'$ satisfies $\sigma'' \acti$.  Since $P'_0$ satisfies the
  given correspondence, there exist $\sigma''_0$ and $j \in \{ 1,
  \ldots, m\}$ such that $\sigma''_0 \acti_j = \sigma'' \acti$ and for all $k
  \in \{ 1, \ldots, l_j \}$, $\trace'$ satisfies $\pasevent(\sigma''_0
  p_{jk})$, so there exists $M''_k$ such that $\env'_1(M''_k) =
  \sigma''_0 p_{jk}$ and $\trace'$ satisfies $\pasevent(M''_k)$. Hence
  $\env'_1(M''_k) = \sigma''_0 \env'_1(M_{jk})$ and $\env'_1(\sigma
  \act) = \sigma'' \acti = \sigma''_0 \acti_j = \sigma''_0
  \env'_1(\act_j)$, that is, $\env'_1((M''_1, \ldots, M''_{l_j},
  \sigma \act)) = \sigma''_0 \env'_1(M_{j1}, \ldots, M_{jl_j},
  \act_j)$. By Lemma~\ref{lem:wl4}, there exists $\sigma_0$ such that
  $(M''_1, \ldots, M''_{l_j}, \sigma \act) = \sigma_0 (M_{j1}, \ldots,
  M_{jl_j}, \act_j)$. So $\sigma \act= \sigma_0 \act_j$ and for all $k \in
  \{ 1, \ldots, l_j \}$, $\trace'$ satisfies $\pasevent(\sigma_0
  M_{jk})$, so $\trace$ also satisfies $\pasevent(\sigma_0 M_{jk})$.
  \proofcomplete
\end{proof}

\section{Proof of Theorem~\ref{th:main}}\label{app:main}

\begin{figure}[t]
\begin{gather}
\frac{\mess(E(M), E(N)) \in \der
\qquad \judg[E]{P}}{\judg[E]{\coutput{M}{N}.P}}\comment{Output}\\[2mm]
\frac{\forall T' \text{ such that }\mess(E(M), T') \in \der, \judg[{E[x \mapsto T']}]{P}}{
\judg[E]{\cinput{M}{x}.P}}\comment{Input}\\[2mm]
\frac{}{\judg[E]{0}}\comment{Nil}\\[2mm]
\frac{\judg[E]{P} \qquad \judg[E]{Q}}{\judg[E]{P \parpop Q}}
\comment{Parallel}\\[2mm]
\frac{\forall \lambda, \judg[{E[i \mapsto \lambda]}]{P}}{\judg[E]{\ReplInstr{i}{P}}}\comment{Replication}\\[2mm]
\frac{\judg[{E[ a\mapsto E(\rlbl)]}]{P}}{\judg[E]{\ResInstr{a}{\rlbl} P}}\comment{Restriction}\\[2mm]
\frac{
\forall T\text{ such that }g(E(M_1), \ldots, E(M_n)) \rightarrow T, \judg[{E[x \mapsto T]}]{P}\qquad \judg[E]{Q}%\\
}{
\judg[E]{\letfun{x}{g(M_1, \ldots, M_n)}{P}{Q}}}\comment{Destructor application}\\[2mm]
\frac{
\pasend(E(M)) \in \der \qquad
\text{if }\pasbegin(E(M)) \in \der\text{ then }\judg[E]{P}}{
\judg[E]{\asevent(M).P}}\comment{Event}
\end{gather}
\caption{Type rules}\label{fig:typing}
\end{figure}

The correctness proof uses a type system as a convenient way of
expressing invariants of processes. This type system can be seen as a
modified version of the type system of~\cite[Section 7]{Abadi04c}, which was
used to prove the correctness of our protocol verifier for
secrecy properties.  In this type system, the types are closed
patterns:
\begin{defn}
\categ{T}{types}\\
\entry{a[T_1, \ldots, T_n, \lambda_1, \ldots, \lambda_k]}{name}\\
\entry{f(T_1, \ldots, T_n)}{constructor application}
\end{defn}
The symbols $\lambda_1, \ldots, \lambda_k$ are constant session identifiers, in a set $S_0$. 
Let $\der$ be the set of closed facts 
derivable from $\rset{P'_0, \rw}\cup \beginset$.

The type rules are defined in Figure~\ref{fig:typing}. The environment
$E$ is a function from names and variables in $V_o$ to types and from
variables in $V_s$ to constant session identifiers.  
The mapping $E$ is extended to all terms 
as a substitution by $E(f(M_1, \ab \ldots, M_n)) = f(E(M_1), \ab
\ldots, \ab E(M_n))$ and to restriction labels
by $E(a[M_1, \ab \ldots,\ab M_n, \ab i_1, \ab \ldots, \ab i_{n'}]) = 
a[E(M_1), \ab \ldots, \ab E(M_n), \ab E(i_1), \ab \ldots, \ab E(i_{n'})]$
and $E(\advnfs[a[i_1, \ab \ldots, \ab i_{n'}]]) = \advnfs[a[E(i_1), \ab 
\ldots, \ab E(i_{n'})]]$, 
so that it maps closed terms and restriction labels to types.
The rules define the judgment $\judg[E]{P}$, which means that
the process $P$ is well-typed in the environment~$E$. We do not consider
the case of conditionals here, since it is a particular case
of destructor applications.

We say that an instrumented semantic configuration $S, \env, \pset$ is
well-typed, and we write $\vdash S, \env, \pset$, when it is
well-labeled and $\judg[\env]{P}$ for all $P \in \pset$.

\begin{proofsk}[of Theorem~\ref{th:main}]
Let $P_0$ be the considered process and $P'_0 = \instr{P_0}$.  
Let $Q$ be an $\rw$-adversary and $Q' = \instradv{Q}$.  
Let $E_0$ such that $\fn(P'_0) \cup \rw
\subseteq \dom(E_0)$ and for all $a\in \dom(E_0)$, $E_0(a) = a[\,]$.

\begin{enumerate}

\item {\it Typability of the adversary:} Let $P'$ be a subprocess of
$Q'$. Let $E$ be an environment such that $\forall a \in \fn(P')$,
$\attacker(E(a)) \in \der$ and $\forall x \in \fv(P')$, $\attacker(E(x))
\in \der$. (In particular, $E$ is defined for all free names and free
variables of $P'$.) We show that $\judg[E]{P'}$, by induction on $P'$.
This result is similar to~\cite[Lemma~5.1.4]{Abadi04c}.  In
particular, we obtain $\judg[E_0]{Q'}$.

\item {\it Typability of $P'_0$:} We prove by induction on the process
$P$, subprocess of $P'_0$, that, if (a) $\rho$ binds all free names
and variables of $P$, (b) $\rset{P'_0, \rw} \supseteq \lp P\rp \rho H$,
(c) $\sigma$ is a closed substitution, and (d) $\sigma H$ can be
derived from $\rset{P'_0, \rw} \cup \beginset$, then $\judg[\envsigmarho]{P}$.
This result is similar to~\cite[Lemma~7.2.2]{Abadi04c}.

In particular, $\rset{P'_0, \rw} \supseteq \lp P_0'\rp \rho 
\emptyseq$, where $\rho = \{ a \mapsto a[\,] \mid a \in
\fn(P'_0) \}$. So, with $E = \envsigmarho = \{ a \mapsto a[\,] \mid a
\in \fn(P'_0) \}$, $\judg[E]{P'_0}$. A fortiori, $\judg[E_0]{P'_0}$.

\item {\it Properties of $P'_0, Q'$:}  By Lemma~\ref{lem:wl1},
$S_0, E_0, \{ P'_0, Q' \}$ is well-labeled. So, using the first two points,
$\vdash S_0, E_0, \{ P'_0, Q' \}$.

\item {\it Substitution lemma:} Let $E' = E[x \mapsto E(M)]$. We show
by induction on $M'$ that $E(M'\{M/x\}) = E'(M')$. We show by
induction on $P$ that, if $\judg[E']{P}$, then $\judg[E]{P\{M/x\}}$.
This result is similar to~\cite[Lemma~5.1.1]{Abadi04c}.

\item {\it Subject reduction:} 
Assume that $\vdash S, \env, \pset$ and $S, \env, \pset \rightarrow S', \env', \pset'$. Furthermore, assume that,
if the reduction $S, \env, \pset \rightarrow S', \env', \pset'$ executes $\asevent(M)$, then $\pasbegin(\env(M)) \in \beginset$.
Then $\vdash S', \env', \pset'$.  This is
proved by cases on the derivation of $S, \env, P \rightarrow S', \env', P'$.
This result is similar to~\cite[Lemma~5.1.3]{Abadi04c}.

\item Consider the trace $\trace = S_0, \env_0, \{ P'_0, Q'\}
\rightarrow^* S', \env', \pset'$.  By the hypothesis of the theorem,
if $\asevent(M)$ has been executed in $\trace$, 
then $\trace$ satisfies $\pasevent(\env'(M))$, so
$\pasbegin(\env'(M)) \in \beginset$. If the reduction that executes
$\asevent(M)$ is $S, \ab \env, \ab \pset \rightarrow S, \ab \env, \ab \pset''$, we have
$\env(M) = \env'(M)$, since $\env'$ is an extension of $\env$, and
$\env$ already contains the names of $M$. Hence we obtain the
hypothesis of subject reduction.  So, by Items~3 and~5, we infer
that all configurations in the trace are well-typed.

When $\acti = \pasevent(p)$, since $\trace$ satisfies
$\pasevent(p)$, there exists $M$ such that $\trace$ satisfies $\pasevent(M)$
and $\env'(M) = p$.
So $\trace$ contains a reduction $S_1, E_1,
\pset_1 \cup \{ \asevent(M).P \} \rightarrow S_1, E_1, \pset_1 \cup \{ P
\}$.  Therefore $\judg[E_1]{\asevent(M).P}$, so $\pasend(E_1(M)) \in
\der$. Moreover, $E_1(M) = \env'(M)$ since $\env'$ is an extension
of $E_1$, therefore $\pasend(\env'(M)) = \pasend(p) = \acti$ is derivable from 
$\rset{P'_0, \rw} \cup \beginset$.

When $\acti = \mess(p,p')$, since $\trace$ satisfies $\mess(p,p')$, 
there exist $M$ and $M'$ such that $\trace$ satisfies $\mess(M,M')$,
$\env'(M) = p$, and $\env'(M') = p'$.
So $\trace$ contains a reduction
$S_1, E_1, \pset_1 \cup \{ \coutput{M}{M'}.P, \cinput{M}{x}.Q \}
\rightarrow S_1, E_1, \pset_1 \cup \{ P, Q\{M/x\} \}$. Therefore
$\judg[E_1]{\coutput{M}{M'}.P}$. This judgment must have been derived by
(Output), so $\mess(E_1(M), E_1(M')) \in \der$. Moreover, $E_1(M) =
\env'(M)$ and $E_1(M') = \env'(M')$ since $\env'$ is an extension of
$E_1$, so $\mess(\env'(M), \env'(M')) = \mess(p,p') = \acti$ is derivable from
$\rset{P'_0, \rw} \cup \beginset$.

When $\acti = \attacker(p')$, $\trace$ also satisfies
$\mess(c[\,], p')$ for some $ c\in \rw$. Therefore, by the previous case,
$\mess(c[\,], p')$ is derivable from $\rset{P'_0, \rw} \cup
\beginset$. Since $c \in \rw$, $\attacker(c[\,])$ is in $\rset{P'_0,
\rw}$. So, by Clause~\eqref{ruleRl}, $\attacker(p') =
\acti$ is derivable from $\rset{P'_0, \rw} \cup \beginset$.
\proofcomplete

\end{enumerate}
\end{proofsk}

\section{Correctness of the Solving Algorithm}\label{app:corr}

In terms of security, the soundness of our analysis means that, if a
protocol is found secure by the analysis, then it is actually
secure. Showing soundness in this sense essentially amounts to showing
that no derivable fact is missed by the resolution algorithm, which,
in terms of logic programming, is the completeness of the resolution
algorithm.
Accordingly, in terms of security, the completeness of our analysis
would mean that all secure protocols can be proved secure by our
analysis.  Completeness in terms of security corresponds, in terms of
logic programming, to the correctness of the resolution algorithm,
which means that the resolution algorithm does not derive false facts.

The completeness of ``binary resolution with free selection'',
which is our basic algorithm, was proved 
in~\cite{Nivelle95,Lynch97,Bachmair01}.
We extend these proofs by showing that completeness still holds
with our simplifications of clauses. (These simplifications are often
specific to security protocols.)

As a preliminary, 
we define a sort system, with three sorts: session identifiers,
ordinary patterns, and environments. 
Name function symbols expect session identifiers as
their last $k$ arguments where $k$ is the number of replications above
the restriction that defines the considered name function symbol, and
ordinary patterns as other arguments. The pattern $a[p_1, \ldots, p_n, i_1,
\ldots, i_k]$ is an ordinary pattern. Constructors $f$ expect ordinary
patterns as arguments and $f(p_1, \ldots, p_n)$ is an ordinary
pattern. The predicates $\attacker$ and $\mess$
expect ordinary patterns as arguments.
The predicate $\pasend$ expects an ordinary pattern and, for injective
events, a session identifier.
The predicate $\pasbegin$ expects an ordinary pattern and, for injective
events, an environment.
We say that a pattern, fact, clause, set of clauses is \emph{well-sorted}
when these constraints are satisfied.

\begin{lemma}\label{lem:varocc}
All clauses manipulated by the algorithm are well-sorted, and if a variable
occurs in the conclusion of a clause and is not a session identifier,
then it also occurs in non-$\pasbegin$ facts 
in its hypothesis.
\end{lemma}
\begin{proof}
It is easy to check that all patterns and facts are well-sorted
in the clause generation algorithm.
One only unifies patterns of the same sort. The environment $\rho$
and the substitutions always map a variable to a pattern of the same sort.
During the building of clauses, the variables in the image of $\rho$
that are not session identifiers also occur 
in non-$\pasbegin$ facts in $H$, and the variables
in the conclusion of generated clauses are in the image of $\rho$.
Hence, the clauses in $\rset{P'_0,\rw}$ satisfy Lemma~\ref{lem:varocc}.

Furthermore, this property is preserved by resolution.
Resolution generates a clause $R'' = \sigma_u H \wedge \sigma_u H'
\rewrite \sigma_u C'$ from clauses $R = H \rewrite C$ and
$R' = H' \wedge F_0 \rewrite C'$ that satisfy Lemma~\ref{lem:varocc}, 
where $\sigma_u$ is the most general
unifier of $C$ and $F_0$.  The substitution $\sigma_u$ unifies
elements of the same sort, so $\sigma_u$ maps each variable to an
element of the same sort, so $R''$ is well-sorted.
If a non-session identifier variable $x$ occurs in $\sigma_u C'$,
then there is a non-session identifier variable $y$ in $C'$
such that $x$ occurs in $\sigma_u y$. Then $y$ occurs in non-$\pasbegin$ facts 
in the hypothesis
of $R'$, $H' \wedge F_0$. First case: $y$ occurs in non-$\pasbegin$ facts 
in $H'$,
so $x$ occurs in $\sigma_u H'$, so $x$ occurs in non-$\pasbegin$ facts 
in the hypothesis of $R''$.
Second case: $y$ occurs in $F_0$, so $x$ occurs in $\sigma_u F_0 = \sigma_u C$,
so there is a non-session identifier variable $z$ such that $z$ occurs in $C$
and $x$ occurs in $\sigma_u z$, so $z$ occurs in non-$\pasbegin$ facts 
in $H$, so $x$ occurs in non-$\pasbegin$ facts 
in $\sigma_u H$, so $x$ occurs in non-$\pasbegin$ facts 
in the hypothesis of $R''$.
In both cases, $x$ occurs in non-$\pasbegin$ facts 
in the hypothesis of $R''$.
Therefore, $R''$ satisfies Lemma~\ref{lem:varocc}.

This property is also preserved by the simplification functions. 
\proofcomplete
\end{proof}

\begin{definition}[Derivation]
Let $F$ be a closed fact. Let $\sattmp$ be
a set of clauses. 
A derivation of $F$ from $\sattmp$ is
a finite tree defined as follows:
\begin{enumerate}
\item Its nodes (except the root) are labeled by clauses $R \in \sattmp$.

\item Its edges are labeled by closed facts.
  (Edges go from a node to each of its sons.)

\item If the tree contains a node labeled by $R$ with one incoming edge
labeled by $F_0$ and $n$ outgoing edges labeled by $F_1, \ldots, F_n$,
then $R \impl \{ F_1, \ldots, F_n \}\rewrite F_0$. 

\item The root has one outgoing edge, labeled by $F$.
The unique son of the root is named the \emph{subroot}.

\end{enumerate}
\end{definition}

In a derivation, if there is a node labeled by $R$ with one incoming edge
labeled by $F_0$ and $n$ outgoing edges labeled by $F_1, \ldots, F_n$,
then the clause $R$ can be used to infer $F_0$ from $F_1,
\ldots, F_n$. Therefore, there exists a derivation of $F$ from $\sattmp$ if
and only if $F$ can be inferred from clauses in $\sattmp$ (in classical logic).

The key idea of the proof of Lemma~\ref{lem:phase1corr} is the
following. Assume that $F$ is derivable from $\satstset \cup \beginset$ and
consider a derivation of $F$ from $\satstset \cup \beginset$. Assume that the
clauses $R$ and $R'$ are applied one after the other in the derivation
of $F$. Also assume that these clauses have been combined by $R
\circ_{F_0} R'$, yielding clause $R''$. In this case, we replace $R$ and 
$R'$ with
$R''$ in the derivation of $F$. When no more replacement can be done,
we show that all remaining clauses have no selected hypothesis. So all
these clauses are in $\satend = \saturate(\satstset)$, and we have built
a derivation of $F$ from $\satend$.
 
To show that this replacement process terminates, we remark that
the total number of nodes of the derivation strictly decreases.

Next, we introduce the notion of data-decomposed derivation. This notion is
useful for proving the correctness of the decomposition of data constructors.
(In the absence of data constructors, all derivations are data-decomposed.)

\begin{definition}
A derivation $D$ is \emph{data-decomposed} if and only if,
for all edges $\node' \rightarrow \node$ in $D$ labeled by
$\attacker(f(p_1, \ldots, p_n))$ for some data constructor $f$,
the node $\node'$ is labeled by a clause $\attacker(f(x_1, \ldots,
x_n)) \rewrite \attacker(x_i)$ for some $i$
or the node $\node$ is labeled by the clause 
$\attacker(x_1) \wedge \ldots \wedge \attacker(x_n) \rewrite
\attacker(f(x_1, \ab \ldots, \ab x_n))$.
\end{definition}

Intuitively, a derivation is data-decomposed when all intermediate
facts proved in that derivation are decomposed as much as possible
using data-destructor clauses $\attacker(f(x_1, \ldots,
x_n)) \rewrite \attacker(x_i)$ before being used to prove other facts.
We are going to transform the initial derivation
into a data-decomposed derivation. Further transformations of the derivation
will keep it data-decomposed.

The next lemma shows that two nodes in a derivation 
can be replaced by one when combining their clauses by resolution.

\begin{lemma}\label{lem:compos}
Consider a data-decomposed 
derivation containing a node $\node'$, labeled $R'$. Let $F_0$
be a hypothesis of $R'$. Then there exists a son $\node$ of $\node'$,
labeled $R$, such that the edge $\node' \rightarrow \node$ is labeled by an
instance of $F_0$, $R \circ_{F_0} R'$ is defined, and, 
if $\sel(R) = \emptyset$ and $F_0 \in \sel(R')$, one
obtains a data-decomposed derivation of the same fact 
by replacing the nodes $\node$ and $\node'$ with a node
$\node''$ labeled $R'' = R \circ_{F_0} R'$.
\end{lemma}%
\begin{proof}
This proof is illustrated in Figure~\ref{fig:lemcompos}.
Let $R' = H' \rewrite C'$, $H'_1$ be the multiset of the labels of the
outgoing edges of $\node'$, and $C'_1$ the label of its incoming edge. 
We have $R'
\impl (H'_1 \rewrite C'_1)$, so there exists $\sigma$ such that
$\sigma H' \subseteq H'_1$ and $\sigma C' = C'_1$. Hence there is an
outgoing edge of $\node'$ labeled $\sigma F_0$, since $\sigma F_0 \in
H'_1$. Let $\node$ be the node at the end of this edge, let $R = H
\rewrite C$ be the label of $\node$. We rename the variables of $R$
such that they are distinct from the variables of $R'$. Let $H_1$ be
the multiset of the labels of the outgoing edges of $\node$. So $R
\impl (H_1 \rewrite \sigma F_0)$. By the above choice of distinct
variables, we can then extend $\sigma$ such that $\sigma H \subseteq
H_1$ and $\sigma C = \sigma F_0$.

The edge $\node' \rightarrow \node$ is labeled $\sigma F_0$, instance of
$F_0$.  Since $\sigma C = \sigma F_0$, the facts $C$ and $F_0$ are unifiable,
so $R \circ_{F_0} R'$ is defined. Let $\sigma'$ be the most general
unifier of $C$ and $F_0$, and $\sigma''$ such that $\sigma = \sigma''
\sigma'$.  We have $R \circ_{F_0} R' = \sigma' (H \cup (H' \setminus \{ F_0\}))
\rewrite \sigma' C'$. Moreover, $\sigma'' \sigma' (H \cup (H' \setminus \{ F_0\}))
\subseteq H_1 \cup (H'_1 \setminus \{ \sigma F_0\})$ and $\sigma'' \sigma' C' =
\sigma C' = C'_1$. Hence $R'' = R \circ_{F_0} R' \impl (H_1 \cup (H'_1 \setminus
\{ \sigma F_0\})) \rewrite C'_1$. The multiset of labels
of outgoing edges of $\node''$ is precisely $H_1 \cup (H'_1 \setminus \{\sigma
F_0\})$ and the label of its incoming edge is $C'_1$, therefore we have
obtained a correct derivation by replacing $\node$ and $\node'$ with $\node''$.

\begin{figure}
\hspace*{2mm}\input{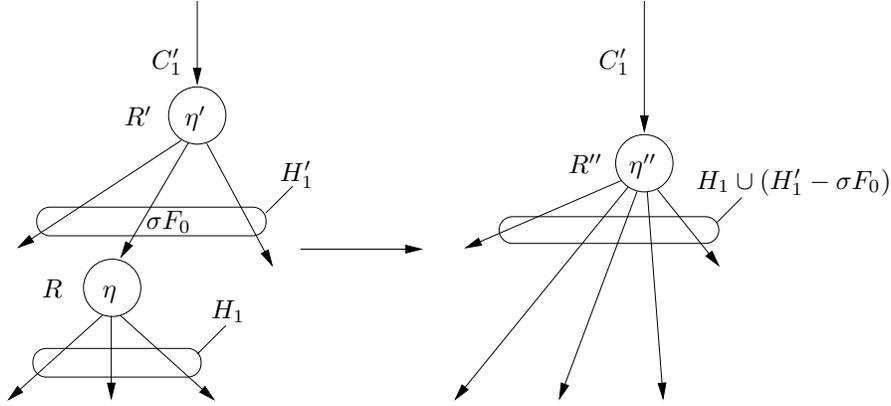}
\caption{Merging of nodes of Lemma~\ref{lem:compos}}\label{fig:lemcompos}
\end{figure}%

Let us show that the obtained derivation is data-decomposed.
Consider an edge $\node'_1 \rightarrow \node_1$ in this derivation,
labeled by $F = \attacker(f(p_1, \ldots, p_n))$, where $f$
is a data constructor.
\begin{itemize}
\item
If $\node'_1$ and $\node_1$ are different from $\node''$,
then the same edge exists in the initial derivation, so it is of the 
desired form.

\item
If $\node'_1 = \node''$, then there is an edge $\node \rightarrow \node_1$
labeled by $F$ in the initial derivation. 
Since the initial derivation is data-decomposed,
$\node$ is labeled by $R = \attacker(f(x_1, \ldots,
x_n)) \rewrite \attacker(x_i)$ or $\node_1$ is labeled by 
$R_1 = \attacker(x_1) \wedge \ldots \wedge \attacker(x_n) \rewrite
\attacker(f(x_1, \ab \ldots, \ab x_n))$.
The former case is impossible because $\sel(R) = \emptyset$.
In the latter case, $\node_1$ is labeled by $R_1$, so we have the 
desired form in the obtained derivation.

\item
If $\node_1 = \node''$, then there is an edge $\node'_1 \rightarrow \node'$
labeled by $F$ in the initial derivation. 
Since the initial derivation is data-decomposed,
$\node'_1$ is labeled by $R'_1 = \attacker(f(x_1, \ldots,
x_n)) \rewrite \attacker(x_i)$ or $\node'$ is labeled by 
$R' = \attacker(x_1) \wedge \ldots \wedge \attacker(x_n) \rewrite
\attacker(f(x_1, \ab \ldots, \ab x_n))$.
The latter case is impossible because $\sel(R) \neq \emptyset$.
In the former case, $\node'_1$
is labeled by $R'_1$, so we have the desired form in the obtained derivation.

\end{itemize}
Hence the obtained derivation is data-decomposed.
\proofcomplete
\end{proof}

\begin{lemma}\label{lem:subsume}
If a node $\node$ of a data-decomposed 
derivation $D$ is labeled by $R$, then one obtains
a data-decomposed derivation $D'$ of the same fact as $D$ by relabeling $\node$ with 
a clause $R'$ such that $R' \impl R$. 
\end{lemma}
\begin{proof}
Let $H$ be the multiset of labels of outgoing edges of the considered 
node $\node$, and $C$ be the label of its incoming edge.
We have $R \impl H \rewrite C$.  By transitivity of $\impl$, 
$R' \impl H \rewrite C$.  So we can relabel $\node$ with $R'$.

Let us show that the obtained derivation $D'$ is data-decomposed.
Consider an edge $\node'_1 \rightarrow \node_1$ in $D'$,
labeled by $F = \attacker(f(p_1, \ldots, p_n))$, where $f$
is a data constructor.
\begin{itemize}
\item
If $\node'_1$ and $\node_1$ are different from $\node$,
then the same edge exists in the initial derivation $D$, so it is of the 
desired form.

\item
If $\node'_1 = \node$, then there is an edge $\node'_1 \rightarrow \node_1$
in $D$, labeled by $F$. Since $D$ is data-decomposed, 
$\node'_1 = \node$ is labeled by 
$R = \attacker(f(x_1, \ldots, x_n)) \rewrite \attacker(x_i)$
or $\node_1$ is labeled by 
$R_1 = \attacker(x_1) \wedge \ldots \wedge \attacker(x_n) \rewrite
\attacker(f(x_1, \ab \ldots, \ab x_n))$ in $D$.
In the latter case, we have the desired form in $D'$.
In the former case, let $R' = H' \rewrite C'$. We have $R' \impl R$,
so there exists $\sigma$ such that 
$\sigma H' \subseteq \{ \attacker(f(x_1, \ldots, x_n)) \}$ and 
$\sigma C' =  \attacker(x_i)$. Hence $C' = \attacker(y)$
where $\sigma y = x_i$, and $H' = \emptyset$ or $H' = \attacker(z)$ with
$\sigma z = f(x_1, \ldots, x_n)$ or $H' =  \attacker(f(y_1, \ldots, y_n))$
with $\sigma y_j = x_j$ for all $j \leq n$.
By Lemma~\ref{lem:varocc}, $y$ occurs in $H'$, so $H' \neq \emptyset$.
If we had $H' = \attacker(z)$, $\sigma z \neq \sigma y$, so $z \neq y$,
so this case is impossible. 
Hence $H' =  \attacker(f(y_1, \ldots, y_n))$. 
Moreover, $\sigma y_j \neq \sigma y$ for all $j \neq i$, so
$y_j \neq y$ for all $j \neq i$. Since $y$ occurs in $H'$, $y = y_i$.
Hence $R' = R$ up to renaming, and we have the desired form in $D'$.
 
\item
If $\node_1 = \node$, then there is an edge $\node'_1 \rightarrow \node_1$
in $D$, labeled by $F$. Since $D$ is data-decomposed, 
$\node'_1$ is labeled by $R'_1 = \attacker(f(x_1, \ldots,
x_n)) \rewrite \attacker(x_i)$ or
$\node_1 = \node$ is labeled by 
$R = \attacker(x_1) \wedge \ldots \wedge \attacker(x_n) \rewrite
\attacker(f(x_1, \ab \ldots, \ab x_n))$ in $D$.
In the former case, we have the desired form in $D'$.
In the latter case, let $R' = H' \rewrite C'$. We have $R' \impl R$,
so there exists $\sigma$ such that 
$\sigma H' \subseteq \{ \attacker(x_1), \ldots, \attacker(x_n) \}$ and 
$\sigma C' = \attacker(f(x_1, \ab \ldots, \ab x_n))$. 
Hence $H' = \bigwedge_{j \in J} \attacker(y_j)$
where $J \subseteq \{ 1, \ldots, n\}$ and $\sigma y_j = x_j$ for all $j \in J$,
and $C' = \attacker(y)$ with $\sigma y = f(x_1, \ab \ldots, \ab x_n)$ or $C'= \attacker(f(y'_1, \ldots, y'_n))$ with $\sigma y'_j = x_j$
for all $j \leq n$.
By Lemma~\ref{lem:varocc}, if $C' = \attacker(y)$, $y$ occurs in $H'$,
but this is impossible because $\sigma y_j \neq \sigma y$ for all $j \in J$.
So $C'= \attacker(f(y'_1, \ldots, y'_n))$. By Lemma~\ref{lem:varocc},
$y'_j$ occurs in $H'$ for all $j \leq n$, so $J = \{ 1, \ldots, n\}$
and $y'_j = y_j$ for all $j \leq n$.
Hence $R' = R$ up to renaming, and we have the desired form in $D'$.

\end{itemize}
Hence the obtained derivation $D'$ is data-decomposed.
\proofcomplete
\end{proof}

\newcommand{\implset}{\impl_{\mathrm{Set}}}
\begin{definition}
We say that $\sattmp \implset \sattmp'$ if, for all clauses $R$ in $\sattmp'$,
$R$ is subsumed by a clause of $\sattmp$.
\end{definition}

\begin{lemma}\label{lem:implsetderiv}
If $\sattmp \implset \sattmp'$ and $D$ is a data-decomposed derivation
containing a node $\node$ labeled by $R \in \sattmp'$, then one can build
a data-decomposed derivation $D'$ of the same fact as $D$ by relabeling
$\node$ with a clause in $\sattmp$.
\end{lemma}
\begin{proof}
Obvious by Lemma~\ref{lem:subsume}.
\proofcomplete
\end{proof}

\begin{lemma}\label{lem:elimcorr}
If $\sattmp \implset \sattmp'$, then $\elim(\sattmp) \implset \sattmp'$.
\end{lemma}
\begin{proof}
This is an immediate consequence of the transitivity of $\impl$.
\proofcomplete
\end{proof}

\begin{lemma}\label{lem:phase1imm}
At the end of $\saturate$, 
$\sattmp$ satisfies the following properties:
\begin{enumerate}

\item For all $R\in \satstset$, $\sattmp \implset \simplify(R)$;

\item Let $R \in \sattmp$ and $R' \in \sattmp$. Assume that $\sel(R) =
\emptyset$ and there exists $F_0 \in \sel(R')$ such that $R
\circ_{F_0} R'$ is defined.  In this case, $\sattmp \implset 
\simplify(R \circ_{F_0} R')$.

\end{enumerate}
\end{lemma}
\begin{proof}
To prove the first property, let $R \in \satstset$. We show that, after the
addition of $R$ to $\sattmp$, $\sattmp \implset \simplify(R)$.

In the first step of $\saturate$, 
we execute the instruction $\sattmp \leftarrow
\elim(\simplify(R) \cup \sattmp)$. We have $\simplify(R) \cup \sattmp
\implset \simplify(R)$, so, by Lemma~\ref{lem:elimcorr}, after
execution of this instruction, $\sattmp \implset \simplify(R)$.

Assume that we execute $\sattmp \leftarrow \elim(\simplify(R'')\cup
\sattmp)$, and before this execution $\sattmp \implset \simplify(R)$.
Hence $\simplify(R'')\cup \sattmp \implset \simplify(R)$, so, by
Lemma~\ref{lem:elimcorr}, after the execution of this instruction,
$\sattmp \implset \simplify(R)$.

The second property simply means that the fixpoint is reached at
the end of $\saturate$, so $\sattmp = \elim(\simplify(R \circ_{F_0} R') 
\cup \sattmp)$. Since $\simplify(R \circ_{F_0} R') \cup \sattmp
\implset \simplify(R \circ_{F_0} R')$, by Lemma~\ref{lem:elimcorr},
$\elim(\simplify(R \circ_{F_0} R') \cup \sattmp)
\implset \simplify(R \circ_{F_0} R')$, so 
$\sattmp \implset \simplify(R \circ_{F_0} R')$.
\proofcomplete
\end{proof}

\begin{lemma}\label{lem:simp:all}
Let $f \in \{ \elimattx$, $\elimtaut$, $\elimnot$, 
$\elimredundanthyp$, $\elimdup$, $\decomp$, $\decomphyp$, 
$\simplify$, $\simplify'\}$.

If the data-decomposed derivation $D$ contains a node $\node$ labeled
$R$, then one obtains a data-decomposed derivation $D'$ of the same fact as
$D$ or of an instance of a fact in $\fnot$ by relabeling $\node$ with
some $R' \in f(R)$ or removing $\node$, and possibly deleting
nodes. Furthermore, if $D'$ is not a derivation of the same fact as
$D$, then $\node$ is removed.

If $D'$ contains a node labeled $R' \in f(R)$, then there exists a
derivation $D$ using $R$, the clauses of $D'$ except $R'$, and the
clauses of $\satstset$ that derives the same fact as $D'$.
\end{lemma}
When $R$ is unchanged by $f$, that is, $f(R) = \{ R \}$, this lemma
is obvious. So, in the proofs below, we consider only the cases
in which $R$ is modified by $f$.

\begin{proof}[for $\elimattx$]
The direct part is obvious: $R'$ is built from $R$ by removing some 
hypotheses, so we just
remove the subtrees corresponding to removed hypotheses of $R$.

Conversely, let $p$ be a closed pattern such that $\attacker(p)$ is
derivable from $\satstset$. (There exists an infinite number of such
$p$.)  We build a derivation $D$ by replacing $R'$ with $R$ in $D$ and 
adding a derivation of $\attacker(p)$ as a subtree of the nodes 
labeled by $R'$ in $D$. 
\proofcomplete
\end{proof}

\begin{proof}[for $\elimtaut$]
Assume that $R$ is a tautology.
For the direct part, we remove $\node$ and replace it with one of its
subtrees.
The converse is obvious since $\elimtaut(R) = \emptyset$.
\proofcomplete
\end{proof}

\begin{proof}[for $\elimnot$]
Assume that $R$ contains as hypothesis an instance $F$ of a fact in
$\fnot$.  Then $\elimnot(R) = \emptyset$. Since $D$ is a
derivation, a son $\node'$ of $\node$ infers an instance of $F$. We
let $D'$ be the sub-derivation with subroot $\node'$. $D'$ is a
derivation of an instance of a fact in $\fnot$, so we obtain the
direct part. The converse is obvious since $\elimnot(R) =
\emptyset$.  
\proofcomplete
\end{proof}

\begin{proof}[for $\elimredundanthyp$]
We have $R = H \wedge H'
\rewrite C$, $\sigma H \subseteq H'$, $\sigma$ does not change the
variables of $H'$ and $C$, and $R' = H' \rewrite C$.

For the direct part, $R'$ is built from $R$ by removing some 
hypotheses, so we just
remove the subtrees corresponding to removed hypotheses of $R$.

For the converse, we obtain a derivation $D$ by
duplicating the subtrees proving instances of elements of $H'$ that
are also in $\sigma H$ and replacing $R'$ with $R$.
\proofcomplete
\end{proof}

\begin{proof}[for $\elimdup$]
For the direct part, $R'$ is built from $R$ by removing some 
hypotheses, so we just
remove the subtrees corresponding to removed hypotheses of $R$.

Conversely, we can form a derivation using $R$ instead of $R'$
by duplicating the subtrees that derive the duplicate hypotheses of $R$.
\proofcomplete
\end{proof}

\begin{proof}[for $\decomp$ and $\decomphyp$]
If $R$ is modified by $\decomp$ or $\decomphyp$, then $R$ is of one of
the following forms:
\begin{itemize}
\item $R = \attacker(f(p_1, \ldots, p_n)) \wedge H \rewrite C$, where $f$ is
a data constructor (for both $\decomp$ and $\decomphyp$). 

For the direct part, let $\node'$ be the son of $\node$ corresponding to 
the hypothesis $\attacker(f(p_1, \ab \ldots, \ab p_n))$. 
The edge $\node \rightarrow \node'$ is labeled by an instance
of $\attacker(f(p_1, \ab \ldots, \ab p_n))$, 
so, since $D$ is data-decomposed,
$\node'$ is labeled by $\attacker(x_1) \wedge \ldots \wedge \attacker(x_n)
\rewrite \attacker(f(x_1, \ab \ldots, \ab x_n))$. 
(The clause $R$ that labels $\node$ cannot be
$\attacker(f(x_1, \ab \ldots, \ab x_n)) 
\rewrite \attacker(x_i)$, 
since this clause would be unmodified by $\decomp$ and $\decomphyp$.) 
Then we build $D'$
by relabeling $\node$ with $R' = \attacker(p_1) \wedge \ldots \wedge 
\attacker(p_n) \wedge H \rewrite C$ and deleting $\node'$.

For the converse, we replace $R' = \attacker(p_1) \wedge \ldots \wedge 
\attacker(p_n) \wedge H \rewrite C$ in $D'$ with
$\attacker(x_1) \wedge \ldots \wedge \attacker(x_n)
\rewrite \attacker(f(x_1, \ab \ldots, \ab x_n))$ and
$R = \attacker(f(p_1, \ab \ldots, \ab p_n)) \wedge H \rewrite C$ 
in $D$.

\item $R = H \rewrite \attacker(f(p_1, \ldots, p_n))$, where $f$ is
a data constructor (for $\decomp$ only). 

For the direct part, let $\node'$ be the father
of $\node$. The edge $\node' \rightarrow \node$ is labeled
by an instance of $\attacker(f(p_1, \ab \ldots, \ab p_n))$,
so, since $D$ is data-decomposed, $\node'$ is labeled by
$\attacker(f(x_1, \ab \ldots, \ab x_n)) \rewrite \attacker(x_i)$ for some $i$.
(The clause $R$ that labels $\node$ cannot be
$\attacker(x_1) \wedge \ldots \wedge \attacker(x_n) \rewrite \attacker(f(x_1, \ldots, x_n))$ since this clause would be unmodified by $\decomp$.)
Then we build $D'$
by relabeling $\node$ with $R' = H \rewrite \attacker(p_i)$ and
deleting $\node'$.

For the converse, we replace $R' = H \rewrite \attacker(p_i)$ in $D'$
with $R = H \rewrite \attacker(f(p_1, \ab \ldots, \ab p_n))$ and 
$ \attacker(f(x_1, \ab \ldots, \ab x_n)) \rewrite \attacker(x_i)$ in $D$.
\proofcomplete

\end{itemize}
\end{proof}

\begin{proof}[for $\simplify$ and $\simplify'$]
For $\simplify$ and $\simplify'$, the result is obtained by 
applying Lemma~\ref{lem:simp:all} for the functions that
compose $\simplify$ and $\simplify'$.
\proofcomplete
\end{proof}

\begin{restate}{Lemma~\ref{lem:phase1corr}}
Let $F$ be a closed fact. If, for all $F' \in \fnot$, no instance of $F'$
is derivable from $\saturate(\satstset) \cup \beginset$, 
then $F$ is derivable from $\satstset \cup \beginset$ if
and only if $F$ is derivable from $\saturate(\satstset) \cup \beginset$.
\end{restate}
\begin{proof}
Assume that $F$ is derivable from $\satstset \cup \beginset$ and consider a derivation of
$F$ from $\satstset \cup \beginset$. 
We show that $F$ or an instance of a fact in $\fnot$ is derivable from
$\saturate(\satstset) \cup \beginset$.

\begin{figure}
\begin{center}
\input{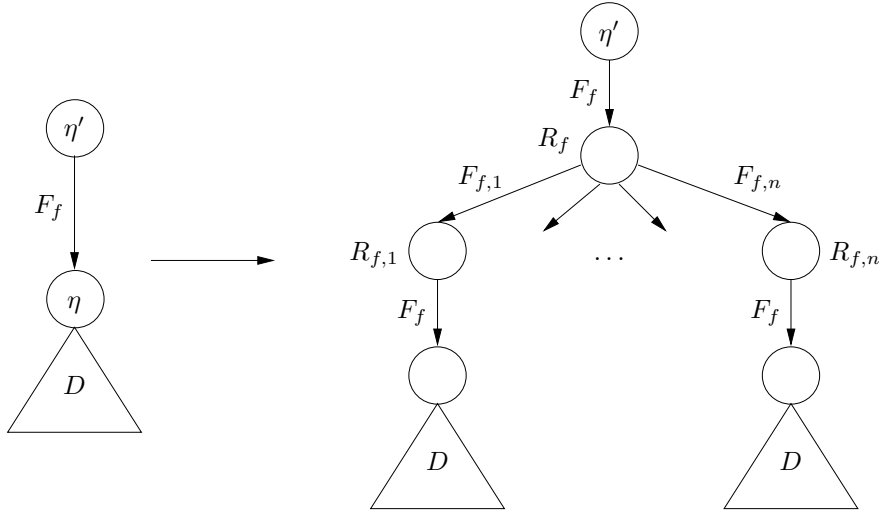}
\end{center}
\caption{Construction of a data-decomposed derivation}\label{fig:datadecomp}
\end{figure}

We first transform the derivation of $F$ into a data-decomposed
derivation. We say that an edge $\node' \rightarrow \node$ 
is \emph{offending} when it is labeled by $F_f = \attacker(f(p_1, \ldots, p_n))$ 
for some data constructor $f$, $\node'$ is not labeled by 
$R_{f,i} = \attacker(f(x_1, \ldots, x_n)) \rewrite \attacker(x_i)$ for some $i$,
and $\node$ is not labeled by 
$R_f = \attacker(x_1) \wedge \ldots \wedge \attacker(x_n) \rewrite
\attacker(f(x_1, \ab \ldots, \ab x_n))$.
We consider an offending edge $\node'
\rightarrow \node$ such that the subtree $D$ of root $\node$
contains no offending edge.
We copy the subtree $D$, which concludes $F_f$, $n$ times and
add the clauses $R_{f,i}$ for $i = 1, \ldots n$, to conclude $F_{f,i} =
\attacker(p_i)$, then use the clause $R_f$ to conclude $F_f$ again,
as in Figure~\ref{fig:datadecomp}.
This transformation decreases the total number of data constructors
at the root of labels of offending edges. Indeed, since there are no offending
edges in $D$, the only edges that may be offending in the new 
subtree of root $\node'$ are those 
labeled by $F_1, \ldots, F_n$. The total number of data constructors
at the root of their labels is the total number of data constructors
at the root of $p_1, \ldots, p_n$, which is one less than the 
total number of data constructors at the root of $f(p_1, \ldots, p_n)$.
Hence, this transformation terminates and, upon termination, 
the obtained derivation contains no offending edge, so it is data-decomposed.

We consider the value of the set of clauses $\sattmp$ at the end of $\saturate$.
For each clause $R$ in $\satstset$, $\sattmp \implset \simplify(R)$ 
(Lemma \ref{lem:phase1imm}, Property 1).  Assume that
there exists a node labeled by $R \in \satstset \setminus \sattmp$ 
in this derivation.  
By Lemma~\ref{lem:simp:all}, we can replace $R$ with some 
$R'' \in \simplify(R)$ or remove $R$. 
(After this replacement, we may obtain a derivation
of an instance of a fact in $\fnot$ instead of a derivation of $F$.) 
If $R$ is replaced with $R''$, 
by Lemma~\ref{lem:implsetderiv}, we can replace $R''$ 
with a clause in $\sattmp$.
This transformation decreases the number of nodes labeled by clauses
not in $\sattmp$.
So this transformation terminates and, upon termination, no node
of the obtained derivation is labeled by a clause in $\satstset \setminus \sattmp$.
Therefore, we obtain a
data-decomposed derivation $D$ of $F$ or of an instance of a fact in $\fnot$
from $\sattmp \cup \beginset$.

Next, we build a data-decomposed derivation of $F$ or of an instance of a fact in $\fnot$
from $\satend \cup \beginset$, where $\satend = \saturate(\satstset)$.
If $D$ contains a node labeled by a clause not in $\satend \cup \beginset$, 
we can transform $D$ as follows.
Let $\node'$ be a lowest node of $D$ labeled by a clause not in $\satend \cup 
\beginset$.
  So all sons of $\node'$ are labeled by elements of $\satend \cup
  \beginset$.  Let $R'$ be the clause labeling $\node'$.
  Since $R'\notin \satend \cup \beginset$, $\sel(R')
  \neq \emptyset$. Take $F_0 \in \sel(R')$. By Lemma~\ref{lem:compos},
  there exists a son of $\node$ of $\node'$ labeled by $R$, such that $R
  \circ_{F_0} R'$ is defined. Since all sons of $\node'$ are labeled by 
  elements of $\satend \cup \beginset$, $R \in \satend \cup
  \beginset$. By definition of the selection function, $F_0$
  is not a $\pasbegin$ fact, so $R \notin \beginset$, so $R \in \satend$. 
  Hence $\sel(R) = \emptyset$.  So, by
  Lemma~\ref{lem:phase1imm}, Property 2, $\sattmp \implset
  \simplify(R \circ_{F_0} R')$. 
  So, by Lemma~\ref{lem:compos}, we can replace $\node$ and
  $\node'$ with $\node''$ labeled by $R \circ_{F_0} R'$. By
  Lemma~\ref{lem:simp:all}, we can replace $R \circ_{F_0} R'$ with
  some $R''' \in \simplify(R \circ_{F_0} R')$ or remove $R \circ_{F_0} R'$. 
\begin{itemize}

\item
  If $R \circ_{F_0} R'$ is replaced with $R'''$, then
  by Lemma~\ref{lem:implsetderiv}, we can replace $R'''$ with 
  a clause in $\sattmp$.
  The total number of nodes strictly decreases since $\node$ and
  $\node'$ are replaced with a single node.

\item If $R \circ_{F_0} R'$ is removed, then the total number of nodes
  strictly decreases since $\node$ and $\node'$ are removed.

\end{itemize}
  So in all cases, we obtain a derivation $D'$ of $F$ or of an
  instance of a fact in $\fnot$ from $\sattmp \cup \beginset$, such
  that the total number of nodes strictly decreases.
Hence, this replacement process 
terminates. Upon termination, all clauses are in $\satend \cup \beginset$.
So we obtain a data-decomposed derivation of $F$ or of an instance of a fact in
$\fnot$ from $\satend \cup \beginset$, which is the expected result.

For the converse implication, notice that if a fact is derivable from
$\satend$ then it is derivable from $\sattmp$, and that all clauses
added to $\sattmp$ do not create new derivable facts: when composing
two clauses $R$ and $R'$, the created clause can derive facts that could
also by derived by $R$ and $R'$.
\proofcomplete
\end{proof}

\begin{restate}{Lemma~\ref{lem:reach}}
Let $F'$ be a closed instance of $F$.  
If, for all $F'' \in \fnot$, $\reach(F'', \derivstset) = \emptyset$, then
$F'$ is derivable from $\derivstset
\cup \beginset$ if and only if there exist a clause $H \rewrite C$ in
$\reach(F, \derivstset)$ and a substitution $\sigma$ such that $\sigma C = F'$
and all elements of $\sigma H$ are derivable from $\derivstset \cup \beginset$.
\end{restate}
\begin{proof}
\newcommand{\fset}{{\cal F}}
\newcommand{\dset}{{\cal D}}
Let us prove the direct implication. 
Let $\fset = \{ (F,F') \} \cup \{ (F'',\sigma F'') \mid F'' \in \fnot, \sigma$ 
any substitution$\}$.
We show that, if $F'$ is derivable from $\derivstset
\cup \beginset$, then there exist a clause $H \rewrite C$ in
$\reach(F_{\mathrm{g}}, \derivstset)$ and a substitution $\sigma$ such that $(F_{\mathrm{g}}, \sigma C) \in \fset$ and all elements of $\sigma H$ are derivable from $\derivstset \cup \beginset$.
(This property proves the desired result.
If, for all $F'' \in \fnot$, $\reach(F'', \derivstset) = \emptyset$ and
$F'$ is derivable from $\derivstset \cup \beginset$, then
there exist a clause $H \rewrite C$ in
$\reach(F_{\mathrm{g}}, \derivstset)$ and a substitution $\sigma$ such that $(F_{\mathrm{g}}, \sigma C) \in \fset$
and all elements of $\sigma H$ are derivable from $\derivstset \cup \beginset$.
Since, for all $F'' \in \fnot$, $\reach(F'', \derivstset) = \emptyset$, we have $F_{\mathrm{g}} = F$ and $F \notin \fnot$.
Since $(F, \sigma C) \in \fset$, we have then $\sigma C = F'$.)

Let $\dset$ be the set of derivations
$D'$ of a fact $F_{\mathrm{i}}$ 
such that, for some $F_{\mathrm{g}}$ and $\derivtmp$,
$(F_{\mathrm{g}}, F_{\mathrm{i}}) \in \fset$,
the clause $R'$ at the subroot of $D'$ satisfies
$\deriv(R', \derivtmp, \derivstset) \subseteq \reach(F_{\mathrm{g}}, 
\derivstset)$
and $\forall R'' \in \derivtmp, R'' \not\impl R'$,
and the other clauses of $D'$ are in $\derivstset \cup \beginset$.

Let $\attacker'$ be a new predicate symbol.
Let $D$ be a derivation. If $D$ is a derivation of $\attacker(p)$,
we let $D'$ be the derivation obtained by replacing the 
clause $H \rewrite \attacker(p_1)$ with $H \rewrite \attacker'(p_1)$
and the fact $\attacker(p)$ derived by $D$ with $\attacker'(p)$.
If $D$ is not a derivation of $\attacker(p)$, we let $D'$ be $D$.
We say that the derivation $D$ is \emph{almost-data-decomposed} when
$D'$ is data-decomposed. 
We first show that all derivations $D$ in $\dset$ are almost-data-decomposed.
Let $D'$ be the transformed derivation as defined above.
Let $\node' \rightarrow \node$ be an edge of $D'$ labeled by
$F = \attacker(f(p_1, \ldots, p_n))$, where $f$ is a data constructor.
This edge is not the outgoing edge of the root of $D'$, because $D'$ does not
conclude $\attacker(p)$ for any $p$. So the clause that labels $\node$
is of the form $R = H \rewrite \attacker(p)$ and it is in $\derivstset$. 
In order to obtain a contradiction, assume that $p$ is a variable $x$.
Since $\sel(R) = \emptyset$, $H$ contains only unselectable facts.
By Lemma~\ref{lem:varocc}, $x$ occurs in non-$\pasbegin$
facts in $H$, so $H$ contains $\attacker(x)$. So $R$ is a tautology.
This is impossible because $R$ would have been removed from $\derivstset$
by $\elimtaut$.
So $p$ is not a variable. Hence $p = f(p'_1, \ldots, p'_n)$.  If $R$ was
different from $\attacker(x_1) \wedge \ldots \wedge \attacker(x_n)
\rewrite \attacker(f(x_1, \ab \ldots, \ab x_n))$, 
$R$ would have been transformed by $\decomp$, so $R$ would
not be in $\derivstset$. Hence 
$R = \attacker(x_1) \wedge \ldots \wedge \attacker(x_n) \rewrite 
\attacker(f(x_1, \ab \ldots, \ab  x_n))$. 
Therefore, $D'$ is data-decomposed, so $D$ is almost-data-decomposed.
Below, when we apply Lemma~\ref{lem:compos}, \ref{lem:simp:all}, or
\ref{lem:subsume}, we first transform the considered derivation $D$
into $D'$, apply the lemma to the data-decomposed derivation $D'$,
and transform it back by replacing $\attacker'$ with $\attacker$. We
obtain the same result as by transforming $D$ directly, because
the simplifications of $\simplify'$ apply in the same way
when the conclusion is $\attacker(p)$ or $\attacker'(p)$,
since $\simplify'$ uses $\decomphyp$ instead of $\decomp$ and 
does not use $\elimtaut$.

Let $D_0$ be a derivation of $F'$ from $\derivstset \cup \beginset$.
Let $D'_0$ be obtained from $D_0$ by adding
a node labeled by $\{ F \} \rewrite F$ at the subroot of $D_0$. By
definition of $\reach$, $\deriv(R', \emptyset, \derivstset) \subseteq
\reach(F, \derivstset)$, and $\forall R'' \in \emptyset, R'' \not\impl
R'$. Hence $D'_0$ is a derivation of $F'$ in $\dset$, so $\dset$ is non-empty.

Now consider a derivation $D_1$ in $\dset$ with the smallest number of nodes.  The
clause $R'$ labeling the subroot $\node'$ of $D_1$ satisfies 
$(F_{\mathrm{g}}, F_{\mathrm{i}}) \in \fset$, 
$\deriv(R', \derivtmp, \derivstset) \subseteq \reach(F_{\mathrm{g}}, \derivstset)$, 
and $\forall R'' \in \derivtmp, R''
\not\impl R'$.  In order to obtain a contradiction, we assume that
$\sel(R') \neq \emptyset$. Let $F_0 \in \sel(R')$. By
Lemma~\ref{lem:compos}, there exists a son $\node$ of $\node'$, labeled by $R$, 
such that $R \circ_{F_0} R'$ is defined. By hypothesis on the
derivation $D_1$, $R \in \derivstset \cup \beginset$. 
By the choice of the
selection function, $F_0$ is not a $\pasbegin$ fact, 
so $R \notin \beginset$, so $R \in \derivstset$. 
Let $R_0 = R \circ_{F_0} R'$. So, by
Lemma~\ref{lem:compos}, we can replace $R'$ with $R_0$, obtaining a
derivation $D_2$ of $F_{\mathrm{i}}$ with fewer nodes than $D_1$. 

By Lemma~\ref{lem:simp:all}, we can either
replace $R_0$ with some $R'_0 \in \simplify'(R_0)$
or remove $R_0$, yielding a derivation $D_3$. 
\begin{itemize}
\item
In the latter case, $D_3$ is a derivation of a fact $F'_{\mathrm{i}}$
which is either $F_{\mathrm{i}}$ or an instance of a fact $F'_{\mathrm{g}}$ in
$\fnot$. If $F'_{\mathrm{i}} = F_{\mathrm{i}}$, we let $F'_{\mathrm{g}} = F_{\mathrm{g}}$.
So $(F'_{\mathrm{g}}, F'_{\mathrm{i}}) \in \fset$.

We replace $R_0$ with $R'_0 = F'_{\mathrm{g}} \rewrite
F'_{\mathrm{g}}$ in $D_2$. Hence we obtain a derivation with fewer nodes than $D_1$ and
such that $\deriv(R'_0, \emptyset, \derivstset) \subseteq
\reach(F'_{\mathrm{g}}, \derivstset)$ and $\forall R_1 \in \emptyset,
R_1 \not\impl R'_0$. So we have a derivation in $\dset$ with fewer nodes than $D_1$, 
which is a contradiction.

\item
In the former case, $D_3$ is a derivation of $F_{\mathrm{i}}$, and
$\deriv(R'_0, \ab \{ R' \} \cup \derivtmp, \ab \derivstset) 
\subseteq
\deriv(R', \ab \derivtmp, \ab \derivstset) \subseteq
\reach(F_{\mathrm{g}}, \ab \derivstset)$ (third case of the definition of
$\deriv(R', \ab \derivtmp, \ab \derivstset)$).

\begin{itemize}
\item
If $\forall R_1 \in \{ R' \} \cup \derivtmp, R_1
\not\impl R'_0$, $D_3$ is a derivation of $F_{\mathrm{i}}$ in $\dset$, 
with fewer nodes than $D_1$, which is a contradiction. 

\item
Otherwise, $\exists R_1 \in \{ R' \} \cup \derivtmp,
R_1 \impl R'_0$. Therefore, by Lemma~\ref{lem:subsume}, 
we can build a derivation $D_4$ by replacing $R'_0$ with $R_1$ in $D_3$. There
is an older call to $\deriv$, of the form $\deriv(R_1, \derivtmpp,
\derivstset)$, such that $\deriv(R_1, \derivtmpp, \derivstset)
\subseteq \reach(F_{\mathrm{g}}, \derivstset)$. Moreover, $R_1$ has been added to
$\derivtmpp$ in this call, since $R_1$ appears in $\{ R' \} \cup
\derivtmp$. Therefore the third case of the definition of $\deriv(R_1,
\derivtmpp, \derivstset)$ has been applied, and not the first
case. So $\forall R_2 \in \derivtmpp, R_2 \not\impl R_1$, so the
derivation $D_4$ is in $\dset$ and has fewer nodes than $D_1$,
which is a contradiction. 

\end{itemize}

\end{itemize}
In all cases, we could find a derivation in $\dset$ that has
fewer nodes than $D_1$. This is a contradiction, so
$\sel(R') = \emptyset$, hence $R' \in \reach(F_{\mathrm{g}}, \derivstset)$. The
other clauses of this derivation are in $\derivstset \cup \beginset$. 
By definition of a derivation, $R' \impl H'
\rewrite F_{\mathrm{i}}$ 
where $H'$ is the multiset of labels of the outgoing edges of the subroot
of the derivation. Taking $R' = H \rewrite C$, there exists
$\sigma$ such that $\sigma C = F_{\mathrm{i}}$ and 
$\sigma H \subseteq H'$, so all
elements of $\sigma H$ are derivable from $\derivstset \cup
\beginset$. We have the result, since $(F_{\mathrm{g}},
F_{\mathrm{i}}) \in \fset$.

The proof of the converse implication is left to the reader.
(Basically, the clause $R \circ_{F_0} R'$ does not generate facts that
cannot be generated by applying $R$ and $R'$.)
\proofcomplete
\end{proof}

\section{Termination Proof}\label{app:termination}

In this section, we give the proof of Proposition~\ref{prop:termination}
stated in Section~\ref{sect:taggedterm}.
We denote by $P_0$ a tagged protocol and let 
$P'_0 = \instr{P_0}$. We have the following properties:
\begin{itemize}
\item
By Condition~\ref{condCpubchannel}, 
the input and output constructs in the protocol
always use a public channel $c$. So the facts $\mess(c,p)$ are
replaced with $\attacker(p)$ in all clauses. The only remaining clauses
containing $\mess$ are~\eqref{ruleRl} and~\eqref{ruleRs}.  Since
$\mess(x,y)$ is selected in these clauses, the only inference with
these clauses is to combine~\eqref{ruleRs} with~\eqref{ruleRl},
and it yields a tautology which is immediately removed. Therefore,
we can ignore these clauses in our termination proof. 

\item
By hypothesis on the queries and Remark~\ref{rem:event},
the clauses do not contain $\pasbegin$ facts.

\end{itemize}

In this section, we use the sort system defined at the beginning of Appendix~\ref{app:corr} (Lemma~\ref{lem:varocc}).

The \emph{patterns} of a fact $\pred(p_1, \ldots, p_n)$ are $p_1, \ldots, p_n$.
The \emph{patterns} of a clause $R$ are the patterns of all facts
in $R$, and we denote the set of
patterns of $R$ by $\patts(R)$.
A pattern is said to be \emph{non-data} when it is not of the form
$f(\ldots)$ with $f$ a data constructor. The set $\subg(S)$ contains
the subterms of patterns in the set~$S$. Below, we use the word ``program''
for a set of clauses (that is, a logic program).

\begin{definition}[Weakly tagged programs]\label{hypProg}
Let $S_0$ be a finite set of closed patterns and $\taggen$ be a set of patterns.

A pattern is \emph{top-tagged} when it is an instance of a pattern in
$\taggen$. 

A pattern is \emph{fully tagged} when all its non-variable
non-data subterms are top-tagged.

Let $\Bp$ be the set of clauses $R$ that satisfy Lemma~\ref{lem:varocc}
and are of one of the following three forms:
\begin{enumerate}

\item $\Bpr$ contains clauses $R$ of the form
$F_1 \wedge \ldots \wedge F_n \rewrite F$
where for all $i$, $F_i$ is of the form $\attacker(p)$ for some $p$,
$F$ is of the form $\attacker(p)$ or $\pasend(p)$ for some $p$, 
there exists a substitution $\sigma$ such
that $\patts(\sigma R) \subseteq \subg(S_0)$, and
the patterns of $R$ are fully-tagged.

\item $\Bpconstr$ contains clauses of the form
$\attacker(x_1) \wedge \ldots \wedge \attacker(x_n)
\rewrite \attacker(f(x_1, \ab \ldots, \ab x_n))$
where $f$ is a constructor.

\item $\Bpdestr$ contains clauses of the form $\attacker(f(p_1,
  \ldots, p_n)) \wedge \attacker(x_1) \wedge \ldots \wedge
  \attacker(x_k) \rewrite \attacker(x)$ where $f$ is a constructor,
  $p_1, \ldots, p_n$ are fully tagged, $x$ is one of $p_1, \ldots,
  p_n$, and $f(p_1, \ldots, p_n)$ is more general than every pattern
  of the form $f(\ldots)$ in $\subg(S_0)$.

\end{enumerate}
A program $\Bzero$ is
\emph{weakly tagged} if there exist a finite set of closed patterns
$S_0$ and a set of patterns $\taggen$ such that
\begin{enumerate}

\pname{W1}\label{condW2}$\Bzero$ is included in $\Bp$.

\pname{W2} If two patterns $p_1$ and $p_2$ in $\taggen$ unify, $p'_1$ is
an instance of $p_1$ in $\subg(S_0)$, and $p'_2$ is an instance of
$p_2$ in $\subg(S_0)$, then $p'_1 = p'_2$. \label{tagcond}

\end{enumerate}
\end{definition}

Intuitively, a pattern is top-tagged when its root function symbol is tagged
(that is, it is of the form $f((\cto, M_1, \ldots, M_n), \ldots)$).
A pattern is fully tagged when all its function symbols are tagged.

We are going to show that all clauses generated by the resolution
algorithm are in $\Bp$. Basically, the clauses in $\Bpr$ satisfy two
conditions: they can be instantiated into clauses whose patterns are
in $\subg(S_0)$ and they are tagged.  Then, all patterns in clauses
of $\Bpr$ are instances of $\taggen$ and have instance in
$\subg(S_0)$. Property~\ref{tagcond} allows us to show that this
property is preserved by resolution: when unifying two patterns that
satisfy the invariant, the result of the unification also satisfies
the invariant, because the instances in $\subg(S_0)$ of those
two patterns are in fact equal. Thanks to this property, we can show that
clauses obtained by resolution from clauses in $\Bpr$ are still in $\Bpr$.
To prove termination, we show that the size of generated clauses decreases,
for a suitable notion of size defined below.
The clauses of $\Bpconstr$ and $\Bpdestr$ are needed for constructors
and destructors. Although they do not satisfy exactly the conditions
for being in $\Bpr$, their resolution with a clause in $\Bpr$ yields
a clause in $\Bpr$.

Let $\params{\pk}$ and $\params{\host}$ be the sets of arguments of
$\pk$ resp.~$\host$ in the terms that occur in the trace of
Condition~\ref{condC5}. Let $\condense(\satstset)$ be the set of clauses 
$\sattmp$ obtained
by $\sattmp \leftarrow \emptyset$; for each $R \in \satstset$, 
$\sattmp \leftarrow \elim(\simplify(R) \cup \sattmp)$.
We first consider the case in which a single long-term key is used, that is,
$\params{\pk}$ and
$\params{\host}$ have at most one element. The results will be generalized
to any number of keys at the end of this section. The next proposition
shows that the initial clauses given to the resolution algorithm 
form a weakly tagged program.

\begin{figure}[t]
\begin{align}
&\env,\pset \cup \{ \,0\, \}, \tset \rightarrow \env, \pset, \tset\tag{Red Nil'}\\
&\env,\pset \cup \{ \,\ReplInstr{i}{P}\, \}, \tset \rightarrow \env[i \mapsto \idc], \pset \cup \{ \,P\{ \idc/ i\}\, \}, \tset \cup \{ \idc \}\tag{Red Repl'}\\
&\env,\pset \cup \{ \,P \parpop Q\, \},\tset \rightarrow \env, \pset \cup \{ \,P, Q\, \}, \tset\tag{Red Par'}\\
&\env,\pset \cup \{ \,\ResInstr{a}{\rlbl}P\, \} \rightarrow \env[a \mapsto \env(\rlbl)], \pset \cup \{ \,P\,\}, \tset \cup \{ M_1, \ldots, M_n, a \}\tag{Red Res'}\\
&\env,\pset \cup \{ \,\coutput{c}{M}.Q \,\}, \tset\rightarrow \env, \pset \cup \{ \,Q\, \}, \tset \cup \{ M \}\tag{Red Out'}\\
&\env,\pset \cup \{\, \cinput{c}{x}.P\, \}, \tset\rightarrow \env[x \mapsto \env(M)], \pset \cup \{ \,P\{ M/x \}\, \}, \tset\text{ if }M \in \tset\tag{Red In'}\\
\begin{split}
&\env,\pset \cup \{ \,\letfun{x}{g(M_1, \ldots, M_n)}{P}{0}\,\}, \tset\rightarrow \notag\\
&\qquad \env[x \mapsto \env(M')], \pset \cup \{ \, P\{ M' / x\}\, \}, \tset \cup \{ M_1, \ldots, M_n, M' \}\\
&\qquad \text{ if $g(M_1, \ldots, M_n) \rightarrow M'$}
\end{split}\tag{Red Destr 1'}\\
&\env,\pset \cup \{ \,\asevent(M).Q \,\}, \tset\rightarrow \env, \pset \cup \{ \,Q\, \}, \tset \cup \{ M \}\tag{Red Event'}
\end{align}
\caption{Special semantics for instrumented processes}\label{fig:reductionins}
\end{figure}

\begin{proposition}\label{prop:tagweak}
If $P_0$ is a tagged protocol such that $\params{\pk}$ and
$\params{\host}$ have at most one element
and $P'_0 = \instr{P_0}$, then
$\condense(\rset{P'_0, \rw})$ is a weakly tagged program.
\end{proposition}

\begin{proofsk}
The fully detailed proof is very long (about 8 pages) so we give only
a sketch here. A similar proof (for strong secrecy instead of secrecy and 
reachability) with more details can be found
in the technical report~\cite[Appendix~C]{Blanchet04c}.

We assume that different occurrences of restrictions and variables
have different identifiers and identifiers different from free names
and variables.  In Figure~\ref{fig:reductionins}, we define a special
semantics for instrumented processes, which is only used as a tool in
the proof.  A semantic configuration consists of three components: an
environment $\env$ mapping names and variables to patterns, a multiset
of instrumented processes $\pset$, and a set of terms $\tset$. The
semantics is defined as a reduction relation on semantic
configurations. 
In this semantics, $\Res{a}$ creates the name $a$, instead of a 
fresh name $a'$. 
Indeed, creating fresh names is useless, since
the replication does not copy processes in this semantics, and the
names are initially pairwise distinct.

Let $\env_0 = \{ a \mapsto a[\,] \mid a \in \fn(P_0) \}$.
We show that $\env_0,
\{ P'_0 \}, \fn(P_0) \rightarrow^* \env', \emptyset, \tset'$, for some
$\env'$ and $\tset'$,
such that the second argument of $\pencrypt$ in $\tset'$ is of the form
$\pk(M)$ and the arguments of $\pk$ and $\host$ in $\tset'$ are atomic
constants in $\params{\pk}$ and $\params{\host}$ respectively.
This result is obtained by simulating in the semantics of 
Figure~\ref{fig:reductionins} the trace of Condition~\ref{condC5}.
Moreover, the
second argument of $\pencrypt$ in $\tset'$ is of the form $\pk(M)$ by
Condition~\ref{HYP2} and the arguments of $\pk$ and $\host$ in
$\tset'$ are atomic constants in $\params{\pk}$ and $\params{\host}$
respectively, by Condition~\ref{HYP5} and definition of $\params{\pk}$
and $\params{\host}$.

Let us define $S_0 = \env'(\tset') \cup \{ \advnfs[\idc] \}$.
If $\params{\pk}$ is empty, we add some key $k$ to it, so that
$\params{\pk} = \{ k \}$. 
Let $c, c', c'',c'''$ be constants.  If $S_0$ contains no
instance of $\sencrypt(x,\ab y)$, we add $\sencrypt((c,c'),\ab c'')$ to
$S_0$. If $S_0$ contains no
instance of $\sencryptp(x,\ab y,\ab z)$, we add $\sencryptp((c,c'),\ab c'', \ab c''')$ to
$S_0$. If $S_0$ contains no instance of $\pencrypt(x,y,z)$, we add
$\pencrypt((c,c'),\ab \pk(k), \ab c'')$ to $S_0$. If $S_0$ contains no instance of
$\sign(x,y)$, we add $\sign((c,c'),\ab k)$ to $S_0$. If $S_0$ contains no
instance of $\nmrsign(x,\ab y)$, we add $\nmrsign((c,c'),\ab k)$ to
$S_0$. So $S_0$ is a finite set of closed patterns.
Intuitively, $S_0$ is the set of patterns
corresponding to closed terms that occur in the trace of
Condition~\ref{condC5}.

Let $\env_t$ be $\env$ in which all patterns $a[\ldots]$ 
are replaced with their corresponding term $a$.
In all reductions 
$\env_0, \{ P'_0 \}, \fn(P_0) 
\rightarrow^* \env, \pset, \tset$, all patterns of the form $a[\ldots]$
in the image of $\env$ are equal to $\env(a)$, so $\env \circ \env_t = \env$.
We show the following result by induction on $P$:
\begin{quote}
Let $P$ be an instrumented process, subprocess of $P'_0$.  Assume that 
$\env_0, \ab \{ P'_0 \}, \ab \fn(P_0) 
\rightarrow^* \env, \pset \cup \{ \env_t(P) \}, \tset \rightarrow^* \env', \emptyset,
\tset'$, and that there exists $\sigma'$ such that
$\env'_{\mid \dom(\rho)} = \sigma' \circ \rho$ and $\patts(\sigma' H)
\subseteq \subg(S_0)$.  Then for all $R\in \lp P\rp \rho H$, there
exists $\sigma''$ such that $\patts(\sigma'' R) \subseteq \subg(S_0)$.
\end{quote}
Let $\rho_0 = \{ a \mapsto a[\,] \mid a \in \fn(P_0) \}$.
By applying this result to $P = P'_0$, we obtain that 
for all clauses $R$ in $\lp
P_0' \rp \rho_0 \emptyseq$, there exists a substitution $\sigma$ such that
$\patts(\sigma R) \subseteq \subg(S_0)$.

Let 
\[\begin{split}
\taggen&= \{ f((\ct{i}, x_1, \ldots, x_n), x'_2, \ldots, x'_{n'}) \mid\\
&\qquad f \in \{ \sencrypt, \sencryptp, \pencrypt, \sign, \nmrsign, \hash, \mac \}\}\\
&\quad\cup \{ a[x_1, \ldots, x_n]
\mid a \text{ name function symbol}\}\\
&\quad \cup \{ \pk(x), \host(x) \} \cup \{ c \mid c\text{ atomic constant}\}
\end{split}\]
We show the following result by induction on $P$:
\begin{quote}
Assume that the patterns of the image of $\rho$ and of $H$ are fully
tagged. Assume that $P$ is an instrumented process, subprocess of
$P'_0$. For all $R \in \lp P\rp \rho H$, $\patts(R)$ are fully tagged.
\end{quote}
This result relies on Condition~\ref{condC3} to show that the created terms are
tagged, and on Condition~\ref{condC4} to show that the tags are checked.
By applying this result to $P = P'_0$, we obtain that 
for all $R \in \lp P'_0 \rp \rho_0 \emptyseq$, 
the patterns of $R$ are fully tagged.

By the previous results, $\lp P'_0 \rp \rho_0 \emptyseq \subseteq \Bpr$.

The clauses~\eqref{ruleRf} are in $\Bpconstr$. The clauses~\eqref{ruleInit} and~\eqref{ruleRn} are in 
$\Bpr$ given the value of $S_0$.
The clauses~\eqref{ruleRg} for $\nth{n}{i}$, $\sdecrypt$, $\sdecryptp$, $\pdecrypt$,
and $\getmess$ are:
\begin{align}
&\attacker((x_1, \ldots, x_n)) \rewrite \attacker(x_i) \tag{$\nth{n}{i}$}\\
&\attacker(\sencrypt(x,y)) \wedge \attacker(y) \rewrite \attacker(x)\tag{$\sdecrypt$}\\
&\attacker(\sencryptp(x,y,z)) \wedge \attacker(y) \rewrite \attacker(x)\tag{$\sdecryptp$}\\
&\attacker(\pencrypt(x,\pk(y),z)) \wedge \attacker(y) \rewrite \attacker(x)\tag{$\pdecrypt$}\\
&\attacker(\sign(x,y)) \rewrite \attacker(x)\tag{$\getmess$}
\end{align}
and they are in $\Bpdestr$ provided that all public-key
encryptions in $S_0$ are of the form $\pencrypt(p_1,\ab \pk(p_2),
\ab p_3)$ (that
is, Condition~\ref{HYP2}).  The clauses for $\checksign$ and $\nmrchecksign$ are
\begin{align}
&\attacker(\sign(x,y)) \wedge \attacker(\pk(y)) \rewrite \attacker(x)\tag{$\checksign$}\\
&\attacker(\nmrsign(x,y)) \wedge \attacker(\pk(y)) \wedge \attacker(x) \rewrite \attacker(\nmrtrue)\tag{$\nmrchecksign$}
\end{align}
These two clauses are subsumed respectively by the clauses for
$\getmess$ (given above) and $\nmrtrue$ (which is simply
$\attacker(\nmrtrue)$ since $\nmrtrue$ is a zero-ary constructor), so
they are eliminated by $\condense$, \emph{i.e.}, they are not in
$\condense(\rset{P'_0, \rw})$. (This is important, because they are not in
$\Bpdestr$.)
Therefore all clauses in $\condense(\rset{P'_0, \rw})$ are in $\Bp$, 
since the set of clauses $\Bp$ is preserved by simplification,
so we have Condition~\ref{condW2}.

Different patterns in $\taggen$ do not unify. Moreover, each pattern in
$\taggen$ has at most one instance in $\subg(S_0)$.  For $\pk(x)$ and
$\host(x)$, this comes from the hypothesis that $\params{\pk}$ and
$\params{\host}$ have at most one element.  For atomic constants, this
is obvious. (Their only instance is themselves.) For other patterns, this
comes from the fact that the trace of Condition~\ref{condC5}
executes each program point at most once, and that patterns created at
different programs points are associated with different symbols
$(f,c)$ for $f((c, \ldots), \ldots)$ and $a$ for $a[\ldots]$. (For
$f((c, \ldots), \ldots)$, this comes from Condition~\ref{condC3}. For
$a[\ldots]$, this is because different restrictions use a different
function symbol by construction of the clauses.)
So we have Condition~\ref{tagcond}.
\proofcomplete
\end{proofsk}

The next proposition shows that saturation terminates for weakly
tagged programs.

\begin{proposition}\label{prop:term}
Let $\Bzero$ be a set of clauses. If $\condense(\Bzero)$ is 
a weakly tagged program
(Definition~\ref{hypProg}), then the computation of
$\saturate(\Bzero)$ terminates.
\end{proposition}
\begin{proof}
This result is very similar to~\cite[Proposition 8]{Blanchet04e},
so we give only a brief sketch and refer the reader to that paper
for details.

We show by induction that all clauses $R$ generated from $\Bzero$
are in $\Bpr \cup \Bpconstr \cup \Bpdestr$ and the patterns of $\attacker$
facts in clauses $R$ in $\Bpr$ are non-data.

First, by hypothesis, all clauses in $\condense(\Bzero)$ satisfy this
property, by definition of weakly tagged programs and because of the
decomposition of data constructors by $\decomp$.

If we combine by resolution two clauses in $\Bpconstr \cup \Bpdestr$,
we in fact combine a clause of $\Bpconstr$ with a clause of $\Bpdestr$.
The resulting clause is a tautology by definition of $\Bpconstr$ and 
$\Bpdestr$, so it is eliminated by $\elimtaut$.

Otherwise, we combine by resolution a clause $R$ in $\Bpr$ with a
clause $R'$ such that $R' \in \Bpr$, $\sel(R') = \emptyset$, and
$\sel(R) \neq \emptyset$, or $R' \in \Bpconstr$, or $R' \in \Bpdestr$.
Let $R''$ be the clause obtained by resolution of $R$ and $R'$.  We
show that the patterns of $R''$ are fully tagged, and for each
$\sigma$ such that $\patts(\sigma R) \subseteq \subg(S_0)$, there
exists $\sigma''$ such that $\patts(\sigma'' R'') \subseteq
\subg(S_0)$ and $\size(\sigma'' R'') < \size(\sigma R)$, where the
size is defined as follows. The size of a pattern $\size(p)$ is defined
as usual, $\size(\attacker(p)) = \size(\pasevent(p)) = \size(p)$, and
$\size(F_1 \wedge \ldots \wedge F_n \rewrite F) = \size(F_1) + \ldots +
\size(F_n) + \size(F)$.

Let $R_s \in \simplify(R'')$. The patterns of $R_s$ are non-data
fully tagged, $\patts(\sigma'' R_s) \subseteq \subg(S_0)$,
and $\size(\sigma'' R_s) \leq \size(\sigma'' R'') < \size(\sigma R)$.
So $R_s \in \Bpr$ and its patterns are non-data. 

Moreover, for all generated clauses $R$, there exists $\sigma$ such
that $\size(\sigma R)$ is smaller than the maximum initial value of
$\size(\sigma R)$ for a clause of the protocol. There is a finite number
of such clauses (since $\size(R) \leq \size(\sigma R)$).
So $\saturate(\Bzero)$ terminates. 
\proofcomplete
\end{proof}

Next, we show that $\reach$ terminates when it is called on the result
of the saturation of a weakly tagged program.

\begin{proposition}
If $F$ is a closed fact and $\derivstset$ is a weakly tagged program
simplified by $\simplify$
such that, for all $R \in \derivstset$, $\sel_0(R) = \emptyset$, 
then $\reach(F, \derivstset)$ terminates. 
\end{proposition}
\begin{proof}
We show the following property:
\begin{quote}
For all calls $\deriv(R, \derivtmp, \derivstset)$,
$R = F \rewrite F$ or $R =
\attacker(p_1) \wedge \ldots \wedge \attacker(p_n) \rewrite F$ where
$p_1, \ldots, p_n$ are closed patterns. 
\end{quote}
This property is proved by induction. 
It is obviously true for the initial call to $\deriv$, $\deriv(F \rewrite F,
\emptyset, \derivstset)$. 
For recursive calls to $\deriv$, $\deriv(R'', \derivtmp, \derivstset)$, 
the clause $R''$ is in $\simplify'(R' \circ_{F_0} R)$, where 
$R' = \attacker(x_1) \wedge
\ldots \wedge \attacker(x_k) \rewrite F'$ since $R' \in \derivstset$
and $R = F \rewrite F$ or $R =
\attacker(p_1) \wedge \ldots \wedge \attacker(p_n) \rewrite F$ where
$p_1, \ldots, p_n$ are closed patterns,  
by induction hypothesis.
After unification of $F'$ and $F_0$, $x_i$ is substituted by a closed
pattern $p'_i$ (subpattern of $F_0$, and $F_0$ is closed since $F_0$ is a hypothesis of $R$), since $x_i$ appears in $F'$.
(If $x_i$ did not appear in $F'$, $\attacker(x_i)$ would have been removed by
$\elimattx$.)

If $R = F \rewrite F$, $R' \circ_{F_0} R = \attacker(p'_1)
\wedge \ldots \wedge \attacker(p'_k) \rewrite F$ has only closed
patterns in its hypotheses, and so has the clause $R''$ in $\simplify'(R' \circ_{F_0} R)$.

Otherwise, $R = \attacker(p_1) \wedge \ldots \wedge
\attacker(p_n) \rewrite F$, $F_0 = \attacker(p_i)$, and $p_i$
is a closed pattern.  We have $R' \circ_{F_0} R = \attacker(p'_1) \wedge \ldots
\wedge \attacker(p'_k) \wedge \attacker(p_1) \wedge \ldots \wedge
\attacker(p_{i-1}) \wedge \attacker(p_{i+1}) \wedge \ldots \wedge
\attacker(p_n) \rewrite F$, which has only closed
patterns in its hypotheses, and so has the clause $R''$ in $\simplify'(R' \circ_{F_0} R)$.
Moreover, $p'_1, \ldots, p'_k$ are disjoint
subterms of $p_i$, therefore the total size of $p'_1, \ldots, p'_k$ is
strictly smaller than the size of $p_i$. (If we had equality, 
$F'$ would be a variable; this variable would occur in the hypothesis
by definition of $\Bp$, so $R'$ would have been removed by $\elimtaut$.) 
Therefore the total
size of the patterns in the hypotheses strictly decreases. 
(The simplification function $\simplify'$ cannot increase this size.) 
This decrease proves termination.
\proofcomplete
\end{proof}

\newcommand{\h}{\mathrm{OneKey}}

From the previous results, we infer the termination of the algorithm
for tagged protocols, when $\params{\pk}$ and $\params{\host}$ have at
most one element. The general case can then be obtained as
in~\cite{Blanchet04e}: we define a function $\h$ which maps
all elements of $\params{\pk}$ and $\params{\host}$ to a single atomic
constant.  When $P_0$ is a tagged protocol, $\h(P_0)$ is a tagged
protocol in which $\params{\pk}$ and $\params{\host}$ are singletons.
We consider a ``less optimized algorithm'' in which elimination of 
duplicate hypotheses and of tautologies are performed only for
facts of the form $\attacker(x)$, elimination of redundant hypotheses
is not performed, and elimination of subsumed clauses
is performed only for eliminating the destructor clauses for 
$\checksign$ and $\nmrchecksign$.
We observe that the previous results still hold for the less
optimized algorithm, with the same proof, so this algorithm
terminates on $\h(P_0)$.
All resolution steps possible for the less optimized algorithm
applied to $P_0$ are possible for the less optimized algorithm
applied to $\h(P_0)$ as well (more patterns are unifiable, and the remaining
simplifications of the less optimized algorithm commute with
applications of $\h$).
Hence, the derivations from $\rset{P'_0,\rw}$ are mapped by $\h$ to
derivations from $\rset{\h(P'_0),\rw}$, which are finite, so
derivations from $\rset{P'_0,\rw}$ are also finite, so the less
optimized algorithm terminates on $P_0$. We can then show that the
original, fully optimized algorithm also terminates on $P_0$.  So we
finally obtain Proposition~\ref{prop:termination}.

\section{General Correspondences}\label{app:recentness}

In this appendix, we prove Theorem~\ref{th:recentcorresp}.  
For simplicity, we assume that the function applications at the
root of events are unary. 

\begin{lemma}\label{lem:diffsessid}
Let $P_0$ be a closed process and $P'_0 = \instrinj{P_0}$.
Let $Q$ be an $\rw$-adversary and $Q' = \instradv{Q}$. 
Assume that, in $P_0$, the arguments of events are function applications.
Let $f$ be a function symbol.  Assume that there is a single occurrence of
$\asevent(f(\_))$ in $P_0$ and this occurrence is
under a replication.  Consider any trace  $\trace = S_0, \env_0, \{ P'_0, Q'\}\rightarrow^* S',
\env', \pset'$. The multiset of session identifiers $\lambda$ of
events $\aseventInstr{\lambda}{f(\_)}$ executed in $\trace$ contains
no duplicates.  
\end{lemma}
\begin{proof}
Let us define the multiset $\clabel(P)$ by
$\clabel(\aseventInstr{\lambda}{f(M)}.P) = \{ \lambda \} \cup
\clabel(P)$ (for the given function symbol $f$),
$\clabel(\ReplInstr{i}{P}) = \emptyset$, and in all other cases,
$\clabel(P)$ is the union of the $\clabel(P')$ for all immediate
subprocesses $P'$ of $P$.
For a trace $\trace$, let $\clabel(\trace)$ be the set of session identifiers
$\lambda$ of events $\aseventInstr{\lambda}{f(\_)}$ executed in the trace $\cal
T$.

We show that, for each trace $\trace = S_0, \env_0, \{
P'_0, Q'\}\rightarrow^* S', \env', \pset'$, $\clabel(\trace) \cup \bigcup_{P \in \pset'}
\clabel(P) \cup S'$ contains no duplicates. 
The proof is by induction on the length of the trace.

For the empty trace $\trace = S_0, \env_0, \{ P'_0, Q'\}\rightarrow^* S_0,
\env_0, \{ P'_0, Q'\}$, $\clabel(\trace) = \emptyset$ and
$\clabel(P'_0)\cup \clabel(Q) = \emptyset$ by definition.

The reduction (Red Repl) moves at most one session identifier from $S'$ to $\bigcup_{P
\in \pset'} \clabel(P)$ (without introducing duplicates since there is
one occurrence of $\aseventInstr{\_}{f(\_)}$). The reduction (Red Event) moves at
most one session identifier from $\bigcup_{P \in \pset'} \clabel(P)$ to $\clabel({\cal
T})$. The other reductions can only remove session identifiers from $\bigcup_{P \in \pset'}
\clabel(P)$ (by removing subprocesses).  
\proofcomplete
\end{proof}

\newcommand{\stepf}{\step_{\mathrm{f}}}
\begin{lemma}\label{lem:moverun}
Let $P_0 = C[\asevent(f(M)).D[\asevent(\fbegin{f}(M,x).P]]$, where no
replication occurs in $D[\,]$ above the hole $[\,]$, and the variables
and names bound in $P_0$ are all pairwise distinct and distinct from
free names. Assume that, in $P_0$, the arguments of events are function 
applications, and that there is a single occurrence of
$\asevent(f(\_))$ and of $\asevent(\fbegin{f}(\_,\_))$ in $P_0$. 

Let $Q$ be an $\rw$-adversary and $Q' = \instradv{Q}$.
Let $P_0' = \instrinj{P_0}$.
Consider a trace of $P_0'$:
$\trace = S_0, \env_0, \pset_0  = \{ P_0', Q' \} \rightarrow^* S_{\stepf}, 
\env_{\stepf}, \pset_{\stepf}$.

Then there exists a function $\inject$ such that
a) if $\aseventInstr{\lambda}{\fbegin{f}(p, p')}$ is executed at step
$\step$ in $\trace$ for some $\lambda, p, p', \step$, 
then $\aseventInstr{\lambda}{f(p)}$ is executed
at step $\inject(\step)$ in $\trace$,
b) $\inject$ is injective, and 
c) if $\inject(\step)$ is defined, then $\inject(\step) < \step$.
\end{lemma}

\begin{proof}
We denote by $S_{\step}, \env_{\step}, \pset_{\step}$ the
configuration at the step $\step$ in the trace $\trace$.
Let
\begin{align*}
&S^1(\step) = \{ (\lambda,p) \mid \aseventInstr{\lambda}{f(p)}
\text{ is executed in the first $\step$ steps of $\trace$} \},\\
&S^2(\step) = \{ (\lambda,p) \mid
\aseventInstr{\lambda}{\fbegin{f}(p,p')}\text{ is executed in the first $\step$ steps of $\trace$}\}\\*
\begin{split}
&S^3(\step)= \{ (\lambda,p) \mid \aseventInstr{\lambda}{\fbegin{f}(M, M')}\text{ occurs not under }\aseventInstr{\lambda}{f(M)}\text{ in}\\*
&\phantom{S^3(\step)= \{} \pset_{\step}\text{ for }\env_{\step}(M) = p \}
\end{split}
\end{align*}
For each $\step$, we show that $S^2(\step) \cup S^3(\step) \subseteq S^1(\step)$.
\begin{itemize}

\item For $\step=0$, the sets $S^1(\step)$, $S^2(\step)$, and $S^3(\step)$ are empty.

\item If $S_\step, \env_\step, \pset_{\step} \rightarrow S_{\step+1},
\env_{\step+1}, \pset_{\step+1}$ using (Red Event) to execute
$\asevent(f(M), \ab \lambda)$, then the same $(\lambda, \ab \env_{\step+1}(M))$ 
is added to
$S^3(\step+1)$ and to $S^1(\step+1)$. Similarly, for (Red Event)
executing $\asevent(\fbegin{f}(M,M'), \ab \lambda)$, a pair
$(\lambda, \ab \env_{\step+1}(M))$
is moved from $S^3(\step)$ to $S^2(\step+1)$. These changes preserve
the desired inclusion.

\item Otherwise, if $S_\step, \env_\step, \pset_{\step} \rightarrow
S_{\step+1}, \env_{\step+1}, \pset_{\step+1}$, then $S^1(\step+1)
= S^1(\step)$, $S^2(\step+1) = S^2(\step)$, and $S^3(\step+1)
\subseteq S^3(\step)$ (because some subprocesses may be removed by
the reduction).

\end{itemize}
In particular, $S^2(\stepf) \subseteq S^1(\stepf)$. By
Lemma~\ref{lem:diffsessid}, there is a bijection $\phi_1$ from the session
labels $\lambda$ of executed $\aseventInstr{\lambda}{f(\_)}$ events in 
$\trace$  to the steps at which these events are executed in $\trace$, 
and similarly $\phi_2$ for $\aseventInstr{\_}{\fbegin{f}(\_,\_)}$ events.
Let $\inject = \phi_1 \circ \phi_2^{-1}$.
\begin{itemize}
\item
If $\aseventInstr{\lambda}{\fbegin{f}(p, p')}$ is executed
at step $\step$, $(\lambda, p) \in S^2(\stepf) \subseteq
S^1(\stepf)$, so $\aseventInstr{\lambda}{f(p)}$ is executed at a certain step
$\step'$.  So $\phi_2(\lambda) =
\step$ and $\phi_1(\lambda) = \step'$, so $\inject(\step)$ is defined
and $\step' = \inject(\step)$. 

\item
Since $\phi_1$ and $\phi_2^{-1}$ are injective, $\inject$ is injective.

\item
If $\inject(\step)$ is defined, the event $\aseventInstr{\lambda}{\fbegin{f}(\sigma y,
\sigma x)}$ is executed at step $\step$ by (Red Event). 
So $(\lambda, \sigma y) \in S^3(\step)$, where $\pset_\step$ corresponds
to the state just before the event $\aseventInstr{\lambda}{\fbegin{f}(\sigma y,
\sigma x)}$ is executed. 
Hence $(\lambda, \sigma y) \in S^1(\step)$ since
$S^2(\step) \cup S^3(\step) \subseteq S^1(\step)$.
So $\aseventInstr{\lambda}{f(\sigma y)}$ is executed at step $\step' < \step$.
We have $\phi_2(\lambda) = \step$ and $\phi_1(\lambda) = \step'$, so
$\inject(\step) = \step' < \step$.
\proofcomplete

\end{itemize}
\end{proof}

\begin{proof}[of Theorem~\ref{th:recentcorresp}]
For each non-empty $\overline{jk}$, when $\injopt_{\overline{jk}} = \inj$,
let $f_{\overline{jk}}$ be the root function symbol of 
$p_{\overline{jk}}$.  We
consider a modified process $P_1$ built from $P_0$ as follows.  For
each $\overline{jk}$ such that $\injopt_{\overline{jk}} = \inj$
and $\asevent(f_{\overline{jk}}(M))$ occurs in $P_0$, we
add another event
$\asevent(\fbegin{f_{\overline{jk}}}(M, x_{\overline{jk}}))$ just
under the definition of variable $x_{\overline{jk}}$ if
$x_{\overline{jk}}$ is defined under
$\asevent(f_{\overline{jk}}(M))$ and just under
$\asevent(f_{\overline{jk}}(M))$ otherwise.
Let $P'_1 = \instrinj{P_1}$.
The process $P'_1$ is built from $P'_0$ as follows.  For
each $\overline{jk}$ such that $\injopt_{\overline{jk}} = \inj$
and $\aseventInstr{i}{f_{\overline{jk}}(M)}$ occurs in $P'_0$, we
add another event
$\aseventInstr{i}{\fbegin{f_{\overline{jk}}}(M, x_{\overline{jk}})}$ just
under the definition of variable $x_{\overline{jk}}$ if
$x_{\overline{jk}}$ is defined under
$\aseventInstr{i}{f_{\overline{jk}}(M)}$ and just under
$\aseventInstr{i}{f_{\overline{jk}}(M)}$ otherwise.
(When $\injopt_{\overline{jk}} = \inj$, $x_{\overline{jk}} \in
\dom(\rho_{\overline{jrk}})$ where $\rho_{\overline{jrk}}$ is the
environment added as argument of $\pasbegin$ facts in the clauses, so
$x_{\overline{jk}}$ is defined either above
$\aseventInstr{i}{f_{\overline{jk}}(M)}$ or under
$\aseventInstr{i}{f_{\overline{jk}}(M)}$ without any replication
between the event and the definition of $x_{\overline{jk}}$, since the
domain of the environment given as argument to $\pasbegin$ is set at 
replications by substituting $\square$ and not modified later.)
We will show that
$P'_1$ satisfies the desired correspondence. It is then clear that
$P'_0$ also satisfies it.

The clauses $\rset{P'_1, \rw}$ can be obtained from $\rsetp{P'_0,\rw}$
by replacing all facts $\pasbegin(p, \ab \rho)$ with 
\[\pasbegin(p, \ab i)
\wedge \bigwedge_{\overline{jk} \text{ such that }p = f_{\overline{jk}}(p')
\text{ and }x_{\overline{jk}} \in \dom(\rho)}
\pasbegin(\fbegin{f_{\overline{jk}}}(p', \ab \rho(x_{\overline{jk}})), \ab i)\]
for some $i$,
and adding clauses that conclude
$\pasend(\fbegin{f_{\overline{jk}}}(\ldots), \ab \ldots)$.

The clauses in $\solve{P'_1, \rw}$ can be obtained in the same way
from $\solvep{P'_0,\rw}$.
So we can define a function $\checkrc'$ like $\checkrc$ with an additional argument $(x_{\overline{jk}j'k'})_{\overline{jk}j'k'}$
by adding $(x_{jk\overline{jk}j'k'})_{\overline{jk}j'k'}$ in the arguments of recursive call of Point~V2.3
and replacing Point~V2.1 with
$\solve{P'_1, \rw}(\pasend(p,i)) \subseteq \{ H \wedge
\bigwedge_{k=1}^{l_j} \pasbegin(\arg_{jrk}, i_{jrk}) \rewrite
\pasend(\sigma_{jr} p'_j, i_{jr})$ for some $H$, $j\in \{1, \ldots,
m\}$, $r$, $i_{jrk}$, and $(\rho_{jrk}, i_{jr}) \in \envset_{jk}$ for all $k \}$
where
$\arg_{jrk} = \sigma_{jr} p_{jk}$ if $\injopt_{jk} \neq \inj$,
and
$\arg_{jrk} = \fbegin{f_{jk}}(\sigma_{jr} p', \rho_{jrk}(x_{jk}))$ 
if $\injopt_{jk} = \inj$ and $p_{jk} = f_{jk}(p')$.
When $\checkrc(q, \ab (\envset_{\overline{jk}})_{\overline{jk}})$ is true,
$\checkrc'(q, \ab (\envset_{\overline{jk}})_{\overline{jk}}, \ab 
(x_{\overline{jk}})_{\overline{jk}})$
is also true.

Let $Q$ be an $\rw$-adversary and $Q' = \instradv{Q}$.
Let $\env_0$ such that $\env_0(a) = a[\,]$ for all $a \in \dom(\env_0)$
and $\fn(P'_1) \cup \rw \subseteq \dom(\env_0)$.
Let us now consider a trace of $P'_1$, $\trace = S_0, \env_0, \{ P'_1, Q' \} 
\rightarrow^* S', \env', \pset'$. 

By Lemma~\ref{lem:moverun}, 
for each non-empty $\overline{jk}$ such that $\injopt_{\overline{jk}} = \inj$,
there exists a function $\inject_{\overline{jk}}$ such that
a) if $\aseventInstr{\lambda}{\fbegin{f_{\overline{jk}}}(p, p')}$ is
executed at step $\step$ in $\trace$ for some $\lambda, p, p',
\step$, then $\aseventInstr{\lambda}{f_{\overline{jk}}(p)}$ is
executed at step $\inject_{\overline{jk}}(\step)$ in $\trace$,
b) $\inject_{\overline{jk}}$ is injective, and 
c) if $\inject_{\overline{jk}}(\step)$ is defined, then
$\inject_{\overline{jk}}(\step) < \step$.

\newcommand{\psicirc}{\psi^{\circ}}
\newcommand{\sigmar}{\sigma_u} %sigma unify
\newcommand{\sigmaq}{\sigma''}
When $\psi_{\overline{jk}}$ is a family of functions from steps
to steps in a trace, we define $\psicirc_{\overline{jk}}$ as follows:
\begin{itemize}

\item
  $\psicirc_{\epsilon}(\step) = \step$ for all $\step$; 

\item
  for all $\overline{jk}$, for all $j$ and $k$,
  $\psicirc_{\overline{jk}jk} = \inject_{\overline{jk}jk} \circ
  \psi_{\overline{jk}jk} \circ \psicirc_{\overline{jk}}$ when
  $\injopt_{\overline{jk}jk} = \inj$ and $\psicirc_{\overline{jk}jk} =
  \psi_{\overline{jk}jk} \circ \psicirc_{\overline{jk}}$ otherwise.

\end{itemize}

We show that, if $\checkrc'(q', \ab
(\envset_{\overline{jk}})_{\overline{jk}}, \ab
(x_{\overline{jk}})_{\overline{jk}})$ is true for 
\begin{align*}
&q' =
\pasevent(p) \Rightarrow \ab \sor_{j=1}^{m} \left(\pasevent(p'_j)
  \rightsquigarrow \sand_{k = 1}^{l_j} \injopt_{jk}q'_{jk}\right)\\
&q'_{\overline{jk}} = \pasevent(p_{\overline{jk}})
\ab \rightsquigarrow \ab \sor_{j=1}^{m_{\overline{jk}}} 
\sand_{k=1}^{l_{\overline{jk}j}} \injopt_{\overline{jk}jk}
q'_{\overline{jk}jk}
\end{align*}
then there exists a function
$\psi_{\overline{jk}}$ for each $\overline{jk}$ such that
\begin{enumerate}

\item[P1.]
  For all $\step$, if the event
  $\aseventInstr{\lambda_{\epsilon}}{\sigma p}$ is
  executed at step $\step$ in $\trace$, then there exist $\sigmaq$ and
  $J = (j_{\overline{k}})_{\overline{k}}$ such that $\sigmaq
  p'_{j_{\epsilon}} = \sigma p$ and, for all non-empty $\overline{k}$,
  $\psicirc_{\indp{\overline{k}}{J}}(\step)$ is defined and
  $\aseventInstr{\lambda_{\overline{k}}}{\sigmaq p_{\indp{\overline{k}}{J}}}$ is
  executed at step $\psicirc_{\indp{\overline{k}}{J}}(\step)$ in $\trace$.

\item[P2.]  
  For all non-empty $\overline{jk}$, if $\injopt_{\overline{jk}} = \inj$ and
  $\psi_{\overline{jk}}(\step)$ is defined, then
  $\aseventInstr{\lambda'_1}{p''_1}$ is executed at step $\step$ in
  $\trace$,
  $\aseventInstr{\lambda'_2}{\fbegin{f_{\overline{jk}}}(p''_2, \theta
  \rho(x_{\overline{jk}})) }$ is executed at step $\psi_{\overline{jk}}(\step)$ in
  $\trace$, and $\theta i = \lambda'_1$ for some $p''_1$, $p''_2$,
  $\lambda'_1$, $\lambda'_2$, $\theta$, and $(\rho, i) \in
  \envset_{\overline{jk}}$, where $f_{\overline{jk}}$ is the
  root function symbol of $p_{\overline{jk}}$.  (This property is used
  for proving injectivity and recentness.)

\item[P3.] 
For all non-empty $\overline{jk}$, if $\psi_{\overline{jk}}(\step)$ is defined, then
$\psi_{\overline{jk}}(\step) \leq \step$.

\end{enumerate}
The proof is by induction on $q'$.
\begin{itemize}

\item If $q' = \pasevent(p)$ (that is, $m=1$, $l_1 = 0$, and $p_1 = p$),
we define $j_{\epsilon} = 1$ and $\sigmaq = \sigma$, so that $\sigmaq
  p'_{j_{\epsilon}} = \sigma p$. All other conditions hold trivially,
since there is no non-empty $\overline{k}$.

\item Otherwise, we define $\psi_{jk}$ as follows.

Using Point~V2.1, by Theorem~\ref{thbeginend3}, $P'_1$ satisfies the correspondence
\begin{equation}
\pasevent(p, i) \Rightarrow \sor_{j=1..m, r} \left(\pasevent(\sigma_{jr} p'_j, i_{jr}) \rightsquigarrow 
\sand_{k = 1}^{l_j} \pasevent(\arg_{jrk}, i_{jrk})\right)
\label{corr:noninj}
\end{equation}
against $\rw$-adversaries.

Assume that $\asevent(\sigma p, \lambda)$ is executed at step $\step$
in $\trace$ for some substitution $\sigma$. Let us consider the trace
$\trace$ cut just after step $\step$.  By
Correspondence~\eqref{corr:noninj}, there exist $\sigma'$, 
$j \in \{ 1, \ldots, m\}$, and $r$ such that $\sigma' \sigma_{jr} p'_j = \sigma
p$, $\sigma' i_{jr} = \sigma \lambda = \lambda$, and for $k \in \{1,
\ldots, l_j\}$, there exists $\lambda_k$ such that
$\asevent(\sigma' \arg_{jrk}, \lambda_k)$ is executed in the
trace $\trace$ cut after step $\step$.  So the event $\asevent(\sigma'
\arg_{jrk}, \lambda_k)$ is executed at step $\step_k \leq \step$ in $\trace$.
In this case, we define $\psi_{jk}(\step) = \step_k$ and $r(\step) = r$.

If $\injopt_{jk} = \inj$, then $\asevent(\sigma' \sigma_{jr} p_{jk},
\lambda_k)$ is executed as step $\inject_{jk}(\psi_{jk}(\step)) =
\psicirc_{jk}(\step)$.

If $\injopt_{jk} \neq \inj$, then $\arg_{jrk} = \sigma_{jr} p_{jk}$,
so $\asevent(\sigma' \sigma_{jr} p_{jk}, \lambda_k)$
is executed as step $\psi_{jk}(\step) = \psicirc_{jk}(\step)$.

By construction, if $\psi_{jk}(\step)$ is defined, then $\psi_{jk}(\step) \leq \step$.

When $\injopt_{\overline{jk}} = \inj$, we let $f_{\overline{jk}}$ be
the root function symbol of $p_{\overline{jk}}$.

By Point~V2.3, for all $j,r,k$, 
$\checkrc'(\sigma_{jr} q'_{jk}, 
\ab (\envset_{jk\overline{jk}})_{\overline{jk}}, 
\ab (x_{jk\overline{jk}})_{\overline{jk}})$
is true.
So, by induction hypothesis, there exist functions $\psi_{jrk,\overline{jk}}$ such that
\begin{itemize}

\item
  For all $\step_k$, if the event
  $\aseventInstr{\lambda_k}{\sigma' \sigma_{jr} p_{jk}}$ is
  executed at step $\step_k$ in $\trace$, then there exist $\sigmaq_{jrk}$ and
  $J = (j_{jrk,\overline{k}})_{\overline{k}}$ such that $\sigmaq_{jrk}
  \sigma_{jr} p_{jk} = \sigma' \sigma_{jr} p_{jk}$ and, for all non-empty $\overline{k}$,
  $\psicirc_{jrk,\indp{\overline{k}}{J}}(\step_k)$ is defined and
  $\aseventInstr{\lambda_{k\overline{k}}}{\sigmaq_{jrk} \sigma_{jr} p_{jk\indp{\overline{k}}{J}}}$ is
  executed at step $\psicirc_{jrk,\indp{\overline{k}}{J}}(\step_k)$ in $\trace$.

\item
  For all non-empty $\overline{jk}$, if $\injopt_{jk\overline{jk}} = \inj$ and
  $\psi_{jrk,\overline{jk}}(\step)$ is defined, then
  $\aseventInstr{\lambda'_1}{p''_1}$ is executed at step $\step$ in
  $\trace$,
  $\asevent(\fbegin{f_{jk\overline{jk}}}(p''_2, \theta
  \rho(x_{jk\overline{jk}})), \ab \lambda'_2)$ is executed at step $\psi_{jrk,\overline{jk}}(\step)$ in
  $\trace$ and $\theta i = \lambda'_1$ for some $p''_1$, $p''_2$,
  $\lambda'_1$, $\lambda'_2$, $\theta$, and $(\rho, i) \in
  \envset_{jk\overline{jk}}$.

\item 
For all non-empty $\overline{jk}$, if $\psi_{jrk,\overline{jk}}(\step)$ is defined, then
$\psi_{jrk,\overline{jk}}(\step) \leq \step$.

\end{itemize}

We define $\psi_{jk\overline{jk}}(\step) = \psi_{jrk,\overline{jk}}(\step)$ for $r = r(\step)$.
Then we have $\psicirc_{jk\overline{jk}}(\step) = \psicirc_{jrk, \overline{jk}}(\psicirc_{jk}(\step))$ for $r = r(\step)$.

Therefore, for all $\step$, if $\asevent(\sigma p, \lambda)$ is
executed at step $\step$ in $\trace$, then 
\begin{itemize}

\item there exist $\sigma'$, $J_{\epsilon} = (j_{\overline{k}})_{\overline{k}}$,
  and $r$ such that $j_{\epsilon} = j \in \{ 1, \ldots, m\}$, $j_{\overline{k}}$
is undefined for all $\overline{k} \neq \epsilon$, $\sigma'
  \sigma_{jr} p'_j = \sigma p$, and, for all $k$,
  $\psicirc_{\indp{k}{J_{\epsilon}}}(\step)$ is defined and $\asevent(\sigma'
  \sigma_{jr} p_{\indp{k}{J_{\epsilon}}}, \lambda_k)$ is executed as step
  $\psicirc_{\indp{k}{J_{\epsilon}}}(\step)$;

\item for all $k$, there exist $\sigma''_{jrk}$ and
$J_k = (j_{k\overline{k}})_{k\overline{k}}$ such that $\sigma''_{jrk}
\sigma_{jr} p_{jk} = \sigma' \sigma_{jr} p_{jk}$ and, for all
non-empty $\overline{k}$, $\psicirc_{\indp{k\overline{k}}{J_k}}(\step)$ is
defined and $\aseventInstr{\lambda_{k\overline{k}}}{\sigmaq_{jrk} \sigma_{jr} p_{\indp{k\overline{k}}{J_k}}}$ is
  executed at step $\psicirc_{\indp{k\overline{k}}{J_k}}(\step)$ in $\trace$.

\end{itemize}
We define a family of indices $J$ by merging $J_{\epsilon}$ and $J_k$
for all $k$, that is, $J = (j_{\overline{k}})_{\overline{k}}$.
Therefore, in order to obtain P1, it is enough to find a substitution
$\sigmaq$ such that $\sigmaq p'_j = \sigma' \sigma_{jr} p'_j$,
$\sigmaq p_{jk} = \sigma' \sigma_{jr} p_{jk}$, and $\sigmaq
p_{jk\overline{jk}} = \sigmaq_{jrk} \sigma_{jr} p_{jk\overline{jk}}$
for all non-empty $\overline{jk}$.
Let us define $\sigmar$ as follows:
\begin{itemize}

\item For all $x \in \fv(\sigma_{jr} p'_j) \cup \bigcup_k \fv(\sigma_{jr} p_{jk})$, $\sigmar x = \sigma' x$.

\item For all $k$, for all $x \in \fv(\sigma_{jr} q'_{jk}) \setminus \fv(\sigma_{jr} p_{jk})$, $\sigmar x = \sigmaq_{jrk} x$.

\end{itemize}
By Point~V2.2, these sets of variables are disjoint, so $\sigmar$ is well defined. Let $\sigmaq = \sigmar \sigma_{jr}$.

We have $\sigmaq p'_j = \sigmar \sigma_{jr} p'_j = \sigma' \sigma_{jr} p'_j$
and $\sigmaq p_{jk} = \sigmar \sigma_{jr} p_{jk} = \sigma' \sigma_{jr} p_{jk}$.
Since $\sigmaq q'_{jk} = \sigmar \sigma_{jr} q'_{jk}$, 
we just have to show that
$\sigmar \sigma_{jr} q'_{jk} = \sigmaq_{jrk} \sigma_{jr} q'_{jk}$.
We have $\sigmar \sigma_{jr} p_{jk} = \sigma' \sigma_{jr} p_{jk} = \sigmaq_{jrk} \sigma_{jr} p_{jk}$.
Therefore, if $x \in \fv(\sigma_{jr} p_{jk})$, then $\sigmar x = \sigmaq_{jrk} x$.\footnote{This property does not hold in the presence of an equational theory (see Section~\ref{sec:DiffieHellman}). In that case, we conclude by the additional hypothesis mentioned in Section~\ref{sec:DiffieHellman}.}
Hence, for all $x \in \fv(\sigma_{jr} q'_{jk})$, $\sigmar x = \sigmaq_{jrk} x$,
which proves that $\sigmar \sigma_{jr} q'_{jk} = \sigmaq_{jrk} \sigma_{jr} q'_{jk}$. Hence we obtain P1.

If $\injopt_{jk} = \inj$ and $\psi_{jk}(\step)$ is defined, then
$\aseventInstr{\lambda'_1}{p''_1} = \asevent(\sigma p, \ab \lambda)$ 
is executed at step $\step$ in
$\trace$, $\asevent(\fbegin{f_{jk}}(p''_2, \ab \theta
\rho(x_{jk})), \ab \lambda'_2) = \asevent(\sigma' \arg_{jrk}, \ab \lambda_k)$ 
is executed at step $\psi_{jk}(\step)$ in $\trace$, and
$\theta i = \lambda'_1$ for some $p''_1 = \sigma p$, $p''_2$, $\lambda'_1 =
\lambda$, $\lambda'_2 = \lambda_k$, $\theta = \sigma'$, 
and $(\rho, i) = (\rho_{jrk}, i_{jr}) \in \envset_{jk}$.
  For all non-empty $\overline{jk}$, if $\injopt_{jk\overline{jk}} = \inj$ and
  $\psi_{jk\overline{jk}}(\step)$ is defined, then
  $\aseventInstr{\lambda'_1}{p''_1}$ is executed at step $\step$ in
  $\trace$,
  $\aseventInstr{\lambda'_2}{\fbegin{f_{jk\overline{jk}}}(p''_2, \theta
  \rho(x_{jk\overline{jk}})) }$ is executed at step $\psi_{jk\overline{jk}}(\step)$ in
  $\trace$, and $\theta i = \lambda'_1$ for some $p''_1$, $p''_2$,
  $\lambda'_1$, $\lambda'_2$, $\theta$, and $(\rho, i) \in
  \envset_{jk\overline{jk}}$.
So we obtain P2.

If $\psi_{jk}(\step)$ is defined, then $\psi_{jk}(\step) \leq \step$.
For all non-empty $\overline{jk}$, if $\psi_{jk\overline{jk}}(\step)$
is defined, then $\psi_{jk\overline{jk}}(\step) \leq \step$.
Therefore, we have P3.

\end{itemize}
Let $q = \pasevent(p) \Rightarrow \sor_{j=1}^{m} \left(\pasevent(p'_j)
  \rightsquigarrow \sand_{k = 1}^{l_j} \injopt_{jk}q_{jk}\right)$,
and $q_{\overline{jk}} = \pasevent(p_{\overline{jk}})
\rightsquigarrow \sor_{j=1}^{m_{\overline{jk}}} 
\sand_{k=1}^{l_{\overline{jk}j}} \injopt_{\overline{jk}jk}
q_{\overline{jk}jk}$.
By Hypothesis~H1, $\checkrc'(q, \ab
(\envset_{\overline{jk}})_{\overline{jk}}, \ab
(x_{\overline{jk}})_{\overline{jk}})$ is true, so there exists a
function $\psi_{\overline{jk}}$ for each $\overline{jk}$ such that P1, P2,
and P3 are satisfied.
Let $\phi_{\overline{jk}} = \psicirc_{\overline{jk}}$.
\begin{itemize}
\item
  By P1, for all $\step$, if the event
  $\aseventInstr{\lambda_{\epsilon}}{\sigma p}$ is
  executed at step $\step$ in $\trace$, then there exist $\sigma'$ and
  $J = (j_{\overline{k}})_{\overline{k}}$ such that $\sigma'
  p'_{j_{\epsilon}} = \sigma p$ and, for all non-empty $\overline{k}$,
  $\phi_{\indp{\overline{k}}{J}}(\step)$ is defined and
  $\aseventInstr{\lambda_{\overline{k}}}{\sigma' p_{\indp{\overline{k}}{J}}}$ is
  executed at step $\phi_{\indp{\overline{k}}{J}}(\step)$ in $\trace$.

Let us show recentness. Suppose that $\injopt_{\indp{\overline{k}}{J}} = \inj$.
We show that the runtimes of $\session(\lambda_{\cutlast{\overline{k}}})$ and 
$\session(\lambda_{\overline{k}})$ overlap.
We have $\phi_{\indp{\overline{k}}{J}}(\step) = \inject_{\indp{\overline{k}}{J}}(\psi_{\indp{\overline{k}}{J}}(\phi_{\indp{\cutlast{\overline{k}}}{J}}(\step)))$.
Let $\step_1 = \phi_{\indp{\cutlast{\overline{k}}}{J}}(\step)$.
Then $\psi_{\indp{\overline{k}}{J}}(\step_1)$
is defined. 
Hence, by P2, $e_1 = \asevent(p''_1, \lambda'_1)$ is executed at step
$\step_1$ in $\trace$,
$e_2 = \asevent(\fbegin{f_{\indp{\overline{k}}{J}}}(p''_2, \ab \theta
\rho(x_{\indp{\overline{k}}{J}})), \ab \lambda'_2)$ is executed at step
$\step_2 = \psi_{\indp{\overline{k}}{J}}(\step_1)$ in
$\trace$ by a reduction
$S_{\step_2}, \env_{\step_2}, \pset_{\step_2} 
\rightarrow S_{\step_2+1}, \env_{\step_2+1}, \pset_{\step_2+1} $,
and $\theta i = \lambda'_1$ for some $p''_1$, $p''_2$,
$\lambda'_1$, $\lambda'_2$, $\theta$, and $(\rho, i) \in
\envset_{\indp{\overline{k}}{J}}$.
Since the event $\aseventInstr{\lambda_{\cutlast{\overline{k}}}}{\sigma'
p_{\indp{\cutlast{\overline{k}}}{J}}}$ is also executed at step $\step_1 =
\phi_{\indp{\cutlast{\overline{k}}}{J}}(\step)$, we have $\lambda'_1 =
\lambda_{\cutlast{\overline{k}}}$.
By the properties of $\inject_{\indp{\overline{k}}{J}}$, 
$\aseventInstr{\lambda'_2}{f_{\indp{\overline{k}}{J}}(p''_2)}$ is executed at
step $\inject_{\indp{\overline{k}}{J}}(\step_2) = 
\phi_{\indp{\overline{k}}{J}}(\step)$.
Moreover, $\aseventInstr{\lambda_{\overline{k}}}{\sigma' p_{\indp{\overline{k}}{J}}}$ is also executed at step $\phi_{\indp{\overline{k}}{J}}(\step)$,
so $\lambda'_2 = \lambda_{\overline{k}}$.
 
By Hypothesis~H2, $\rho(x_{\indp{\overline{k}}{J}})\{ \lambda/i \}$
does not unify with $\rho(x_{\indp{\overline{k}}{J}})\{ \lambda'/i \}$ 
when $\lambda \neq \lambda'$, so $i$ occurs in $\rho(x_{\indp{\overline{k}}{J}})$,
so $\lambda_{\cutlast{\overline{k}}} = \lambda'_1 = \theta i$
occurs in $\theta \rho(x_{\indp{\overline{k}}{J}})$, so $\lambda_{\cutlast{\overline{k}}}$ occurs in $e_2$.

So $e_2$ is executed after the rule
$S, \env,
\pset \cup \{ \ReplInstr{i'}{P'} \} \rightarrow S \setminus \{ \lambda_{\cutlast{\overline{k}}} \}, E,
\pset \cup \{ P'\{ \lambda_{\cutlast{\overline{k}}}/i'\}, \ReplInstr{i'}{P'} \}$ in $\trace$.
Indeed, since $\lambda_{\cutlast{\overline{k}}}$ occurs in the event
$e_2$ executed at step $\step_2$,
$\lambda_{\cutlast{\overline{k}}} \in \clabel'(\env_{\step_2}) \cup
\clabel'(\pset_{\step_2})$ where $\clabel'(\pset)$
(resp. $\clabel'(\env)$) is the set of session identifiers $\lambda$ that
occur in $\pset$ (resp. $\env$). Moreover, $\clabel'(\env_0) \cup
\clabel'(\{ P'_1, Q'\}) = \emptyset$, and the only rule that
increases $\clabel'(\env) \cup \clabel'(\pset)$ is $S, \env, \pset
\cup \{ \ReplInstr{i}{P'} \} \rightarrow S \setminus \{ \lambda \}, \env,
\pset \cup \{ P'\{\lambda/i\}, \ReplInstr{i}{P'} \}$, which adds
$\lambda$ to $\clabel'(\env) \cup \clabel'(\pset)$. Therefore,
$e_2$ is executed after the beginning of the runtime of
$\session(\lambda_{\cutlast{\overline{k}}})$.

Moreover, $e_2$ is executed at step $\step_2 =
\psi_{\indp{\overline{k}}{J}}(\step_1)$ and $e_1$ is executed at step
$\step_1$ in $\trace$, with $\psi_{\indp{\overline{k}}{J}}(\step_1) \leq
\step_1$, so $e_2$ is executed before $e_1 = \aseventInstr{\lambda_{\cutlast{\overline{k}}}}{p''_1}$.

So $e_2 = \aseventInstr{\lambda_{\overline{k}}}{\fbegin{f_{\indp{\overline{k}}{J}}}(p''_2,
\theta \rho(x_{\indp{\overline{k}}{J}}))}$ is executed during
the runtime of $\session(\lambda_{\cutlast{\overline{k}}})$, therefore the runtimes of
$\session(\lambda_{\cutlast{\overline{k}}})$ and $\session(\lambda_{\overline{k}})$ overlap.

\item Let us show that, for all non-empty $\overline{jk}$, 
if $\injopt_{\overline{jk}} = \inj$, then 
$\psi_{\overline{jk}}$ is injective. 
Let $\step_1$ and $\step_2$ such that $\psi_{\overline{jk}}(\step_1) = \psi_{\overline{jk}}(\step_2)$.
By P2, $\aseventInstr{\lambda'_1}{p''_1}$ is executed at step $\step_1$ in
  $\trace$,
  $\aseventInstr{\lambda'_3}{\fbegin{f_{\overline{jk}}}(p''_3, \theta_1
  \rho_1(x_{\overline{jk}})) }$ is executed at step $\psi_{\overline{jk}}(\step_1)$ in
  $\trace$, and $\theta_1 i_1 = \lambda'_1$ for some $p''_1$, $p''_3$,
  $\lambda'_1$, $\lambda'_3$, $\theta_1$, and $(\rho_1, i_1) \in
  \envset_{\overline{jk}}$.
Also by P2,
$\aseventInstr{\lambda'_2}{p''_2}$ is executed at step $\step_2$ in
  $\trace$,
  $\aseventInstr{\lambda'_4}{\fbegin{f_{\overline{jk}}}(p''_4, \theta_2
  \rho_2(x_{\overline{jk}})) }$ is executed at step $\psi_{\overline{jk}}(\step_2)$ in
  $\trace$, and $\theta_2 i_2 = \lambda'_2$ for some $p''_1$, $p''_4$,
  $\lambda'_2$, $\lambda'_4$, $\theta_2$, and $(\rho_2, i_2) \in
  \envset_{\overline{jk}}$.
Since $\psi_{\overline{jk}}(\step_1) = \psi_{\overline{jk}}(\step_2)$,
$\theta_1 \rho_1(x_{\overline{jk}}) = \theta_2
  \rho_2(x_{\overline{jk}})$.
By Hypothesis~H2, this implies that $\theta_1 i_1 = \theta_2 i_2$,
so $\lambda'_1 = \lambda'_2$. By Lemma~\ref{lem:diffsessid}, 
$\step_1 = \step_2$, which proves the
injectivity of $\psi_{\overline{jk}}$.

\item Let us show that, for all non-empty $\overline{jk}$, 
if $\injopt_{\overline{jk}} = \inj$, 
then $\phi_{\overline{jk}}$ is injective, by induction on
the length of the sequence of indices $\overline{jk}$.

For all $j$ and $k$, if $\injopt_{jk} = \inj$, then $\phi_{jk}$ is injective
since $\inject_{jk}$, $\psi_{jk}$, and $\phi_{\epsilon}$ are injective.

For all non-empty $\overline{jk}$, for all $j$ and $k$, 
if $\injopt_{\overline{jk}jk} =\inj$, then, by hypothesis,
$\injopt_{\overline{jk}} = \inj$, so, by induction hypothesis,
$\phi_{\overline{jk}}$ is injective. The functions
$\inject_{\overline{jk}jk}$ and $\psi_{\overline{jk}jk}$ are
injective, so $\phi_{\overline{jk}jk}$ is also injective.

\item 
For all $\overline{jk}$, for all $j$ and $k$,
if $\phi_{\overline{jk}jk}(\step)$ is defined, then
$\phi_{\overline{jk}}(\step)$ is defined, and
$\phi_{\overline{jk}jk}(\step) \leq \phi_{\overline{jk}}(\step)$, since
$\inject_{\overline{jk}jk}(\step'') \leq \step''$ and
$\psi_{\overline{jk}jk}(\step') \leq \step'$ by P3, when they are
defined.

In particular, for all $j$ and $k$, if $\phi_{jk}(\step)$ is defined,
then $\phi_{jk}(\step) \leq \phi_{\epsilon}(\step) = \step$.

\end{itemize}
This concludes the proof of the desired recent correspondence.
\proofcomplete
\end{proof}

\begin{proof}[of Proposition~\ref{prop:recinjprop}]
We have $\checkrc(q, (\envset_{\overline{jk}})_{\overline{jk}})$ with 
$\envset_{jk} = \{ (\rho_{jrk}, i_{jr}) \mid r \in \{ 1, \ab \ldots, \ab n_j \}\}$,
because the first item implies V2.1, 
V2.2 holds trivially since $q_{jk}$ reduces to $\pasevent(p_{jk})$, and 
V2.3 also holds since $q_{jk}$ reduces to $\pasevent(p_{jk})$, 
so $\checkrc(\sigma_{jr}q_{jk}, \ab (\envset_{jk\overline{jk}})_{\overline{jk}})$ 
holds by V1.
The second item implies H2. So we have the result by
Theorem~\ref{th:recentcorresp}.  
\proofcomplete
\end{proof}

\end{document}